\documentclass[aps,prab,reprint,amsmath,amssymb,superscriptaddress,longbibliography,floatfix]{revtex4-2}

\usepackage{graphicx}
\usepackage{xcolor}
\usepackage{amsmath,amssymb}
\usepackage{siunitx}
\usepackage{booktabs}
\usepackage{array}
\usepackage{multirow}
\usepackage{makecell}
\usepackage{subcaption}
\usepackage{tabularx}
\usepackage{tikz}
\usetikzlibrary{positioning,arrows.meta,calc,fit}
\usepackage{listings}
\usepackage[colorlinks=true,linkcolor=blue,citecolor=magenta,urlcolor=blue]{hyperref}

\newcommand{\dd}{\mathop{}\!\mathrm{d}}

\newcommand{\safeincludegraphics}[2][]{%
\IfFileExists{#2}{\includegraphics[#1]{#2}}{%
\fbox{\parbox[c][0.22\textheight][c]{0.75\linewidth}{\centering Missing figure file:\\ \texttt{\detokenize{#2}}}}}%
}

\begin{document}

\title{Thermal and electromechanical response of ultra-thin carbon-strip polarimeter targets in relativistic bunched beams}

\author{F. Rathmann}
\affiliation{Brookhaven National Laboratory, Upton, New York 11973, USA}
\author{O. Eyser}
\affiliation{Brookhaven National Laboratory, Upton, New York 11973, USA}
\author{M. Sangroula}
\affiliation{Brookhaven National Laboratory, Upton, New York 11973, USA}
\author{P. Shanmuganathan}
\affiliation{Brookhaven National Laboratory, Upton, New York 11973, USA}
\author{V. Shmakova}
\affiliation{Brookhaven National Laboratory, Upton, New York 11973, USA}

\date{\today}

\begin{abstract}
		Thin carbon-strip targets provide fast relative hadron beam polarimetry, but their response in intense relativistic bunched beams is not governed by local stopping-power heating alone. We develop a coupled response model that combines beam-target overlap, secondary-electron escape, retained heat, target motion, transient heat transport, RF-induced strip-end heating, beam-induced forces, resistance changes, and slack-strip deformation. RHIC target observations constrain the relevant motion, force, and nonlocal-heating scales and show that target survival depends on both beam-center heating and electromagnetic boundary conditions near the strip ends. Applying the model to Booster, AGS, RHIC, and EIC proton and $^{3}\mathrm{He}$ cases shows that the RHIC proton lifetime scale is reproduced at the order-of-magnitude level, while the RHIC target-holder fin results require the additional RF/end-heating mechanism. For EIC proton flattop operation, carbon-strip polarimetry may remain viable only with reduced dwell time, sufficient detector acceptance, and suppression of RF-induced end heating. For cooled-emittance $^{3}\mathrm{He}$, the calculated sublimation-loss scale is far beyond a straightforward RHIC-like carbon-strip extrapolation. Conventional carbon strips are therefore unlikely to remain viable for the most demanding EIC light-ion cases without major changes in target motion, target technology, or diagnostic concept.
\end{abstract}

\maketitle
\tableofcontents

\section{Introduction}
\label{sec:introduction}

Precise hadron beam polarimetry is essential for the RHIC spin program and for the future Electron-Ion Collider (EIC)~\cite{osti_1765663,ABDULKHALEK2022122447}. At RHIC, fast proton-carbon polarimeters based on elastic scattering in the Coulomb-nuclear-interference region provided high-rate relative polarization measurements and transverse beam-profile information using thin carbon micro-ribbon targets~\cite{Bravar2005ProtonPolarimetryRHIC,Lozowski2008CarbonMicroRibbon,Steski2014CarbonMicroRibbons,Steski2018TargetLifetime}. The polarization-profile information obtained from such scans is important for relating polarimeter measurements to the polarization relevant for colliding-beam experiments~\cite{Fischer2012PolarizationProfiles}. The polarized hydrogen jet target supplied the absolute polarization normalization, and its EIC upgrade path has been analyzed in the context of high-precision proton beam polarimetry~\cite{Rathmann2026HJET}. This combination of an absolute hydrogen-jet polarimeter and fast carbon polarimeters established the operational basis for high-energy hadron polarimetry with thin internal targets.

The EIC places more demanding requirements on hadron polarimetry. The physics program requires precise polarization measurements for polarized protons and light ions, with systematic uncertainties at the percent level~\cite{ABDULKHALEK2022122447,Rathmann2026HJET}. In addition to absolute polarization calibration, fast relative polarimeters are needed to monitor polarization loss during stores, provide bunch-resolved or time-resolved information, and measure transverse beam and polarization profiles. The broader polarized-ion program also anticipates operation with polarized deuteron, $^{3}\mathrm{He}$, and eventually $^{6,7}\mathrm{Li}$ beams~\cite{Atoian2025PolarizedIonBeamsEIC}. These requirements make the response and survivability of thin carbon targets a central technical issue.

\begin{figure}[t]
	\centering
	\includegraphics[width=0.9\columnwidth]{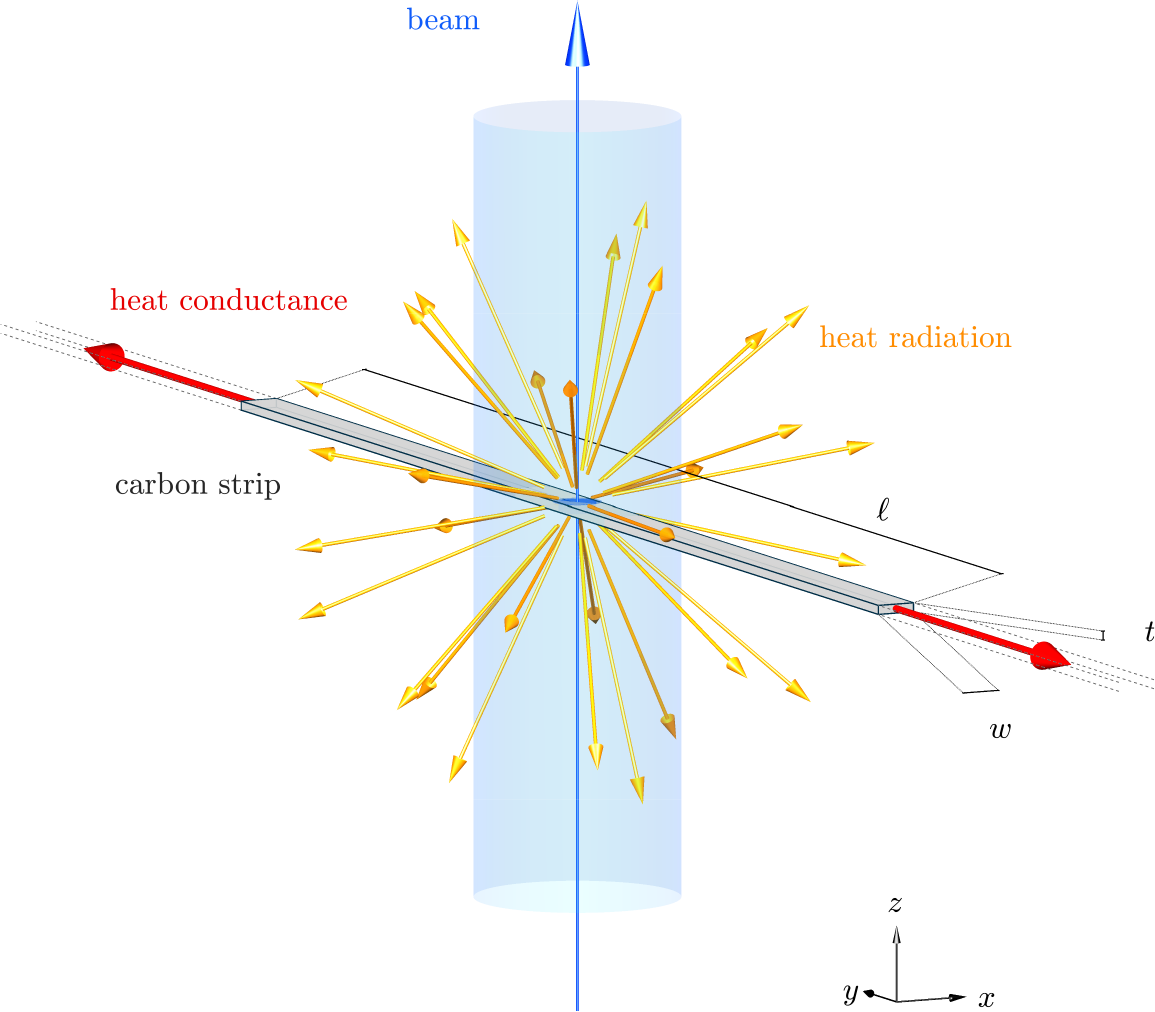}
	\caption{ Sketch of the carbon-strip target geometry. The strip has length $\ell$, width $w$, and thickness $t$. The beam travels along the $z$ direction and intercepts the $\ell\times w$ surface of the strip. Red arrows indicate heat conduction along the strip toward the supports, and orange arrows indicate radiative heat loss from the heated region. }
	\label{fig:carbon_strip_geometry}
\end{figure}

The basic target geometry considered in this paper is shown in Fig.~\ref{fig:carbon_strip_geometry}. A thin carbon strip of length $\ell$, width $w$, and thickness $t$ is suspended between two supports and moved transversely through the circulating beam. Direct beam interception produces a localized heat source near the beam-crossing point. Heat is conducted along the strip toward the supports, and thermal radiation is emitted from the exposed surfaces. This static geometry defines the reference coordinate system, heat-flow picture, and boundary conditions used throughout the model.

RHIC operation showed, however, that a carbon micro-ribbon in a bunched high-intensity beam is not only a passive absorber with one localized thermal source. Target lifetime studies and operational observations indicated beam-induced deformation, localized strip-end glow, sensitivity to RF conditions, changes in target resistance after beam exposure, and improved lifetime when nearby metal structures were added~\cite{Steski2014CarbonMicroRibbons,Steski2018TargetLifetime}. These observations motivate a model that includes not only direct stopping-power heating, but also target motion, secondary-electron escape, retained heat, wakefield coupling, RF-induced end heating, resistance evolution, electrostatic forces, and mechanical response of a slack strip. The observed time sequence of the target response is discussed in Sec.~\ref{sec:observed-beam-target-response-rhic}.

The purpose of this paper is to organize these effects into a common physical model for ultra-thin carbon-strip targets in relativistic bunched beams. The model is then used to compare RHIC reference operation with EIC proton and light-ion benchmark cases, and with Booster and AGS cases relevant to the injector chain. Particular attention is given to the EIC cooled-emittance $^{3}\mathrm{He}$ case, which provides the most demanding thermal scenario among the cases considered here.

The paper is organized as follows. Section~\ref{sec:observed-beam-target-response-rhic} summarizes the RHIC beam-target observations that motivate the model. Section~\ref{sec:carbon-strip-target-description} defines the carbon-strip geometry, material parameters, electrical state, and mechanical state. Section~\ref{sec:beam-target-geometry} constructs the beam-target overlap and heat-source terms. Section~\ref{sec:thermal-response-carbon-strip} develops the thermal response model. Section~\ref{sec:wakefield-coupling-rf-heating-resistance} discusses wakefield coupling, RF-induced end heating, and resistance changes. Section~\ref{sec:beam-induced-forces} estimates beam-induced forces and mechanical response. Section~\ref{sec:carbon-estimates-protons_light-ions} compares temperature estimates, sublimation margins, and operating trends for Booster, AGS, RHIC, and EIC proton and $^{3}\mathrm{He}$ cases. Section~\ref{sec:conclusion} summarizes the implications for EIC proton operation, cooled-emittance $^{3}\mathrm{He}$ operation, and possible future light-ion scenarios.

\section{Observed beam--target response at RHIC}
\label{sec:observed-beam-target-response-rhic}

\begin{figure*}[t]
	\centering
	\includegraphics[width=\textwidth]{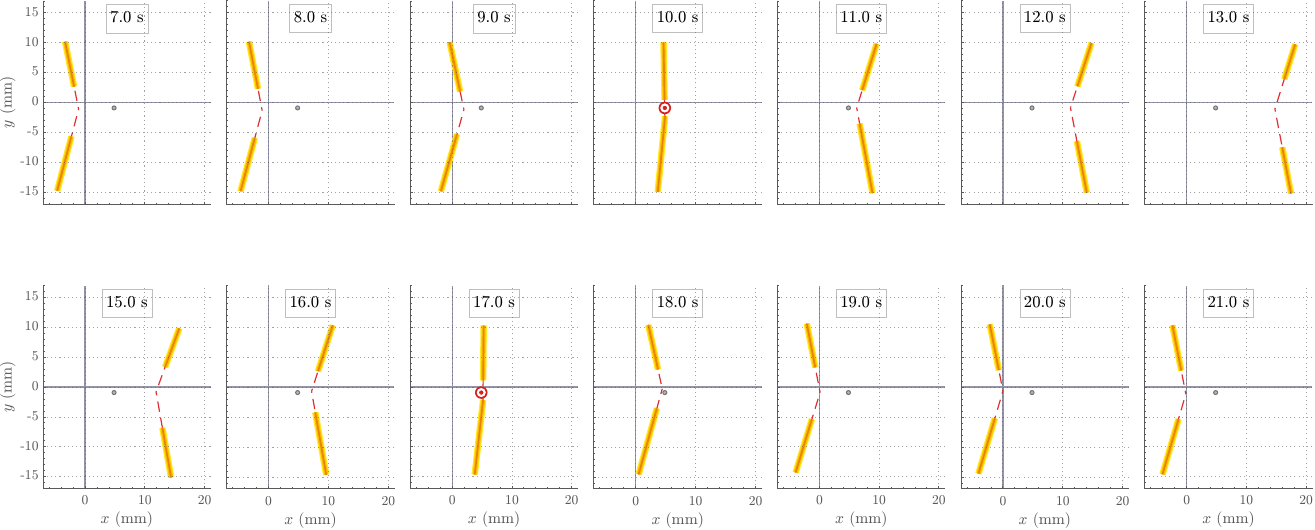}
\caption{Schematic digest of the observed carbon-strip response during a horizontal target scan through the beam, reconstructed from selected video frames. Yellow segments indicate visibly glowing strip ends, red dashed lines indicate the inferred continuation of the strip, and beam-position markers identify the fixed beam location (grey dots) and the frames with visible beam interception (red dots). Time stamps are relative to an arbitrary start time of the selected image sequence.}
	\label{fig:carbon_strip_scan_schematic}
\end{figure*}

This section collects the RHIC observations that the model must reproduce or parameterize. The relevant observations are target deformation toward the beam, center heating during direct beam interception, carbon-strip ends glowing before interception, suppression of the end glow when the \SI{200}{\mega\hertz} cavity voltage was reduced, mitigation by metal fins, resistance changes after beam exposure, and twisted or multiply rotated carbon micro-ribbons. A schematic digest of one observed scan sequence is shown in Fig.~\ref{fig:carbon_strip_scan_schematic}. The figure is not intended as a geometric reconstruction of the full strip, but as a compact representation of the visible features in the video frames: the glowing upper and lower strip ends, the inferred continuation between them, and the frames in which the beam spot was visible. The two brightest beam-interception frames are at $t=\SI{10}{s}$ and $t=\SI{17}{s}$; the dull-grey markers denote the (invisible) beam position on the other frames.

\begin{table*}[t]
	\centering
	\small
	\renewcommand{\arraystretch}{1.15}
	\setlength{\tabcolsep}{5pt}
	\begin{tabular}{llp{5cm}}
		\hline
		\textbf{Quantity} & \textbf{Effective value / observation} & \textbf{Use in model} \\
		\hline
		
		Visible strip length scale 
		& $\ell_{\mathrm{vis}} \simeq \SI{25}{\milli\meter}$ 
		& video-to-mm calibration \\
		
		Beam-visible interval 
		& $\Delta t_{\mathrm{beam}} \sim \SI{1.5}{\second}$
		& constrains scan timing near beam \\
		
		Representative scan speed near beam 
		& $v_t \sim \SI{4}{\milli\meter\per\second}$ 
		& input for moving-source estimates \\
		
		Projected strip displacement 
		& $\Delta x_{\mathrm{proj}} \sim \SI{3}{\milli\meter}$ 
		& force-scale estimate \\
		
		Visible end-glow length 
		& upper/lower regions of order \SIrange{6}{12}{\milli\meter} 
		& RF-induced end-heating scale \\
		
		End glow before direct beam hit 
		& observed 
		& requires nonlocal heat source \\
		
		Bright beam spot at strip center 
		& observed only in selected frames 
		& direct beam-interception marker \\
		
		Apparent reversal of strip bend 
		& observed during passage through beam 
		& constraint on force symmetry \\
		
		Out-of-plane motion 
		& not measured 
		& projected quantities are lower-bound estimates \\
		
		\hline
	\end{tabular}
	\caption{Effective observational quantities extracted from the video sequence summarized in Fig.~\ref{fig:carbon_strip_scan_schematic}. The values are approximate and should be interpreted as projected quantities in the camera plane. They are used to anchor the thermal, electrical, and mechanical estimates developed in the following sections.}
	\label{tab:video_observation_parameters}
\end{table*}

RHIC target-lifetime studies showed that target survival depends sensitively on beam intensity, beam size, and the electromagnetic environment near the target \cite{Steski2014CarbonMicroRibbons,Steski2018TargetLifetime}. Video observations during target scans showed target deformation toward the beam and associated this deformation with the electric field of the beam. The scan sequence summarized in Fig.~\ref{fig:carbon_strip_scan_schematic} illustrates several features that are important for the present model. The strip is visibly displaced toward the beam region during the scan, the upper and lower strip ends glow before the beam spot is visible on the target, and the apparent strip geometry changes as the target passes through the beam. In the beam-hit frames, the center region appears as bright white light, while the glowing end regions remain spatially separated from the beam spot.

The schematic sequence in Fig.~\ref{fig:carbon_strip_scan_schematic} can be reduced to a small set of effective observational quantities, summarized in Table~\ref{tab:video_observation_parameters}, that are useful for the later estimates. These quantities should not be interpreted as precision measurements of the full three-dimensional strip shape. They are projected, camera-plane observables extracted from the selected video sequence and are used only to set the scale of the thermal, electrical, and mechanical response model.

The end-glow observation is important because it requires a heat source other than direct beam energy deposition at the beam-spot location. In Fig.~\ref{fig:carbon_strip_scan_schematic}, the yellow end segments are already visible in frames before the direct beam-hit frames, showing that heating is not confined to the instantaneous beam-intercept point. Steski et al. interpreted this as beam-induced high-frequency electric fields moving electrons along the target, causing resistive heating. They also reported that lowering the voltage of the RHIC \SI{200}{\mega\hertz} RF system caused the end glow to cease, and that rounded metal fins near the target ends reduced the induced electric field and improved target lifetime \cite{Steski2018TargetLifetime}.

The electrical state of the strip is also history dependent. Steski et al. reported that unexposed carbon micro-ribbons had resistances of order \SIrange{200}{800}{\mega\ohm}, whereas beam-exposed targets had resistances of order \SI{1}{\mega\ohm}, possibly due to graphitization from extreme beam heating \cite{Steski2014CarbonMicroRibbons}. This observation motivates treating the resistance along the strip as nonuniform and time/history dependent.

\section{Carbon-strip target model}
\label{sec:carbon-strip-target-description}

This section defines the geometry, material properties, electrical state, and mechanical state of the carbon-strip target used throughout the model. The goal is to provide a machine-independent description of the target that can be combined with different beam conditions in later sections.

The target consists of a thin carbon strip of length $\ell$, width $w$, and thickness $t$, suspended between two supports (see Fig.\,\ref{fig:carbon_strip_geometry}). The coordinate system is defined such that the strip extends along the $y$ direction ($\ell$), the transverse width is along $x$ ($w$), and the thickness is along $z$ ($t$).

\subsection{Geometric and material definition}

The strip geometry is characterized by
\begin{equation}
	A = w t,
\end{equation}
where $A$ is the cross-sectional area relevant for heat conduction and mechanical tension.

The intrinsic material properties of the carbon film are described by the density $\rho$, specific heat $c_p$, thermal conductivity $\kappa$, and emissivity $\varepsilon$. These quantities depend on the deposition method, microstructure, and thermal history, and are treated as input parameters to the model.

The strip is connected at both ends to a support structure, which serves as a thermal and electrical boundary. In the simplest approximation, the strip ends are held at a fixed temperature
\begin{equation}
	T(y = \pm \ell/2) = T_0,
\end{equation}
with $T_0$ representing the holder temperature. More refined descriptions may include finite thermal and electrical contact resistances at the attachment points.

\subsection{Electrical state}

The electrical properties of the strip are described by an effective resistance $R_{\text{strip}}$, which may evolve under beam exposure due to structural modification, graphitization, or damage. In general, the resistance may be spatially non-uniform,
\begin{equation}
	R_{\text{strip}} \;\rightarrow\; R(y),
\end{equation}
and time dependent.

This effective resistance enters directly in the description of RF-induced heating and must therefore be treated as a state variable rather than a fixed material constant.

\subsection{Mechanical state}

The mechanical response of the strip is governed by its effective axial tension,
\begin{equation}
	T_{\text{eff}} = \sigma_{\text{eff}} A,
\end{equation}
where $\sigma_{\text{eff}}$ is the effective stress in the mounted strip.

It is important to distinguish $\sigma_{\text{eff}}$ from the intrinsic residual film stress $\sigma_{\text{res}}$ of the deposited carbon layer. The effective stress can differ significantly from $\sigma_{\text{res}}$ due to cutting, handling, mounting, thermal cycling, and beam-induced modifications. As a result, $\sigma_{\text{eff}}$ is treated as an independent model parameter.

\subsection{Target parameters}
\label{sec:tgt-parameters}

The main target input parameters used by the model are summarized in Table~\ref{tab:target_parameters}. The table is intended as a compact model-input table, not as a complete material-property compilation. It contains three types of quantities: geometric inputs, intrinsic material properties, and effective state variables of the mounted strip. In particular, $R(y)$ and $\sigma_{\mathrm{eff}}$ are not intrinsic material constants; they describe the electrical and mechanical state of the mounted target and may change after handling, twisting, heating, graphitization, or beam exposure. Detailed material-property ranges and references are given in Appendix~\ref{sec:carbon-material-properties}. Specific parameter sets for individual machines or operating scenarios are defined separately.

\begin{table*}[htb]
	\centering
	\small
	\renewcommand{\arraystretch}{1.2}
	\setlength{\tabcolsep}{6pt}
	\begin{tabular}{lll}
		\hline
		\textbf{Parameter} & \textbf{Symbol} & \textbf{Typical range / comment} \\
		\hline
		Length & $\ell$ & geometry dependent \\
		Width & $w$ & geometry dependent \\
		Thickness & $t$ & geometry dependent \\
		Density & $\rho$ & $\sim 1700$--$2200\ \mathrm{kg\,m^{-3}}$ \\
		Specific heat & $c_p$ & $700$--$900\ \mathrm{J\,kg^{-1}\,K^{-1}}$ \\
		Thermal conductivity & $\kappa$ & $0.1$--$10\ \mathrm{W\,m^{-1}\,K^{-1}}$ \\
		Emissivity & $\epsilon$ & $0.75$--$0.95$ \\
		Strip resistance & $R_{\mathrm{strip}}$ & state dependent (graphitization, damage) \\
		Resistance profile & $R(y)$ & may be non-uniform \\
		Residual film stress & $\sigma_{\mathrm{res}}$ & $10^2$--$10^3\ \mathrm{MPa}$ (process dependent) \\
		Effective stress & $\sigma_{\mathrm{eff}}$ & model parameter (mounted strip) \\
		Holder temperature & $T_0$ & boundary condition \\
		\hline
	\end{tabular}
	\caption{Compact target-input parameters used in the carbon-strip model.}
	\label{tab:target_parameters}
\end{table*}

We do not assign a unique sublimation temperature as an intrinsic target parameter. For a nanometer-scale carbon strip in vacuum, carbon loss depends on the temperature-dependent recession speed and on the exposure time, not on a sharp material threshold. This distinction is important because graphite vapor-pressure data, graphite sublimation-rate measurements, and accelerator carbon-foil experience do not define a single universal survival temperature~\cite{Brewer1948GraphiteVaporPressure,HainesTsai2002GraphiteSublimation,Tahir2014CarbonFoilStripper}. The quantitative treatment is therefore deferred to Sec.~\ref{sec:carbon-sublimation-vacuum-mass-loss}, where the relevant damage measure is the fractional removed thickness, $\Delta h_{\mathrm{sub}}/t$.

\section{Beam-target interaction and heat-source model}
\label{sec:beam-target-geometry}

This section defines the beam and target quantities needed to construct the source terms for the carbon-strip response model. The goal is to separate the beam--target interaction geometry from the later thermal and mechanical response calculations. We first define the transverse beam distribution, current density, particle flux, and benchmark beam-parameter scenarios. We then describe the geometric beam--target overlap, convert the particle flux into nominal energy loss in the carbon strip, and distinguish this energy loss from the heat retained locally after secondary-electron escape. Finally, target motion is introduced as the step that makes the overlap and source terms time dependent.

\subsection{Beam optics, current density, and particle flux}
\label{sec:beam-optics-current-density-particle-flux}

The beam-size definitions used here follow the standard accelerator-optics relation between normalized and geometric rms emittance~\cite{lee2004accelerator}. The notation for the RHIC and EIC benchmark quantities follows Ref.~\cite{Rathmann2026HJET}, whose Table~II provides the reference beam-parameter set used below. In the present section, these quantities are used only to construct the transverse particle distribution, the time-averaged current density, and the particle flux density at the carbon-strip location; the conversion of this flux into energy loss and retained heat is treated in the following subsections.

The transverse beam distribution at the target location is described in terms of the local beta functions, normalized rms emittances, bunch population, bunch number, and revolution frequency. For a particle species with rest mass $m$, charge state $Z$, and kinetic energy $T$, the relativistic factors are
\begin{equation}
	\gamma = 1 + \frac{T}{m c^2}\,,
	\qquad
	\beta = \sqrt{1-\frac{1}{\gamma^2}}\,.
	\label{eq:beam-relativistic-factors}
\end{equation}
If the tabulated beam energy is given as total energy rather than kinetic energy, the first relation is replaced by $\gamma = E_{\text{tot}}/(m c^2)$. In the following, $\beta$ and $\gamma$ denote the relativistic beam parameters, while $\beta_x$ and $\beta_y$ denote the horizontal and vertical lattice beta functions at the target location.

For normalized rms emittances $\varepsilon^n_x$ and $\varepsilon^n_y$, the geometric rms emittances are
\begin{equation}
	\varepsilon_{x,y} = \frac{\varepsilon^n_{x,y}}{\beta\gamma}\,,
	\label{eq:geometric-emittance}
\end{equation}
The corresponding rms beam sizes at the target location are
\begin{equation}
	\sigma_{x,y} =
	\sqrt{\beta_{x,y} \varepsilon_{x,y}}
	=
	\sqrt{\frac{\beta_{x,y} \varepsilon^n_{x,y}}{\beta\gamma}}\,,
	\label{eq:beam-sigmas}
\end{equation}
All quantities in Eq.~\eqref{eq:beam-sigmas} are rms quantities. When $\varepsilon^n_{x,y}$ is given in $\si{\micro\meter}$, it is understood as $\si{\micro\meter\,rad}$, i.e., $1~\si{\micro\meter}=10^{-6}~\si{\meter\,rad}$.

The normalized transverse particle density of the circulating beam is approximated by a two-dimensional Gaussian,
\begin{equation}
	f(x,y)
	=
	\frac{1}{2\pi\sigma_x\sigma_y}
	\exp\!\left[
	-\frac{x^2}{2\sigma_x^2}
	-\frac{y^2}{2\sigma_y^2}
	\right]\,,
	\label{eq:beam-transverse-density}
\end{equation}
with
\begin{equation}
	\int_{-\infty}^{+\infty}
	\int_{-\infty}^{+\infty}
	f(x,y)\,\dd x\,\dd y
	=
	1\,.
	\label{eq:beam-density-normalization}
\end{equation}
Here $x$ and $y$ denote the horizontal and vertical coordinates transverse to the beam direction. The function $f(x,y)$ has units of inverse area and describes the probability density for particles in the transverse plane.

For $N_b$ bunches, $N_p$ particles per bunch, charge state $Z$, and revolution frequency $f_{\text{rev}}$, the time-averaged beam current is
\begin{equation}
	I_{\text{avg}}
	=
	N_b\,N_p\,Z e\,f_{\text{rev}}\,,
	\label{eq:average-current}
\end{equation}
where $e$ is the elementary charge. The corresponding time-averaged transverse current density is
\begin{equation}
	j(x,y)
	=
	I_{\text{avg}}\,f(x,y)
	=
	\frac{I_{\text{avg}}}{2\pi\sigma_x\sigma_y}
	\exp\!\left[
	-\frac{x^2}{2\sigma_x^2}
	-\frac{y^2}{2\sigma_y^2}
	\right]\,.
	\label{eq:current-density}
\end{equation}
The particle flux density, expressed as particles per unit area per unit time, is obtained by dividing the current density by the particle charge $Ze$,
\begin{equation}
	\Phi(x,y)
	=
	\frac{j(x,y)}{Ze}
	=
	\Phi_0
	\exp\!\left[
	-\frac{x^2}{2\sigma_x^2}
	-\frac{y^2}{2\sigma_y^2}
	\right]\,,
	\label{eq:particle-flux}
\end{equation}
with peak particle flux
\begin{equation}
	\Phi_0
	=
	\frac{I_{\text{avg}}}{Ze\,2\pi\sigma_x\sigma_y}
	=
	\frac{N_b\,N_p\,f_{\text{rev}}}{2\pi\sigma_x\sigma_y}\,.
	\label{eq:peak-particle-flux}
\end{equation}
This form is useful because the heat-source calculation depends on the number of particles crossing the carbon strip, while the beam-current specification is naturally expressed in terms of transported charge.

For later comparison of beam envelopes, the 95\% transverse size of a two-dimensional Gaussian beam is defined by the ellipse
\begin{equation}
	\frac{x^2}{\sigma_x^2}
	+
	\frac{y^2}{\sigma_y^2}
	=
	-\;2\ln(1-0.95)
	=
	5.991\,,
	\label{eq:beam-95-ellipse}
\end{equation}
so that the corresponding semi-axes are
\begin{equation}
	\sigma_{x,y}^{95}
	=
	\sqrt{5.991}\,\sigma_{x,y}
	\simeq
	2.45\,\sigma_{x,y}\,,
	\label{eq:beam-95-sizes}
\end{equation}
The factor $\sqrt{5.991}$ is the standard rms-to-95\% conversion used for Gaussian transverse beam envelopes~\cite{lee2004accelerator,Rathmann2026HJET}. These quantities are used only as convenient beam-envelope measures. The source terms themselves are constructed from the full Gaussian flux density in Eq.~\eqref{eq:particle-flux}.

\subsection{RHIC and EIC beam parameters}
\label{sec:benchmark-beam-parameter-scenarios}

The RHIC and EIC benchmark beam scenarios used in the present model are summarized in Table~\ref{tab:beam_parameters_rhic_eic}. The table is intended as an input table for constructing the beam--target interaction geometry, not as a complete machine-parameter summary. It collects the quantities needed to calculate the transverse beam size, bunch structure, average current, current density, and particle flux density at the carbon-strip location.

The benchmark values follow the notation and parameter set of Table~II in Ref.~\cite{Rathmann2026HJET}. The broader EIC accelerator and polarimetry context is described in the EIC Yellow Report~\cite{ABDULKHALEK2022122447}. In the present work, RHIC flattop, EIC injection, and EIC flattop are treated as fixed reference cases. Booster and AGS operation, by contrast, involve beam parameters that evolve during the acceleration cycle. These cases are therefore treated separately in Sec.~\ref{sec:carbon-estimates-protons_light-ions}, using representative points along the ramp rather than single machine columns in the present table.

	\begin{table*}[p]
		\centering
		\scriptsize
		\renewcommand{\arraystretch}{1.15}
		\setlength{\tabcolsep}{3.5pt}
		\resizebox{\linewidth}{!}{%
			\begin{tabular}{
					lll
					S[table-format=4.6e2]
					S[table-format=4.6e2]
					S[table-format=4.6e2]
					S[table-format=4.6e2]
					S[table-format=4.6e2]
				}
				\hline
				\textbf{Quantity} &
				\textbf{Symbol / definition} &
				\textbf{Unit} &
				\multicolumn{1}{c}{\textbf{RHIC}} &
				\multicolumn{1}{c}{\textbf{EIC}} &
				\multicolumn{1}{c}{\textbf{EIC}} &
				\multicolumn{1}{c}{\textbf{EIC}} &
				\multicolumn{1}{c}{\textbf{EIC}} \\
				& & &
				\multicolumn{1}{c}{\textbf{flattop}} &
				\multicolumn{1}{c}{\textbf{injection}} &
				\multicolumn{1}{c}{\textbf{injection}} &
				\multicolumn{1}{c}{\textbf{flattop}} &
				\multicolumn{1}{c}{\textbf{flattop}} \\
				& & &
				\multicolumn{1}{c}{\textbf{reference}} &
				\multicolumn{1}{c}{\textbf{before cooling}} &
				\multicolumn{1}{c}{\textbf{after cooling}} &
				\multicolumn{1}{c}{\textbf{large emittance}} &
				\multicolumn{1}{c}{\textbf{cooled emittance}} \\
				\hline
				Location & -- & -- &
				\multicolumn{1}{c}{IP12 pC target} &
				\multicolumn{1}{c}{IP4 polarimetry} &
				\multicolumn{1}{c}{IP4 polarimetry} &
				\multicolumn{1}{c}{IP4 polarimetry} &
				\multicolumn{1}{c}{IP4 polarimetry} \\
				Total beam energy & $E_{\text{beam}}$ & \si{GeV} & 250 & 23.5 & 23.5 & 275 & 275 \\
				Relativistic velocity factor & $\beta$ & $1$ & 0.999993 & 0.999203 & 0.999203 & 0.999994 & 0.999994 \\
				Lorentz factor & $\gamma$ & $1$ & 266.447 & 25.046 & 25.046 & 293.092 & 293.092 \\
				Particles per bunch & $N_p$ & $10^{10}$ & 20 & 27.6 & 27.6 & 6.9 & 6.9 \\
				Bunch charge & $Q_b=N_p Z e$ & \si{nC} & 32.0435 & 44.2201 & 44.2201 & 11.055 & 11.055 \\
				Number of bunches & $N_b$ & $1$ & 120 & 290 & 290 & 1160 & 1160 \\
				Circumference & $C$ & \si{m} & 3834 & 3833.85 & 3833.85 & 3834 & 3834 \\
				Bunch length rms & $\sigma_L$ & \si{m} & 0.55 & 0.24 & 0.24 & 0.06 & 0.06 \\
				Temporal bunch width rms & $\sigma_t=\sigma_L/(\beta c)$ & \si{ns} & 1.83462 & 0.801193 & 0.801193 & 0.20014 & 0.20014 \\
				Peak current per bunch & $I_b^{\text{pk}}=Q_b/(\sqrt{2\pi}\sigma_t)$ & \si{A} & 6.96796 & 22.0187 & 22.0187 & 22.0362 & 22.0362 \\
				Revolution time & $\tau_{\text{rev}}=C/(\beta c)$ & \si{\micro s} & 12.7889 & 12.7986 & 12.7986 & 12.7889 & 12.7889 \\
				Revolution frequency & $f_{\text{rev}}=1/\tau_{\text{rev}}$ & \si{kHz} & 78.1926 & 78.1338 & 78.1338 & 78.1927 & 78.1927 \\
				Bunch spacing & $\tau_b=\tau_{\text{rev}}/N_b$ & \si{ns} & 106.574 & 44.1329 & 44.1329 & 11.0249 & 11.0249 \\
				Bunch frequency & $f_b=1/\tau_b$ & \si{MHz} & 9.38311 & 22.6588 & 22.6588 & 90.7035 & 90.7035 \\
				Average beam current & $I_{\text{avg}}=N_b N_p Z e f_{\text{rev}}$ & \si{A} & 0.300668 & 1.00197 & 1.00197 & 1.00273 & 1.00273 \\
				\hline
				Normalized rms emittance, horizontal & $\varepsilon_x^n$ & \si{\micro m} & 2.5 & 3.3 & 3.3 & 3.3 & 3.3 \\
				Normalized rms emittance, vertical & $\varepsilon_y^n$ & \si{\micro m} & 2.5 & 3.3 & 0.3 & 3.3 & 0.3 \\
				Normalized average rms emittance & $\varepsilon_{\text{avg}}^n=\sqrt{\varepsilon_x^n\varepsilon_y^n}$ & \si{\micro m} & 2.5 & 3.3 & 0.994987 & 3.3 & 0.994987 \\
				Beta function, horizontal & $\beta_x$ & \si{m} & 25.41 & 93.6 & 93.6 & 230.3 & 230.3 \\
				Beta function, vertical & $\beta_y$ & \si{m} & 28.43 & 39.59 & 39.59 & 69.9 & 69.9 \\
				Average beta function & $\beta_{\text{avg}}=\sqrt{\beta_x\beta_y}$ & \si{m} & 26.8776 & 60.8738 & 60.8738 & 126.878 & 126.878 \\
				Transverse rms beam size, horizontal & $\sigma_x=\sqrt{\beta_x\varepsilon_x^n/(\beta\gamma)}$ & \si{mm} & 0.488279 & 3.51317 & 3.51317 & 1.61029 & 1.61029 \\
				Transverse rms beam size, vertical & $\sigma_y=\sqrt{\beta_y\varepsilon_y^n/(\beta\gamma)}$ & \si{mm} & 0.516481 & 2.28483 & 0.688901 & 0.887146 & 0.267484 \\
				Transverse 95\% beam size, horizontal & $\sigma_x^{95}=\sqrt{5.991}\,\sigma_x$ & \si{mm} & 1.19534 & 8.60044 & 8.60044 & 3.94208 & 3.94208 \\
				Transverse 95\% beam size, vertical & $\sigma_y^{95}=\sqrt{5.991}\,\sigma_y$ & \si{mm} & 1.26438 & 5.5934 & 1.68647 & 2.17179 & 0.654818 \\
				Radial rms beam size & $\sigma_r=\sqrt{\sigma_x\sigma_y}$ & \si{mm} & 0.502182 & 2.83319 & 1.55571 & 1.19522 & 0.656298 \\
				Radial 95\% beam size & $\sigma_r^{95}=\sqrt{5.991}\,\sigma_r$ & \si{mm} & 1.22937 & 6.93583 & 3.80846 & 2.92598 & 1.60666 \\
				Peak particle flux & $\Phi_0=I_{\text{avg}}/(Z e\,2\pi\sigma_x\sigma_y)$ & \si{m^{-2}.s^{-1}} & 1.18433e24 & 1.23998e23 & 4.11255e23 & 6.97261e23 & 2.31255e24 \\
				\hline
			\end{tabular}%
		}
		\caption{Reference RHIC and EIC proton beam-parameter input cases used to construct the beam--target interaction geometry for the carbon-strip model. The RHIC optics correspond to the pC target location in IP12, with $\beta_x=\SI{25.41}{m}$ and $\beta_y=\SI{28.43}{m}$ received from V.\ Shmakova on 12.04.2025. The EIC injection after-cooling case uses the injection reference parameters from Ref.~\cite{Rathmann2026HJET}. The EIC injection before-cooling case is a provisional sensitivity case using the same injection optics and bunch structure, but with $\varepsilon_y^n$ set equal to $\varepsilon_x^n$. The EIC flattop large-emittance case carries the same normalized emittances used for the injection before-cooling case, while the flattop cooled-emittance case uses $\varepsilon_x^n=\SI{3.3}{\micro m}$ and $\varepsilon_y^n=\SI{0.3}{\micro m}$. The EIC flattop optics use the IP4 polarimetry-section values from H.\ Lovelace; the pC and HJET locations are assumed sufficiently close for the present benchmark. Derived quantities are recalculated from the definitions shown in the table. The peak particle flux $\Phi_0$ is added for the present carbon-strip model and denotes the time-averaged particle flux density at the center of the transverse Gaussian beam; for the proton cases listed here $Z=1$. AGS and Booster ramp scenarios are
			treated separately in Sec.~\ref{sec:carbon-estimates-protons_light-ions}.}
		\label{tab:beam_parameters_rhic_eic}
	\end{table*}

\begin{figure*}[htb]
	\centering
	
	\begin{subfigure}{0.49\textwidth}
		\centering
		\safeincludegraphics[width=\textwidth]{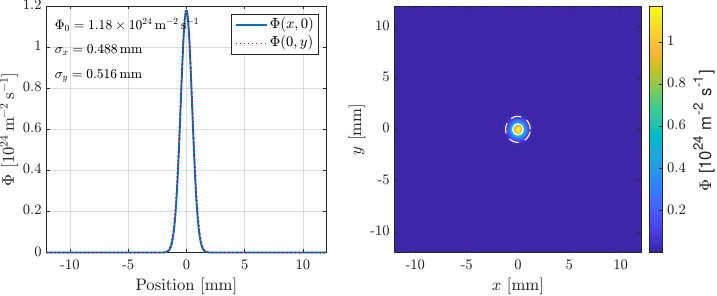}
		\caption{RHIC flattop.}
		\label{fig:beam-flux-rhic-flattop}
	\end{subfigure}
	\hfill
	\begin{subfigure}{0.49\textwidth}
		\centering
		\safeincludegraphics[width=\textwidth]{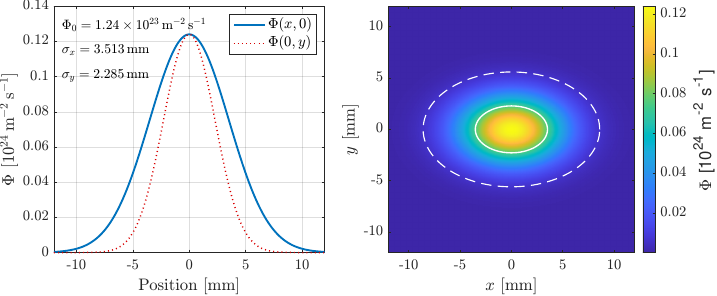}
		\caption{EIC injection before cooling.}
		\label{fig:beam-flux-eic-injection-before-cooling}
	\end{subfigure}
	
	\vspace{0.5em}
	
	\begin{subfigure}{0.49\textwidth}
		\centering
		\safeincludegraphics[width=\textwidth]{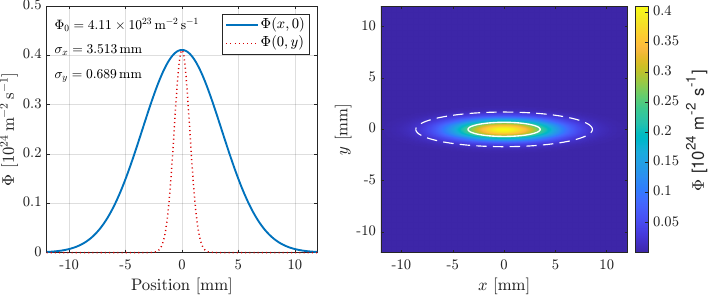}
		\caption{EIC injection after cooling.}
		\label{fig:beam-flux-eic-injection-after-cooling}
	\end{subfigure}
	\hfill
	\begin{subfigure}{0.49\textwidth}
		\centering
		\safeincludegraphics[width=\textwidth]{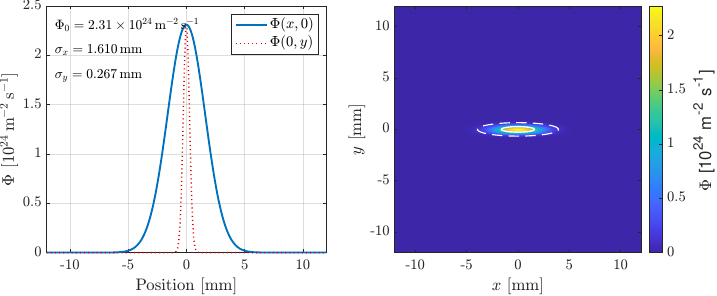}
		\caption{EIC flattop cooled-emittance case.}
		\label{fig:beam-flux-eic-flattop-cooled}
	\end{subfigure}
	
	\caption{Visualization of representative transverse particle-flux distributions from Table~\ref{tab:beam_parameters_rhic_eic}. Each panel shows the horizontal and vertical one-dimensional flux cuts on the left and the corresponding two-dimensional transverse flux distribution on the right. The cases illustrate the RHIC flattop reference, the broad EIC injection beam before cooling, the asymmetric EIC injection beam after cooling, and the EIC flattop cooled-emittance stress case. The white contours in the two-dimensional maps indicate the rms and 95\% transverse beam envelopes.}
	\label{fig:beam-distributions}
\end{figure*}

\subsection{Beam--target overlap and active target length}
\label{sec:beam-target-overlap}

The beam--target overlap specifies what fraction of the transverse beam distribution intersects the carbon strip for a given static target position. At the same time, the active strip length $\ell$ must be large enough to cover the relevant beam envelope along the long direction of the target. This length margin is important for extrapolating from the existing RHIC carbon targets, with active lengths of about \SI{25}{\milli\meter}, to possible EIC targets. Since much longer free carbon strips become increasingly difficult to fabricate and mount, \SI{50}{\milli\meter} is used below as a practical EIC benchmark length. Target motion is not introduced here; it is added later in Sec.~\ref{sec:target-motion-through-beam} by allowing the target-center coordinates to become time dependent.

Let the center of the carbon strip be located at $(x,y)$ relative to the beam center. For a strip whose long direction is parallel to $y$, with physical width $w$ in the $x$ direction and length $\ell$ in the $y$ direction, the intercepted fraction of the normalized transverse beam density is
\begin{equation}
	f_{\text{ov}}^{(y)}(x,y)
	=
	\int_{x-w/2}^{x+w/2}
	\int_{y-\ell/2}^{y+\ell/2}
	f(x',y')\,\dd y'\,\dd x'\,,
	\label{eq:overlap-general-ystrip}
\end{equation}
where $f(x',y')$ is the normalized Gaussian beam density from Eq.~\eqref{eq:beam-transverse-density}. Since the Gaussian distribution factorizes in $x$ and $y$, Eq.~\eqref{eq:overlap-general-ystrip} can be written as
\begin{equation}
	\begin{split}
		f_{\text{ov}}^{(y)}(x,y)
		&=
		\frac{1}{2}
		\left[
		\operatorname{erf}\!\left(
		\frac{x+w/2}{\sqrt{2}\sigma_x}
		\right)
		-
		\operatorname{erf}\!\left(
		\frac{x-w/2}{\sqrt{2}\sigma_x}
		\right)
		\right]
		\\
		&\quad\times
		\frac{1}{2}
		\left[
		\operatorname{erf}\!\left(
		\frac{y+\ell/2}{\sqrt{2}\sigma_y}
		\right)
		-
		\operatorname{erf}\!\left(
		\frac{y-\ell/2}{\sqrt{2}\sigma_y}
		\right)
		\right]\,.
	\end{split}
	\label{eq:overlap-erf-ystrip}
\end{equation}
For the usual case in which the strip length covers the relevant beam envelope in the long direction, $\ell\gg\sigma_y$, the second factor is approximately unity. The overlap then reduces to the one-dimensional expression
\begin{equation}
	f_{\text{ov}}^{(y)}(x)
	\simeq
	\frac{1}{2}
	\left[
	\operatorname{erf}\!\left(
	\frac{x+w/2}{\sqrt{2}\sigma_x}
	\right)
	-
	\operatorname{erf}\!\left(
	\frac{x-w/2}{\sqrt{2}\sigma_x}
	\right)
	\right]\,.
	\label{eq:overlap-erf-ystrip-long}
\end{equation}
If the target is centered on the beam and narrow compared with the beam size, $w\ll\sigma_x$, this becomes
\begin{equation}
	f_{\text{ov}}^{(y)}(0)
	\simeq
	\frac{w}{\sqrt{2\pi}\sigma_x}\,.
	\label{eq:overlap-narrow-ystrip}
\end{equation}

For a strip whose long direction is parallel to $x$, with physical width $w$ in the $y$ direction and length $\ell$ in the $x$ direction, the corresponding expression is obtained by interchanging $x$ and $y$,
\begin{equation}
	f_{\text{ov}}^{(x)}(x,y)
	=
	\int_{x-\ell/2}^{x+\ell/2}
	\int_{y-w/2}^{y+w/2}
	f(x',y')\,\dd y'\,\dd x'\,.
	\label{eq:overlap-general-xstrip}
\end{equation}
For $\ell\gg\sigma_x$, this reduces to
\begin{equation}
	f_{\text{ov}}^{(x)}(y)
	\simeq
	\frac{1}{2}
	\left[
	\operatorname{erf}\!\left(
	\frac{y+w/2}{\sqrt{2}\sigma_y}
	\right)
	-
	\operatorname{erf}\!\left(
	\frac{y-w/2}{\sqrt{2}\sigma_y}
	\right)
	\right]\,,
	\label{eq:overlap-erf-xstrip-long}
\end{equation}
and for a centered narrow strip,
\begin{equation}
	f_{\text{ov}}^{(x)}(0)
	\simeq
	\frac{w}{\sqrt{2\pi}\sigma_y}\,.
	\label{eq:overlap-narrow-xstrip}
\end{equation}

The overlap factor $f_{\text{ov}}$ is dimensionless and represents the fraction of circulating particles that geometrically intercept the carbon strip. It is used later to estimate the total number of particles crossing the target, while the local heat-source density is constructed from the full spatial flux density $\Phi(x,y)$.

The EIC beam sizes in Table~\ref{tab:beam_parameters_rhic_eic} are strongly different in the two transverse directions for several operating cases. Since the beta functions are fixed by the machine optics, the available geometric control is the orientation of the carbon strip in the transverse plane. For a strip with long axis at an angle $\theta$ with respect to the horizontal beam axis, the rms beam size projected along the strip is
\begin{equation}
	\sigma_{\parallel}^2(\theta)
	=
	\sigma_x^2\cos^2\theta
	+
	\sigma_y^2\sin^2\theta\,,
	\label{eq:sigma-parallel-theta}
\end{equation}
and the rms beam size projected perpendicular to the strip is
\begin{equation}
	\sigma_{\perp}^2(\theta)
	=
	\sigma_x^2\sin^2\theta
	+
	\sigma_y^2\cos^2\theta\,.
	\label{eq:sigma-perp-theta}
\end{equation}
The parallel projection controls the required active target length, while the perpendicular projection controls the beam width sampled when the target is moved normal to its long axis. For $\theta=\pm45^\circ$, both target orientations have the same projected beam sizes,
\begin{equation}
	\sigma_{\parallel}^2
	=
	\sigma_{\perp}^2
	=
	\frac{\sigma_x^2+\sigma_y^2}{2}\,,
	\label{eq:sigma-45deg}
\end{equation}
for an untilted Gaussian beam with no $x$--$y$ covariance. Thus, two EIC targets mounted at $\theta=+45^\circ$ and $\theta=-45^\circ$ provide two diagonal scan directions while avoiding the most unfavorable purely horizontal length requirement.

The length margin is defined as
\begin{equation}
	M_\ell^{95}
	=
	\frac{\ell}{2\sigma_{\parallel}^{95}(\theta)}\,,
	\label{eq:length-margin-95}
\end{equation}
where $\ell$ is the active target length and $\sigma_{\parallel}^{95}(\theta)$ is the projected 95\% semi-axis along the strip. The projected 95\% beam sizes in Table~\ref{tab:target_length_margin_45deg} are calculated from the transverse rms beam sizes listed in Table~\ref{tab:beam_parameters_rhic_eic}, using Eqs.~\eqref{eq:sigma-parallel-theta} and \eqref{eq:sigma-perp-theta} and the same conversion factor $\sigma^{95}=\sqrt{5.991}\,\sigma$ used there.

For RHIC the existing \SI{25}{\milli\meter} horizontal and vertical target orientations are used as the reference, while for EIC a benchmark active length of \SI{50}{\milli\meter} is evaluated for horizontal, diagonal, and vertical orientations.

\begin{table*}[htb]
	\centering
	\small
	\renewcommand{\arraystretch}{1.15}
	\setlength{\tabcolsep}{4.0pt}
	\begin{tabular}{llllccccc}
		\hline
		\textbf{Machine} &
		\textbf{Beam} &
		\textbf{Cooling} &
		\textbf{Target orientation} &
		$\theta$ &
		$\ell$ &
		$\sigma_{\parallel}^{95}(\theta)$ &
		$M_\ell^{95}$ &
		$\sigma_{\perp}^{95}(\theta)$ \\
		& & & &
		\si{\degree} &
		\si{\milli\meter} &
		\si{\milli\meter} &
		$1$ &
		\si{\milli\meter} \\
		\hline
		\multirow{2}{*}{RHIC} &
		\multirow{2}{*}{flattop} &
		\multirow{2}{*}{--} &
		horizontal &
		\num{0} &
		\num{25} &
		\num{1.195} &
		\num{10.5} &
		\num{1.264} \\
		& & &
		vertical &
		\num{90} &
		\num{25} &
		\num{1.264} &
		\num{9.89} &
		\num{1.195} \\
		\hline
		\multirow{12}{*}{EIC} &
		\multirow{6}{*}{injection} &
		\multirow{3}{*}{no cooling} &
		horizontal &
		\num{0} &
		\num{50} &
		\num{8.600} &
		\num{2.91} &
		\num{5.593} \\
		& & &
		diagonal &
		$\pm\num{45}$ &
		\num{50} &
		\num{7.254} &
		\num{3.45} &
		\num{7.254} \\
		& & &
		vertical &
		\num{90} &
		\num{50} &
		\num{5.593} &
		\num{4.47} &
		\num{8.600} \\
		\cline{3-9}
		& &
		\multirow{3}{*}{cooling} &
		horizontal &
		\num{0} &
		\num{50} &
		\num{8.600} &
		\num{2.91} &
		\num{1.686} \\
		& & &
		diagonal &
		$\pm\num{45}$ &
		\num{50} &
		\num{6.197} &
		\num{4.03} &
		\num{6.197} \\
		& & &
		vertical &
		\num{90} &
		\num{50} &
		\num{1.686} &
		\num{14.8} &
		\num{8.600} \\
		\cline{2-9}
		&
		\multirow{6}{*}{flattop} &
		\multirow{3}{*}{no cooling} &
		horizontal &
		\num{0} &
		\num{50} &
		\num{3.942} &
		\num{6.34} &
		\num{2.172} \\
		& & &
		diagonal &
		$\pm\num{45}$ &
		\num{50} &
		\num{3.183} &
		\num{7.85} &
		\num{3.183} \\
		& & &
		vertical &
		\num{90} &
		\num{50} &
		\num{2.172} &
		\num{11.5} &
		\num{3.942} \\
		\cline{3-9}
		& &
		\multirow{3}{*}{cooling} &
		horizontal &
		\num{0} &
		\num{50} &
		\num{3.942} &
		\num{6.34} &
		\num{0.655} \\
		& & &
		diagonal &
		$\pm\num{45}$ &
		\num{50} &
		\num{2.826} &
		\num{8.85} &
		\num{2.826} \\
		& & &
		vertical &
		\num{90} &
		\num{50} &
		\num{0.655} &
		\num{38.2} &
		\num{3.942} \\
		\hline
	\end{tabular}
	\caption{Target-length and scan-direction margins for the RHIC reference targets and candidate EIC target orientations. The rms beam sizes are taken from Table~\ref{tab:beam_parameters_rhic_eic}. The projected 95\% beam sizes are calculated from Eqs.~\eqref{eq:sigma-parallel-theta} and \eqref{eq:sigma-perp-theta}, and the length margin is defined in Eq.~\eqref{eq:length-margin-95}.}
	\label{tab:target_length_margin_45deg}
\end{table*}

The $\theta=\pm\SI{45}{\degree}$ geometry is used here as a simple symmetric benchmark because it provides two orthogonal diagonal scans. It is not necessarily the optimum choice for active-length margin. For the present EIC parameters, steeper diagonal orientations improve the margin in the cases with large horizontal beam size: for example, $\theta\simeq\SI{75}{\degree}$ gives $M_\ell^{95}\simeq\num{9.1}$ for injection after cooling, while $\theta\simeq\SI{70}{\degree}$ gives $M_\ell^{95}\simeq\num{10.2}$ for the flattop large-emittance case. The corresponding scan crossing angles would be about \SI{30}{\degree} and \SI{40}{\degree}, respectively, so the polarization-profile reconstruction would be less symmetric than for $\pm\SI{45}{\degree}$. For the following estimates we therefore keep $\theta=\pm\SI{45}{\degree}$ as the reference geometry, while treating the final target angle as an optimization parameter.

\subsection{Energy loss in the carbon strip}
\label{sec:energy-loss-carbon-strip}

This subsection converts the particle flux defined in Sec.~\ref{sec:beam-optics-current-density-particle-flux} into the nominal energy loss of the beam particles in the carbon strip. The required carbon material inputs, including the density $\rho$, the proton linear energy transfer (LET), and the corresponding line energy loss in SI units, are collected in Sec.~\ref{sec:carbon-material-properties} and Table~\ref{tab:carbon_properties}. Here, LET refers to the tabulated mass stopping power, usually given in units of \si{MeV.cm^2.g^{-1}}, while $\text{LET}_{\text{SI}}$ denotes the resulting line energy loss after multiplication by the material density and conversion to \si{J.m^{-1}}. At this stage the quantity calculated is the energy lost by the incident beam particles in the target material. It is not yet assumed that all of this energy remains as heat in the strip; secondary-electron escape and the retained heating fraction are treated in the following subsections.

The relevant stopping-power input is the mass stopping power,
\begin{equation}
	S_{\text{m}}(E)
	=
	\left(\frac{\dd E}{\rho\,\dd s}\right)_{\text{C}}\,,
	\label{eq:mass-stopping-power}
\end{equation}
usually tabulated in units of \si{MeV.cm^2.g^{-1}}. For a carbon density $\rho$ expressed in \si{g.cm^{-3}}, the corresponding linear stopping power is
\begin{equation}
	S_{\text{C}}(E)
	=
	\left(\frac{\dd E}{\dd s}\right)_{\text{C}}
	=
	S_{\text{m}}(E)\,\rho
	\quad
	[\si{MeV.cm^{-1}}]\,.
	\label{eq:linear-stopping-power-mev}
\end{equation}
In SI units this becomes
\begin{equation}
	S_{\text{C}}^{\text{SI}}(E)
	=
	S_{\text{m}}(E)\,\rho\,
	\left(1.602176634\times10^{-13}\right)\,100
	\quad
	[\si{J.m^{-1}}]\,,
	\label{eq:linear-stopping-power-si}
\end{equation}
where the factor $1.602176634\times10^{-13}$ converts \si{MeV} to \si{J}, and the factor 100 converts \si{cm^{-1}} to \si{m^{-1}}. This quantity is denoted $\text{LET}_{\text{SI}}$ in Table~\ref{tab:carbon_properties}. For graphitic or chemical-vapor-deposited (CVD) carbon at 250 GeV proton energy, the table gives
\begin{equation}
\text{LET}_{\text{SI}}
=
\num{5.123e-11}\ \si{J.m^{-1}}\,.
	\label{eq:let-si-graphitic}
\end{equation}

The nominal local volumetric energy-loss density in the carbon strip is then
\begin{equation}
	Q_{\text{loss}}(x,y)
	=
	\Phi(x,y)\,S_{\text{C}}^{\text{SI}}(E)\,,
	\label{eq:q-loss-local}
\end{equation}
with units of \si{W.m^{-3}}. Using Eq.~\eqref{eq:particle-flux}, this can be written as
\begin{equation}
	Q_{\text{loss}}(x,y)
	=
	Q_{\text{loss},0}
	\exp\!\left[
	-\frac{x^2}{2\sigma_x^2}
	-\frac{y^2}{2\sigma_y^2}
	\right]\,,
	\label{eq:q-loss-gaussian}
\end{equation}
where the peak nominal energy-loss density at the beam center is
\begin{equation}
	Q_{\text{loss},0}
	=
	\Phi_0\,S_{\text{C}}^{\text{SI}}(E)\,.
	\label{eq:q-loss-peak}
\end{equation}
This is the quantity that sets the local scale of the beam-induced source term before corrections for escaping secondary particles are applied.

For a strip of thickness $t$ along the beam direction, the nominal energy lost by one beam particle traversing the carbon is
\begin{equation}
	\Delta E_{\text{loss}}
	=
	S_{\text{C}}^{\text{SI}}(E)\,t\,.
	\label{eq:delta-e-loss}
\end{equation}
The corresponding total nominal beam power loss in the carbon strip is obtained by integrating Eq.~\eqref{eq:q-loss-local} over the target volume. For a strip with transverse overlap fraction $f_{\text{ov}}$, this can be written compactly as
\begin{equation}
	P_{\text{loss}}
	=
	\left(N_b N_p f_{\text{rev}}\right)
	f_{\text{ov}}\,
	S_{\text{C}}^{\text{SI}}(E)\,t\,.
	\label{eq:p-loss-overlap}
\end{equation}
Equivalently, using the spatial flux density,
\begin{equation}
	P_{\text{loss}}
	=
	t\,S_{\text{C}}^{\text{SI}}(E)
	\int_{\text{strip}}
	\Phi(x,y)\,\dd x\,\dd y\,.
	\label{eq:p-loss-integral}
\end{equation}
Equations~\eqref{eq:p-loss-overlap} and \eqref{eq:p-loss-integral} are identical because the overlap integral gives
\begin{equation}
	\int_{\text{strip}}
	\Phi(x,y)\,\dd x\,\dd y
	=
	N_b N_p f_{\text{rev}}\,f_{\text{ov}}\,.
	\label{eq:flux-integral-overlap}
\end{equation}

For the graphitic/CVD carbon reference case with $t=\SI{50}{\nano\meter}$,
Eq.~\eqref{eq:delta-e-loss} gives
\begin{equation}
	\Delta E_{\text{loss}}
	=
	\num{2.56e-18}\ \si{J}
	=
	\num{15.99}\ \si{eV}
\end{equation}
per incident proton. Representative nominal energy-loss scales are summarized in Table~\ref{tab:nominal_energy_loss}. These values are computed before secondary-electron escape is applied.

\begin{table*}[htb]
	\centering
	\small
	\renewcommand{\arraystretch}{1.15}
	\setlength{\tabcolsep}{5pt}
	\begin{tabular}{llllcccc}
		\hline
		\textbf{Machine} &
		\textbf{Beam} &
		\textbf{Cooling} &
		\textbf{Target orientation} &
		$f_{\text{ov}}$ &
		$Q_{\text{loss},0}$ &
		$P_{\text{loss}}$ &
		$Q_{\text{loss},0}/Q_{\text{loss},0}^{\text{RHIC}}$ \\
		& & & &
		$10^{-3}$ &
		$10^{13}\,\si{W.m^{-3}}$ &
		$10^{-2}\,\si{W}$ &
		$1$ \\
		\hline
		\multirow{2}{*}{RHIC} &
		\multirow{2}{*}{flattop} &
		\multirow{2}{*}{--} &
		horizontal &
		\num{7.724} &
		\num{6.068} &
		\num{3.713} &
		\num{1.000} \\
		& & &
		vertical &
		\num{8.170} &
		\num{6.068} &
		\num{3.928} &
		\num{1.000} \\
		\hline
		\multirow{4}{*}{EIC} &
		\multirow{2}{*}{injection} &
		no cooling &
		diagonal &
		\num{1.346} &
		\num{0.635} &
		\num{2.157} &
		\num{0.105} \\
		& &
		cooling &
		diagonal &
		\num{1.576} &
		\num{2.107} &
		\num{2.525} &
		\num{0.347} \\
		\cline{2-8}
		&
		\multirow{2}{*}{flattop} &
		no cooling &
		diagonal &
		\num{3.069} &
		\num{3.572} &
		\num{4.920} &
		\num{0.589} \\
		& &
		cooling &
		diagonal &
		\num{3.456} &
		\num{11.85} &
		\num{5.541} &
		\num{1.953} \\
		\hline
	\end{tabular}
	\caption{Nominal stopping-power energy-loss scales for graphitic/CVD carbon with $t=\SI{50}{\nano\meter}$, before secondary-electron escape is applied. The overlap factors are calculated from Sec.~\ref{sec:beam-target-overlap}. The peak energy-loss density $Q_{\text{loss},0}$ follows the local peak particle flux, while $P_{\text{loss}}$ also depends on the target orientation through $f_{\text{ov}}$.}
	\label{tab:nominal_energy_loss}
\end{table*}

The table separates two effects. The peak quantity $Q_{\text{loss},0}$ determines the local source strength at the beam center and follows the peak particle flux in Table~\ref{tab:beam_parameters_rhic_eic}. The total power $P_{\text{loss}}$ additionally depends on the geometric overlap with the strip. Thus, changing the target orientation can reduce the total intercepted power, but it does not reduce the local peak energy-loss density experienced by a strip element passing through the beam center.

The distinction between $Q_{\text{loss}}$ and the actual heat source is essential for ultra-thin carbon targets. In a thick target, the stopping-power loss can usually be treated as locally deposited heat. In a strip with thickness of order tens of nanometers, however, energetic secondary electrons may escape before depositing their energy locally. The heat source used in the thermal model is therefore introduced only after the secondary-electron escape mechanism has been considered.

\subsection{Secondary-electron escape}
\label{sec:secondary-electron-escape}

The stopping-power loss defined in Sec.~\ref{sec:energy-loss-carbon-strip} is the energy transferred from the incident proton to the carbon target material. In an ultra-thin carbon strip, this transferred energy is not necessarily deposited locally as heat. A fraction of the energy can be carried away by secondary electrons, including energetic delta electrons, that leave the strip before losing their energy inside the target volume.

This effect is especially relevant because the target thickness is only of order tens of nanometers. The electron escape length can then be comparable to, or larger than, the physical thickness of the strip. In this regime the usual thick-target assumption, where the stopping-power loss is identified with local heat deposition, is not automatically valid. The escaping-electron channel therefore has to be treated before defining the heat source used in the thermal model.

The energy balance can be written schematically as
\begin{equation}
	S_{\text{C}}^{\text{SI}}(E)
	=
	S_{\text{heat}}(E,t)
	+
	S_{\text{esc}}(E,t)\,,
	\label{eq:stopping-power-balance}
\end{equation}
where $S_{\text{C}}^{\text{SI}}$ is the nominal line energy loss from Eq.~\eqref{eq:linear-stopping-power-si}, $S_{\text{heat}}$ is the part retained in the strip and ultimately converted into heat, and $S_{\text{esc}}$ is the part carried away by electrons that escape from the target. Both terms may depend on the incident particle energy, target thickness, carbon density, and secondary-electron spectrum.

The corresponding escape fraction is defined as
\begin{equation}
	f_{\text{esc}}(E,t)
	=
	\frac{S_{\text{esc}}(E,t)}
	{S_{\text{C}}^{\text{SI}}(E)}\,,
	\label{eq:secondary-electron-escape-fraction}
\end{equation}
so that the retained fraction is
\begin{equation}
	f_{\text{heat}}(E,t)
	=
	1-f_{\text{esc}}(E,t)\,.
	\label{eq:retained-fraction-from-escape}
\end{equation}
In the thick-target limit, $f_{\text{esc}}\rightarrow 0$ and $f_{\text{heat}}\rightarrow 1$. For a nanometer-scale carbon strip, however, $f_{\text{esc}}$ may be non-negligible and must either be estimated from an electron-transport calculation or constrained experimentally.

For the first estimate used here, the escaping-electron correction is evaluated with a simple thin-film model. The secondary-electron energy spectrum is represented by a phenomenological distribution $g(E_e;E)$, where $E_e$ is the secondary-electron energy and $E$ is the incident proton energy. Electrons are assumed to be produced uniformly through the carbon layer and emitted isotropically. For each electron energy, an approximate range-energy relation in carbon is used to estimate whether the electron can reach a surface before losing its energy locally.

With these assumptions, the escape fraction can be written as
\begin{equation}
	f_{\text{esc}}(E,t)
	=
	\frac{
		\displaystyle
		\int_{E_{\min}}^{E_{\max}}
		E_e\,g(E_e;E)\,
		P_{\text{esc}}(E_e,t)\,
		\dd E_e
	}{
		\displaystyle
		\int_{E_{\min}}^{E_{\max}}
		E_e\,g(E_e;E)\,
		\dd E_e
	}\,.
	\label{eq:escape-fraction-spectrum}
\end{equation}
Here $P_{\text{esc}}(E_e,t)$ is the energy-escape probability for an electron of energy $E_e$ in a carbon layer of thickness $t$. Electrons with ranges much shorter than the film thickness are effectively trapped and contribute to local heating, whereas electrons with ranges comparable to or larger than the distance to a surface can remove part of their energy from the strip.

For a strip of thickness $t$, the energy carried away by escaping electrons per incident beam particle is
\begin{equation}
	\Delta E_{\text{esc}}
	=
	S_{\text{esc}}(E,t)\,t
	=
	f_{\text{esc}}(E,t)\,
	S_{\text{C}}^{\text{SI}}(E)\,t\,.
	\label{eq:delta-e-esc}
\end{equation}
Together with Eq.~\eqref{eq:delta-e-loss}, this defines the escaping part of the nominal stopping-power loss. The retained part is introduced in Sec.~\ref{sec:heat-retained-target}.

The isotropic angular assumption in Eq.~\eqref{eq:escape-fraction-spectrum} should be regarded as a first approximation. It is most appropriate for low-energy cascade secondary electrons whose directions are randomized by transport in the solid. Higher-energy delta electrons are correlated with the incident proton direction and can be more forward-directed. A more complete treatment would combine an energy-dependent delta-electron angular distribution with a diffuse low-energy secondary component.

The geometry of the real carbon strip is also more complicated than an ideal flat slab. The calculation shown below assumes an ideal carbon layer of thickness $t$ (see Table\,\ref{tab:target_parameters}). If the strip is curled, twisted, or locally rotated in the beam-overlap region, the carbon path length and the escape distances to the nearest surfaces can differ from this ideal case. The corresponding roll-angle dependence should be treated as a model uncertainty and addressed with a rectangular-strip escape model.

For this purpose, a second escape estimate is evaluated with the physical rectangular cross section of the carbon strip retained. The strip has width $w$ and thickness $t$, and the local roll angle $\psi$ of this cross section with respect to the incoming beam direction is averaged over. For each roll angle, the proton chord through the rectangular cross section determines the stopping-power loss, while the electron escape probability is evaluated from the distance to the nearest boundary of the same rectangle. Thus the roll-averaged model does not replace the strip by a slab of width $w$; it keeps the actual rectangular geometry and averages over local orientations.

The roll-averaged retained fraction is formed from the averaged energies,
\begin{equation}
	\begin{split}
	\left\langle f_{\text{heat}}(E)\right\rangle_{\psi}
	& =
	\frac{
		\left\langle \Delta E_{\text{heat}}(E,\psi)\right\rangle_{\psi}
	}{
		\left\langle \Delta E_{\text{loss}}(E,\psi)\right\rangle_{\psi}
	}\,,\\
	\left\langle f_{\text{esc}}(E)\right\rangle_{\psi}
	& =
	\frac{
		\left\langle \Delta E_{\text{esc}}(E,\psi)\right\rangle_{\psi}
	}{
		\left\langle \Delta E_{\text{loss}}(E,\psi)\right\rangle_{\psi}
	}\,.
	\label{eq:roll-averaged-fractions}
	\end{split}
\end{equation}
This is important because different roll angles correspond to different chord lengths and therefore to different energy losses per intercepted proton. Averaging the fractions alone would not give the correct retained heat.

For the nominal RHIC/EIC carbon-strip geometry used in the numerical calculations (see Table\,\ref{tab:target_parameters}), the geometric chord length through the rectangular carbon cross section is obtained by averaging the proton chord over the roll angle $\psi$. For a uniform distribution of $\psi$ and of impact parameter across the projected strip width, the mean chord at fixed $\psi$ is the ratio of the cross-sectional area to the projected width, $w\,t/(w\cos\psi + t\sin\psi)$, and the roll-averaged chord is the integral
\begin{equation}
	\left\langle \ell_{\text{chord}} \right\rangle_{\psi}
	= \frac{2}{\pi}\int_0^{\pi/2}
	\frac{w\,t}{w\cos\psi + t\sin\psi}\,\dd\psi\,.
	\label{eq:chord-integral}
\end{equation}
The integrand rises steeply toward edge-on incidence ($\psi\to\pi/2$, where the chord approaches the full width $w$), so the average is evaluated by numerical integration. For the nominal geometry this gives
\begin{equation}
	\left\langle \ell_{\text{chord}} \right\rangle_{\psi}
	=
	\SI{0.19087}{\micro\meter}
	=
	3.817\,t\,,
	\label{eq:roll-averaged-chord-nominal}
\end{equation}
which is larger than the physical thickness because many roll angles expose a longer diagonal path through the rectangular cross section.

The heat source used in the thermal model must be normalized to the flat-layer stopping-power loss used in Eq.~\eqref{eq:delta-e-loss}. The corresponding roll-averaged heat scale is therefore
\begin{equation}
	\eta_{\text{roll}}
	=
	\frac{
		\left\langle \Delta E_{\text{heat}} \right\rangle_{\psi}
	}{
		\Delta E_{\text{loss}}(t)
	}
	=
	\left\langle f_{\text{heat}}\right\rangle_{\psi}
	\frac{
		\left\langle \ell_{\text{chord}} \right\rangle_{\psi}
	}{
		t
	}\,.
	\label{eq:roll-heat-scale}
\end{equation}

For RHIC and EIC proton energies, the numerical model gives
\begin{equation}
	\eta_{\text{roll}}
	\simeq
	1.134\,.
	\label{eq:roll-heat-scale-nominal}
\end{equation}
Thus the roll-averaged rectangular-strip model retains somewhat more heat than the ideal flat-layer stopping-power estimate, about 13\% more for RHIC and EIC proton energies, because the longer mean proton chord through the rolled rectangular cross section more than offsets the energy carried away by escaping electrons.

\begin{figure*}[htb]
	\centering
 	\safeincludegraphics[width=\textwidth]{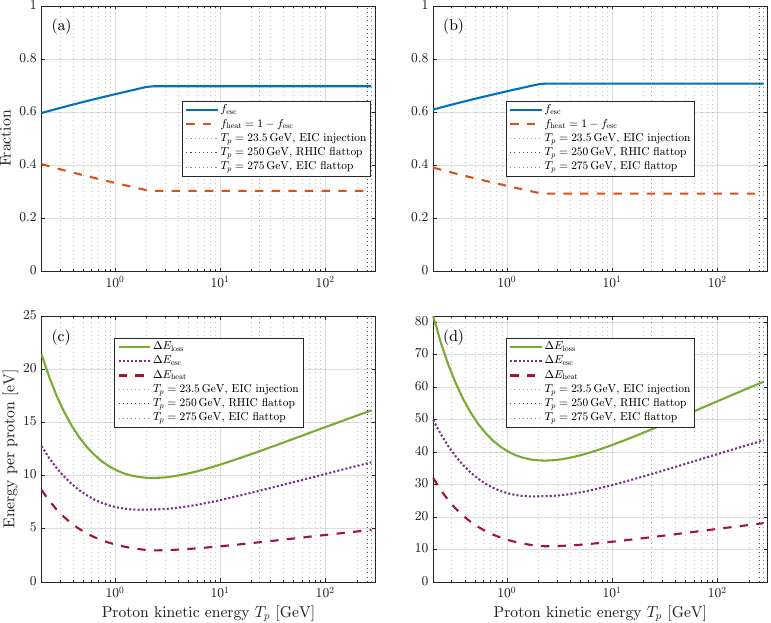}
\caption{Comparison of the secondary-electron escape correction for an ideal flat graphitic/CVD carbon layer and for the roll-averaged rectangular-strip model. Panels (a) and (c) show the ideal flat-layer calculation, where the beam-direction path length is the physical film thickness. Panels (b) and (d) show the corresponding calculation for the physical rectangular strip, averaged over roll angle using the effective chord defined in Eq.~\eqref{eq:roll-averaged-chord-nominal}. The upper panels show the escape and retained fractions, while the lower panels show the corresponding energy components per proton. Vertical guide lines indicate representative EIC injection, RHIC flattop, and EIC flattop proton energies.}
	\label{fig:f_escape_vs_proton_energy}
\end{figure*}

The comparison between the ideal flat-layer model and the roll-averaged rectangular-strip model is shown in Fig.~\ref{fig:f_escape_vs_proton_energy}. Panels (a) and (c) show the ideal \SI{50}{\nano\meter} flat-layer calculation, for which the beam-direction carbon path length is equal to the physical film thickness. Panels (b) and (d) show the corresponding roll-averaged calculation for the same physical strip dimensions, $w=\SI{10}{\micro\meter}$ and $t=\SI{50}{\nano\meter}$.

The upper panels show that the escape fraction is large at RHIC and EIC energies in both models. This means that a substantial part of the stopping-power energy transferred from the incident proton to secondary electrons leaves the target volume before it is thermalized locally.

The difference between the two geometries appears mainly in the lower panels. In the ideal flat-layer case, panel (c), the beam-direction path length is fixed by the physical film thickness. In the roll-averaged rectangular-strip case, panel (d), the longer mean chord increases the nominal stopping-power loss before the escape correction is applied. As a result, the retained heat per proton is larger than in the ideal flat-layer model even though the retained fraction is similar.

The roll-averaged result should therefore be regarded as the more relevant first estimate for a curled target, while the ideal flat-layer result provides a lower-bound heating case.

\subsection{Heat retained in the target}
\label{sec:heat-retained-target}

After secondary-electron escape has been separated from the nominal stopping-power loss, the thermal model must use the heat actually retained in the carbon strip. The nominal source term \(Q_{\text{loss}}\) defined in Sec.~\ref{sec:energy-loss-carbon-strip} corresponds to the ideal flat-layer reference in which the proton path length through carbon is the physical film thickness \(t=\SI{50}{\nano\meter}\). The heat source used in the thermal calculation is therefore written as
\begin{equation}
	Q_{\text{heat}}(x,y)
	=
	\eta_{\text{heat}}(E)\,
	Q_{\text{loss}}(x,y)\,,
	\label{eq:q-heat-from-q-loss}
\end{equation}
where \(\eta_{\text{heat}}\) is the effective heat-source scale relative to the nominal flat-layer stopping-power source.

For the ideal flat-layer escape model, this scale is simply
\begin{equation}
	\eta_{\text{heat}}(E)
	=
	f_{\text{heat}}(E,t)\,.
	\label{eq:eta-heat-flat}
\end{equation}
For the roll-averaged rectangular-strip model, however, the average proton chord length is larger than the physical film thickness. In that case the correct scale is formed from the retained heat per proton,
\begin{equation}
	\eta_{\text{heat}}(E)
	=
	\frac{
		\left\langle \Delta E_{\text{heat}}(E,\psi)\right\rangle_{\psi}
	}{
		\Delta E_{\text{loss}}^{50\,\text{nm}}(E)
	}\,.
	\label{eq:eta-heat-roll}
\end{equation}
This distinction is important: \(f_{\text{heat}}\) is the retained fraction within a given geometry, whereas \(\eta_{\text{heat}}\) is the factor that converts the nominal flat-layer source \(Q_{\text{loss}}\) into the heat source used in the thermal model.

With the Gaussian flux density from Eq.~\eqref{eq:particle-flux}, the retained heat source becomes
\begin{equation}
	Q_{\text{heat}}(x,y)
	=
	Q_{\text{heat},0}
	\exp\!\left[
	-\frac{x^2}{2\sigma_x^2}
	-\frac{y^2}{2\sigma_y^2}
	\right]\,,
	\label{eq:q-heat-gaussian}
\end{equation}
where
\begin{equation}
	Q_{\text{heat},0}
	=
	\eta_{\text{heat}}(E)\,
	Q_{\text{loss},0}\,.
	\label{eq:q-heat-peak}
\end{equation}

The corresponding retained heat per intercepted proton is
\begin{equation}
	\Delta E_{\text{heat}}^{\text{model}}(E)
	=
	\eta_{\text{heat}}(E)\,
	\Delta E_{\text{loss}}^{50\,\text{nm}}(E)\,,
	\label{eq:delta-e-heat-effective}
\end{equation}
and the total retained beam power in the strip is
\begin{equation}
	P_{\text{heat}}
	=
	\eta_{\text{heat}}(E)\,
	P_{\text{loss}}\,.
	\label{eq:p-heat-overlap}
\end{equation}
Equivalently,
\begin{equation}
	P_{\text{heat}}
	=
	t
	\int_{\text{strip}}
	Q_{\text{heat}}(x,y)\,\dd x\,\dd y\,.
	\label{eq:p-heat-integral}
\end{equation}

\begin{table}[htb]
	\centering
	\small
	\renewcommand{\arraystretch}{1.15}
	\setlength{\tabcolsep}{5pt}
	\begin{tabular}{lcccc}
		\hline
		\textbf{Case} &
		$E_{\text{tot}}$ &
		$T_p$ &
		$\Delta E_{\text{heat}}^{\text{roll}}$ &
		$\eta_{\text{heat}}$ \\
		&
		\si{\giga\electronvolt} &
		\si{\giga\electronvolt} &
		\si{\electronvolt/proton} &
		$1$ \\
		\hline
		EIC injection &
		\num{23.5} &
		\num{22.56} &
		\num{13.9} &
		\num{0.87} \\
		RHIC flattop &
		\num{250} &
		\num{249.06} &
		\num{18.1} &
		\num{1.13} \\
		EIC flattop &
		\num{275} &
		\num{274.06} &
		\num{18.3} &
		\num{1.14} \\
		\hline
	\end{tabular}
	\caption{Representative retained-heat scales for the roll-averaged rectangular-strip model. The quantity $\eta_{\text{heat}}$ converts the nominal flat-layer stopping-power source for a \SI{50}{\nano\meter} beam path into the retained heat source used in the thermal model.}
	\label{tab:retained-heat-scale}
\end{table}

The values of $\eta_{\text{heat}}$ are larger than the retained fraction $f_{\text{heat}}$ itself. This is because $\eta_{\text{heat}}$ is defined relative to the nominal flat-layer loss for a \SI{50}{\nano\meter} beam path, whereas the roll-averaged model has a larger mean proton chord length through the rectangular strip.

\begin{figure}[htb]
	\centering
	\safeincludegraphics[width=\columnwidth]{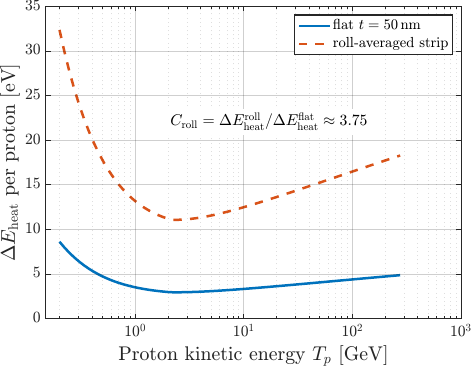}
	\caption{Deposited heat per proton, $\Delta E_{\text{heat}}$, as a function of total proton energy for two target-geometry models. The solid curve shows the ideal flat-layer case with beam-direction thickness $t=\SI{50}{\nano\meter}$. The dashed curve shows the roll-averaged strip model for a rectangular strip with $w=\SI{10}{\micro\meter}$ and $t=\SI{50}{\nano\meter}$, averaged over roll angles from $0$ to $2\pi$. The deposited heat is defined by $\Delta E_{\text{heat}}=(1-f_{\text{esc}})\Delta E_{\text{loss}}$, with $\Delta E_{\text{loss}}$ from Eq.~\eqref{eq:delta-e-loss} and $f_{\text{esc}}$ from Eq.~\eqref{eq:secondary-electron-escape-fraction}. The quantity $C_{\mathrm{roll}}=\Delta E_{\text{heat}}^{\mathrm{roll}}/\Delta E_{\text{heat}}^{\mathrm{flat}}$ gives the enhancement of the deposited heat in the roll-averaged model. 
	}
	\label{fig:heat_retention_correction}
\end{figure}

To illustrate the practical impact of the escape correction on the thermal model, Fig.~\ref{fig:heat_retention_correction} compares the deposited heat per proton, $\Delta E_{\text{heat}}$, for two geometrical descriptions of the carbon strip. The first is the ideal flat-layer limit, in which the beam-direction thickness is fixed at $t=\SI{50}{\nano\meter}$. The second is the roll-averaged strip model, in which the physical rectangular cross section with $w=\SI{10}{\micro\meter}$ and $t=\SI{50}{\nano\meter}$ is averaged over local roll angles from $0$ to $2\pi$. This comparison is useful because real carbon strips are not perfectly flat in the beam-overlap region, but can curl or twist, as discussed in Refs.\,\cite{Steski2014CarbonMicroRibbons,Steski2018TargetLifetime}, so that the local chord length through the material is larger than the nominal \SI{50}{\nano\meter} thickness.

Figure~\ref{fig:heat_retention_correction} shows that the deposited heat per proton in the roll-averaged model is systematically larger than in the ideal flat-layer limit. The reason is not primarily a change in the retained fraction itself, but the larger effective proton path length through the rolled or twisted strip, which increases the nominal stopping-power loss before the escape correction is applied. Over the RHIC and EIC energy range, this leads to an enhancement of the deposited heat by a factor of about $C_{\mathrm{roll}}\approx 3.75$, as indicated in the figure. For the subsequent thermal model, the roll-averaged geometry therefore provides the more realistic reference case, while the ideal flat-layer result serves as a lower-bound estimate for beam-induced heating.

In the following thermal calculations, the roll-averaged rectangular-strip model is used as the default heat-source estimate. The ideal flat-layer result provides a lower-bound heating case. For the present beam energies, the roll-averaged model gives \(\eta_{\text{heat}}\) of order unity, because secondary-electron escape is partly compensated by the longer mean proton chord through the rolled rectangular strip.

\subsection{Target motion through the beam}
\label{sec:target-motion-through-beam}

The preceding subsections define the static beam-target overlap and the retained heat source at a given target position. During pC operation, however, the carbon strip is moved through the beam. The heat source seen by a given material element of the strip is therefore transient, even if the beam current and transverse beam sizes are constant.

\begin{figure}[htb]
	\centering
	\safeincludegraphics[width=\columnwidth]{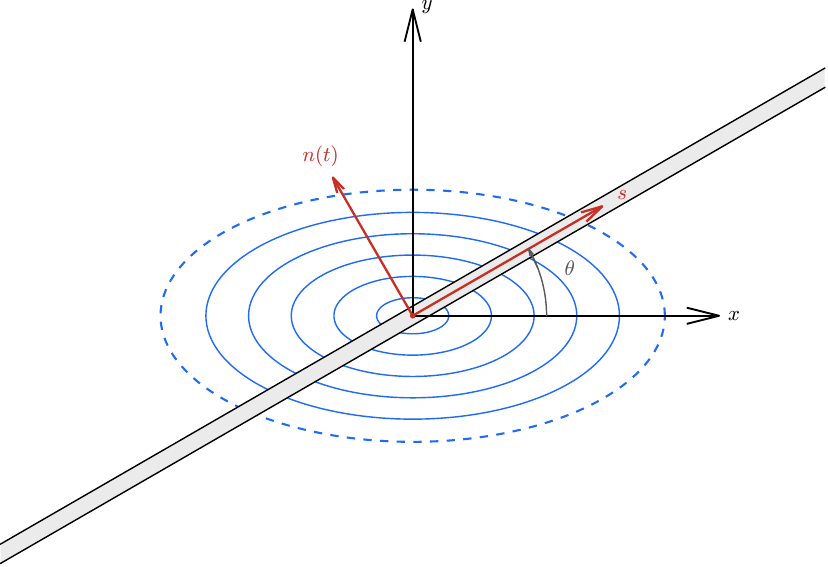}
\caption{Geometry used to describe target motion through the beam. The carbon strip is oriented at an angle $\theta$ with respect to the horizontal beam axis $x$. The coordinate $s$ runs along the strip, while $n(t)$ is the scan coordinate normal to the strip.}
	\label{fig:target_motion_geometry}
\end{figure}

The geometry used to describe this motion is shown in Fig.~\ref{fig:target_motion_geometry}. The transverse beam coordinates are denoted by \(x\) and \(y\). The carbon strip is described by a coordinate \(s\) along the strip and a coordinate \(n\) normal to the strip. The angle \(\theta\) specifies the orientation of the strip with respect to the horizontal beam axis.

In the simplest model the target moves at constant speed through the beam,
\begin{equation}
	n(t)
	=
	n_0 + v_{\text{t}}(t-t_0)\,,
	\label{eq:constant-speed-target-motion}
\end{equation}
where
\begin{equation}
	v_{\text{t}}
	=
	\frac{\dd n}{\dd t}
	\label{eq:target-speed-normal-coordinate}
\end{equation}
is the target speed normal to the strip and \(t_0\) is the beam-center crossing time. More detailed implementations may include acceleration and deceleration of the stepper motor, finite positioning resolution, or deviations between commanded and actual target position.

For a Gaussian beam, the retained heat source at the moving target position can be written schematically as
\begin{equation}
	Q_{\text{heat}}(s,t)
	=
	Q_{\text{heat},0}\,
	\exp\!\left[
	-\frac{s^2}{2\sigma_s^2}
	-\frac{n^2(t)}{2\sigma_n^2}
	\right]\,,
	\label{eq:moving-heat-source-strip-coordinates}
\end{equation}
where \(\sigma_s\) and \(\sigma_n\) are the beam sizes projected onto the strip-parallel and strip-normal directions, respectively. The peak scale \(Q_{\text{heat},0}\) is the retained heat source defined in Sec.~\ref{sec:heat-retained-target}. Target motion changes the exposure time of a material element, but it does not reduce the instantaneous peak source scale when the strip passes through the beam center.

For a fixed material element near the strip center, the time dependence of the source is therefore approximately a Gaussian pulse,
\begin{equation}
	\frac{Q_{\text{heat}}(0,t)}
	{Q_{\text{heat},0}}
	=
	\exp\!\left[
	-\frac{
		v_{\text{t}}^2(t-t_0)^2
	}
	{2\sigma_n^2}
	\right]\,.
	\label{eq:heat-source-pulse-fixed-element}
\end{equation}
A useful estimate of the exposure time is
\begin{equation}
	\tau_{\text{exp}}
	\sim
	\frac{\sigma_n}{v_{\text{t}}}\,.
	\label{eq:exposure-time-estimate}
\end{equation}
Thus faster target motion reduces the duration of heating for a given strip element, while the peak heat source remains set by the beam flux, the beam-target overlap, and the retained heat per proton.

\begin{figure}[htb]
	\centering
	\safeincludegraphics[width=\columnwidth]{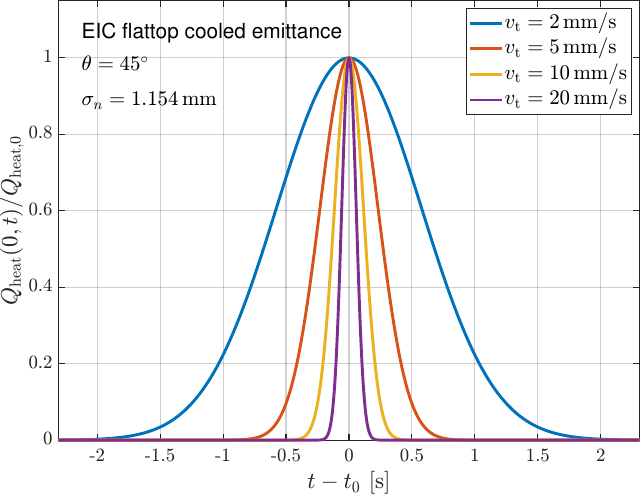}
	\caption{Normalized heat-source pulse at the strip position passing through the beam center for the EIC flattop cooling case with a diagonal target orientation. The peak source strength is unchanged by the target speed $v_{\text{t}}$, while the exposure time decreases as $v_{\text{t}}$ increases. The plotted source is normalized to $Q_{\text{heat},0}$.}
	\label{fig:heat_source_pulse_eic_flattop_cooled}
\end{figure}

Figure~\ref{fig:heat_source_pulse_eic_flattop_cooled} illustrates the distinction between peak heating and exposure time. The target speed is therefore an operational control parameter for the transient temperature response: it does not change \(Q_{\text{heat},0}\), but it changes how long a given part of the strip remains exposed to the high-flux beam core. This time-dependent source term is the input to the thermal response model developed in Sec.~\ref{sec:thermal-response-carbon-strip}.

\begin{figure*}[t]
	\centering
	\begin{tikzpicture}[
		>=Latex,
		font=\small,
		box/.style={
			draw,
			rounded corners=3pt,
			thick,
			align=center,
			minimum height=8mm,
			text width=33mm,
			inner sep=4pt
		},
		widebox/.style={
			draw,
			rounded corners=3pt,
			thick,
			align=center,
			minimum height=8mm,
			text width=45mm,
			inner sep=4pt
		},
		smallbox/.style={
			draw,
			rounded corners=3pt,
			thick,
			align=center,
			minimum height=8mm,
			text width=30mm,
			inner sep=4pt
		},
		arrow/.style={-Latex, thick},
		line/.style={thick}
		]
		
		\node[box]     (beam)   at (0,0.00)    {Beam parameters};
		
		\node[box]     (flux)   at (0,-1.45)   {Particle flux\\$\Phi(x,y)$};
		
		\node[box]     (direct) at (-3.8,-3.45) {Stopping power /\\direct heating};
		
		\node[box]     (wake)   at ( 3.8,-3.45) {Wake fields /\\RF drive};
		
		\node[box]     (escape) at (-3.8,-5.20) {Secondary-electron\\escape / retained heat};
		
		\node[box]     (rfend)  at ( 3.8,-5.20) {RF end heating};
		
		\node[widebox] (thermal) at (0,-6.95)   {Thermal response\\$T(s,t)$};
		
		\node[smallbox] (forces) at (7.6,-3.35) {Beam-induced forces};
		
		\node[smallbox] (estate) at (7.6,-4.90) {Electrical state /\\contact impedance};
		
		\node[smallbox] (mech)   at (7.6,-6.95) {Mechanical\\response};
		
		\node[smallbox] (center) at (-4.3,-8.95) {Center glow};
		
		\node[smallbox] (end)    at (0,-8.95)    {End glow};
		
		\node[smallbox] (deform) at (4.3,-8.95)  {Target deformation};
		
		\draw[arrow] (beam.south) -- (flux.north);
		
		\coordinate (splitstem) at (0,-2.25);
		\coordinate (splitleft) at (-3.8,-2.25);
		\coordinate (splitright) at (3.8,-2.25);
		
		\draw[line]  (flux.south) -- (splitstem);
		\draw[line]  (splitleft) -- (splitright);
		\draw[arrow] (splitleft) -- (direct.north);
		\draw[arrow] (splitright) -- (wake.north);
		
		\draw[arrow] (direct.south) -- (escape.north);
		\draw[arrow] (wake.south)   -- (rfend.north);
		
		\coordinate (joinleft)  at (-3.8,-5.85);
		\coordinate (joinright) at ( 3.8,-5.85);
		\coordinate (joinmid)   at ( 0.0,-5.85);
		
		\draw[line]  (escape.south) -- (joinleft);
		\draw[line]  (rfend.south)  -- (joinright);
		\draw[line]  (joinleft) -- (joinright);
		\draw[arrow] (joinmid) -- (thermal.north);
		
		\draw[arrow] (forces.south) -- (estate.north);
		\draw[arrow] (estate.south) -- (mech.north);
		\draw[arrow] (thermal.east) -- (mech.west);
		
		\coordinate (outstem)   at (0,-7.80);
		\coordinate (outleft)   at (-4.3,-7.80);
		\coordinate (outright)  at ( 4.3,-7.80);
		
		\draw[line]  (thermal.south) -- (outstem);
		\draw[line]  (outleft) -- (outright);
		\draw[arrow] (outleft)  -- (center.north);
		\draw[arrow] (outstem)  -- (end.north);
		\draw[arrow] (outright) -- (deform.north);
		
	\end{tikzpicture}
	\caption{Schematic structure of the carbon-strip response model.
		Beam parameters define the particle flux $\Phi(x,y)$ at the target.
		The direct-heating branch describes stopping-power energy loss and the retained heat after secondary-electron escape. In parallel, wakefields provide an RF drive that can lead to strip-end heating. Both source terms enter the thermal response $T(s,t)$. Beam-induced forces, together with the electrical state and contact impedance, determine the mechanical response. The coupled model is intended to describe the observed center glow, end glow, and target deformation.}
	\label{fig:carbon_response_model_scheme}
\end{figure*}
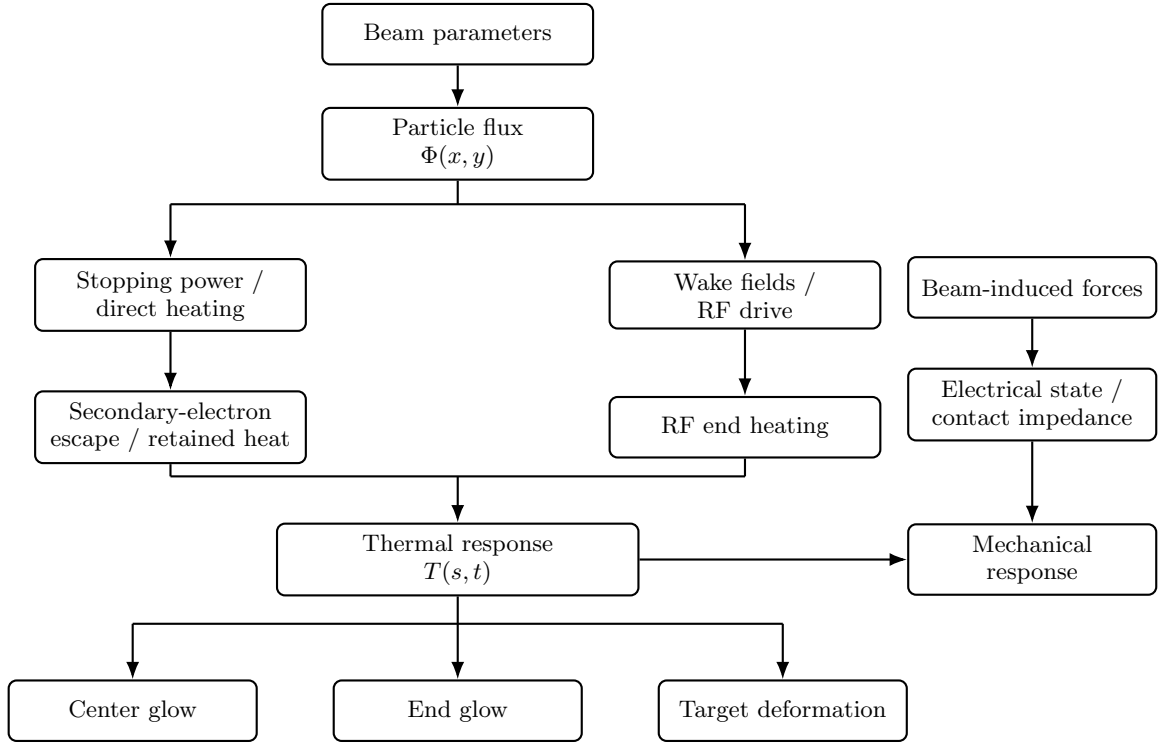

Figure~\ref{fig:carbon_response_model_scheme} summarizes how the source terms constructed in Sec.~\ref{sec:beam-target-geometry} enter the response model. The retained beam-heating source provides the input to the thermal calculation, while wakefield coupling, RF-induced strip-end heating, resistance changes, and beam-induced forces provide additional channels needed to describe the observed RHIC target response.

\section{Thermal response of the carbon strip}
\label{sec:thermal-response-carbon-strip}

The previous section constructed the retained, time-dependent beam-heating source \(Q_{\text{heat}}(s,t)\) for a moving carbon strip. This is the driving term in the thermal model: it represents the part of the beam energy loss that remains in the target after secondary-electron escape and is available to raise the strip temperature. This section describes how \(Q_{\text{heat}}(s,t)\) is converted into the transient temperature response of the strip. The model is deliberately kept one-dimensional in this first implementation: heat conduction is treated along the strip, while the width and thickness enter through the cross-sectional area and radiating surface.

The goal is to understand how the carbon strip responds thermally to the retained beam-heating source: how heat is stored, conducted, and radiated during a scan, and how the peak temperature depends on target motion. This provides a controlled basis for comparing RHIC operation with EIC benchmark cases using common assumptions for material properties, target geometry, heat-source model, and scan motion.

\subsection{Purpose of the thermal model}

The main output of the thermal model is the transient temperature field \(T(s,t)\) and, in particular, the maximum temperature reached during a target scan. The temperature response depends on the beam parameters, the effective heat-source scale \(\eta_{\text{heat}}\), the material properties, the target geometry, and the target speed.

The first numerical implementation is intended to provide three main outputs: the thermal input parameter set, a comparison of static and moving-source heating, and the maximum temperature as a function of target speed. The thermal input parameters used for this first implementation are summarized in Table~\ref{tab:thermal_model_inputs}.  No fixed sublimation temperature is used as a model input; vacuum carbon loss is evaluated in Sec.~\ref{sec:carbon-sublimation-vacuum-mass-loss} in terms of $\Delta h_{\mathrm{sub}}/t$.
\begin{table*}[htb]
	\centering
	\small
	\renewcommand{\arraystretch}{1.15}
	\setlength{\tabcolsep}{6pt}
	\begin{tabular}{llll}
		\hline
		\textbf{Quantity} &
		\textbf{Symbol} &
		\textbf{Value} &
		\textbf{Unit} \\
		\hline
		Carbon scenario &
		-- &
		graphitic/CVD carbon &
		-- \\
		Density &
		$\rho$ &
		\num{1700} &
		\si{kg.m^{-3}} \\
		Specific heat capacity &
		$c_p$ &
		\num{750} &
		\si{J.kg^{-1}.K^{-1}} \\
		Thermal conductivity &
		$\kappa$ &
		\num{5} &
		\si{W.m^{-1}.K^{-1}} \\
		Emissivity &
		$\varepsilon$ &
		\num{0.85} &
		-- \\
		Strip width &
		$w$ &
		\num{10} &
		\si{\micro\meter} \\
		Strip thickness &
		$t$ &
		\num{50} &
		\si{\nano\meter} \\
		Reference RHIC active length &
		$\ell_{\text{RHIC}}$ &
		\num{25} &
		\si{\milli\meter} \\
		Benchmark EIC active length &
		$\ell_{\text{EIC}}$ &
		\num{50} &
		\si{\milli\meter} \\
		Reference target scan speed &
		$v_t$ &
		\num{4} &
		\si{mm.s^{-1}} \\
		Holder temperature &
		$T_0$ &
		\num{300} &
		\si{K} \\
		Effective heat-source scale &
		$\eta_{\text{heat}}$ &
		from Sec.~\ref{sec:heat-retained-target} &
		-- \\
		\hline
	\end{tabular}
	\caption{Thermal-model input parameters used for the first carbon-strip temperature calculations. The material values correspond to the graphitic/CVD carbon scenario from Table~\ref{tab:carbon_properties}; the geometric values define the RHIC reference and EIC benchmark target lengths used in Sec.~\ref{sec:beam-target-overlap}. The reference target scan speed is based on the video-derived near-beam scale summarized in Table~\ref{tab:video_observation_parameters}. The effective heat-source scale $\eta_{\text{heat}}$ converts the nominal flat-layer stopping-power source into the retained heat source used in the thermal model and is taken from the secondary-electron escape and roll-averaged geometry model described in Sec.~\ref{sec:heat-retained-target}.}
	\label{tab:thermal_model_inputs}
\end{table*}
Table~\ref{tab:thermal_model_inputs} lists the material, geometry, boundary-condition, and scan-speed inputs; the heat-source scale $\eta_{\mathrm{heat}}$ is taken from the retained-source construction in Sec.~\ref{sec:heat-retained-target}. Table~\ref{tab:thermal_source_scales} summarizes the corresponding beam-heating source scales used in the thermal equation.

\begin{table*}[htb]
	\centering
	\small
	\renewcommand{\arraystretch}{1.15}
	\setlength{\tabcolsep}{5pt}
	\begin{tabular}{llllcccc}
		\hline
		\textbf{Machine} &
		\textbf{Beam} &
		\textbf{Cooling} &
		\textbf{Target orientation} &
		$f_{\text{ov}}$ &
		$Q_{\text{heat},0}$ &
		$P_{\text{heat}}$ &
		$Q_{\text{heat},0}/Q_{\text{heat},0}^{\text{RHIC}}$ \\
		&
		&
		&
		&
		$10^{-3}$ &
		$10^{13}\,\si{W.m^{-3}}$ &
		$10^{-2}\,\si{W}$ &
		$1$ \\
		\hline
		
		\multirow{2}{*}{RHIC} &
		\multirow{2}{*}{flattop} &
		\multirow{2}{*}{--} &
		horizontal &
		\num{7.724} &
		\num{6.881} &
		\num{4.211} &
		\num{1.000} \\
		&
		&
		&
		vertical &
		\num{8.170} &
		\num{6.881} &
		\num{4.455} &
		\num{1.000} \\
		\hline
		
		\multirow{4}{*}{EIC} &
		\multirow{2}{*}{injection} &
		no cooling &
		diagonal &
		\num{1.346} &
		\num{0.551} &
		\num{1.870} &
		\num{0.080} \\
		&
		&
		cooling &
		diagonal &
		\num{1.576} &
		\num{1.827} &
		\num{2.189} &
		\num{0.265} \\
		\cline{2-8}
		&
		\multirow{2}{*}{flattop} &
		no cooling &
		diagonal &
		\num{3.069} &
		\num{4.089} &
		\num{5.631} &
		\num{0.594} \\
		&
		&
		cooling &
		diagonal &
		\num{3.456} &
		\num{13.56} &
		\num{6.343} &
		\num{1.971} \\
		\hline
	\end{tabular}
	\caption{Retained beam-heating source scales used as input to the one-dimensional thermal model. The source \(Q_{\text{heat},0}\) is obtained from \(Q_{\text{heat},0}=\eta_{\text{heat}}Q_{\text{loss},0}\), using the roll-averaged secondary-electron escape correction. The retained power \(P_{\text{heat}}\) also includes the finite beam-target overlap and therefore depends on target orientation. The ratios are normalized to the RHIC flattop value of \(Q_{\text{heat},0}\).}
	\label{tab:thermal_source_scales}
\end{table*}

\subsection{One-dimensional reduction}

The length of the carbon strip is much larger than its width or thickness (\(\ell \gg w, t\)). The first thermal model therefore treats the temperature as a one-dimensional field along the active strip coordinate \(s\),
\begin{equation}
	T = T(s,t)\,,
	\label{eq:temperature-1d-definition}
\end{equation}
where \(s\) measures distance along the strip. The cross-sectional area of the strip reduces the temperature rise through axial conduction, while its surface area allows heat radiation, through the cross-sectional area \(A\) and radiating perimeter \(p_{\text{rad}}\),
\begin{equation}
	A = wt\,,
	\qquad
	p_{\text{rad}} = 2(w+t)\,.
	\label{eq:strip-area-radiating-perimeter}
\end{equation}

This reduction assumes that the temperature can be represented by an average over the strip cross section, so that transverse temperature structure is not resolved explicitly. The approximation is appropriate as a first model because \(w\) and \(t\) are small compared with the active length \(\ell\), while the dominant thermal transport to the holder occurs along the strip. Possible corrections from transverse gradients, nonuniform surface emission, or local geometric distortions of a curled strip can be treated later as refinements of the effective heat source and boundary conditions.

\subsection{Heat-balance equation}

The one-dimensional thermal model is a local energy balance for a short element of the strip. In compact form, the balance is
\begin{widetext}
\begin{equation}
	\underbrace{\rho c_p A \frac{\partial T}{\partial t}}_{\text{heat storage}}
	=
	\underbrace{\frac{\partial}{\partial s}
		\left(
		\kappa A \frac{\partial T}{\partial s}
		\right)}_{\text{axial conduction}}
	-
	\underbrace{\varepsilon \sigma_{\text{SB}} p_{\text{rad}}
		\left[
		T^4(s,t)-T_0^4
		\right]}_{\text{radiative cooling}}
	+
	\underbrace{A Q_{\text{heat}}(s,t)}_{\text{beam-heating source}}
	\,.
	\label{eq:thermal-balance-1d-conservative}
\end{equation}
\end{widetext}
Before division by the heat capacity per unit length, $\rho c_p A$, each term in Eq.~\eqref{eq:thermal-balance-1d-conservative} has units of \si{W.m^{-1}}, i.e., power per unit strip length. After division by $\rho c_p A$, the equation gives the local temperature-change rate in \si{K.s^{-1}}. Here $T(s,t)$ is the strip temperature, $s$ is the coordinate along the strip, $\rho$ is the carbon density, $c_p$ is the specific heat capacity, $\kappa$ is the thermal conductivity, $\varepsilon$ is the emissivity, and $\sigma_{\text{SB}}=\SI{5.670374419e-8}{W.m^{-2}.K^{-4}}$ is the Stefan-Boltzmann constant~\cite{becker,ozisik}. The cross-sectional area $A$ and radiating perimeter $p_{\text{rad}}$ are defined in Eq.~\eqref{eq:strip-area-radiating-perimeter}. The right-hand side contains axial conduction, radiative cooling to an environment at temperature $T_0$, and the beam-heating source $A Q_{\text{heat}}$, which converts the retained volumetric heat source into power per unit strip length.

For constant material parameters and uniform strip geometry, Eq.~\eqref{eq:thermal-balance-1d-conservative} becomes
\begin{widetext}
\begin{equation}
	\frac{\partial T(s,t)}{\partial t}
	=
	\alpha
	\frac{\partial^2 T(s,t)}{\partial s^2}
	-
	\frac{\varepsilon \sigma_{\text{SB}} p_{\text{rad}}}{\rho c_p A}
	\left[
	T^4(s,t)-T_0^4
	\right]
	+
	\frac{Q_{\text{heat}}(s,t)}{\rho c_p}\,,
	\label{eq:thermal-balance-1d}
\end{equation}
\end{widetext}
where
\begin{equation}
	\alpha
	=
	\frac{\kappa}{\rho c_p}
	\label{eq:thermal-diffusivity}
\end{equation}
is the thermal diffusivity. In the first implementation, \(\rho\), \(c_p\), \(\kappa\), and \(\varepsilon\) are treated as constant input parameters. Temperature-dependent material properties, local variations in the strip geometry, or nonideal thermal contact to the holder can be added later.

\subsection{Beam heating in the one-dimensional strip model}

The one-dimensional heat equation requires the peak retained source \(Q_{\text{heat},0}\) and its dependence on the strip coordinate \(s\) and scan coordinate \(n(t)\). The peak source is taken from Sec.~\ref{sec:heat-retained-target},
\begin{equation}
	Q_{\text{heat},0}
	=
	\eta_{\text{heat}}(E)\,
	Q_{\text{loss},0}\,,
	\label{eq:qheat0-from-qloss0-thermal}
\end{equation}
where \(\eta_{\text{heat}}\) is the effective heat-source scale. For the default roll-averaged strip model used here, it is obtained from Eq.~\eqref{eq:eta-heat-roll}, and the ideal flat-layer limit is given by Eq.~\eqref{eq:eta-heat-flat}.

The projected beam sizes along and normal to the strip are
\begin{equation}
	\sigma_s
	=
	\sigma_{\parallel}(\theta)\,,
	\qquad
	\sigma_n
	=
	\sigma_{\perp}(\theta)\,,
	\label{eq:sigma-s-n-thermal}
\end{equation}
with \(\sigma_{\parallel}\) and \(\sigma_{\perp}\) defined in Eqs.~\eqref{eq:sigma-parallel-theta} and \eqref{eq:sigma-perp-theta}. Using the scan coordinate \(n(t)\) from Eq.~\eqref{eq:constant-speed-target-motion}, the source term used in the thermal equation is
\begin{equation}
	Q_{\text{heat}}(s,t)
	=
	Q_{\text{heat},0}
	\exp\left[
	-\frac{s^2}{2\sigma_s^2}
	-\frac{n^2(t)}{2\sigma_n^2}
	\right]\,.
	\label{eq:qheat-moving-strip-thermal}
\end{equation}
This is the \(Q_{\text{heat}}(s,t)\) term that appears in Eq.~\eqref{eq:thermal-balance-1d-conservative}.

For the static centered benchmark, \(n(t)=0\), so
\begin{equation}
	Q_{\text{heat}}(s)
	=
	Q_{\text{heat},0}
	\exp\left[
	-\frac{s^2}{2\sigma_s^2}
	\right]\,.
	\label{eq:qheat-static-along-strip-thermal}
\end{equation}
Target motion changes the exposure time of a material element, but it does not change the peak source scale \(Q_{\text{heat},0}\), as illustrated in Fig.~\ref{fig:heat_source_pulse_eic_flattop_cooled}.

The static beam-centered benchmark is obtained by setting \(n(t)=0\) in Eq.~\eqref{eq:qheat-moving-strip-thermal}. This gives continuous exposure of the same material element and therefore provides an upper-limit reference for the moving-target calculation. In the constant-speed case of Eq.~\eqref{eq:constant-speed-target-motion}, faster target motion does not reduce the peak source scale \(Q_{\text{heat},0}\), but reduces the characteristic exposure time of a material element,
\begin{equation}
	\tau_{\text{exp}}
	\sim
	\frac{\sigma_n}{v_{\text{t}}}\,,
	\label{eq:thermal-exposure-time-estimate}
\end{equation}
where \(\sigma_n=\sigma_{\perp}(\theta)\). The resulting pulse narrowing with increasing \(v_{\text{t}}\) is illustrated in Fig.~\ref{fig:heat_source_pulse_eic_flattop_cooled}.

\subsection{Boundary and initial conditions}

The active strip is taken to extend from \(s=-\ell/2\) to \(s=+\ell/2\). In the first implementation, both strip ends are assumed to be held at the holder temperature,
\begin{equation}
	T\left(-\frac{\ell}{2},t\right)
	=
	T\left(+\frac{\ell}{2},t\right)
	=
	T_0\,.
	\label{eq:thermal-boundary-fixed-ends}
\end{equation}
This fixed-temperature boundary condition represents ideal thermal contact to a holder with large heat capacity compared with the suspended carbon strip.

The initial condition is taken as a uniform strip temperature,
\begin{equation}
	T(s,0)
	=
	T_0\,.
	\label{eq:thermal-initial-condition}
\end{equation}
This corresponds to a target initially in thermal equilibrium with the holder before the beam-induced heat source is applied.

More refined boundary conditions may include finite thermal contact resistance at the strip attachment points, holder heating, or nonuniform end temperatures. Such effects would reduce the ability of the holder to remove heat and can increase the transient peak temperature. They are not included in the first implementation, but can be added later if RHIC observations or dedicated measurements require them.

\subsection{Static steady-state temperature profile}
\label{sec:static-steady-state-temperature-profile}

Before considering target motion, it is useful to establish the static beam-centered temperature profile. In this limit the heat source remains fixed at the beam center and the same material element is exposed continuously. This case is not intended to represent normal target operation, but it provides a useful upper-limit benchmark for the moving-target calculation.

For the static centered source, the scan coordinate is set to \(n(t)=0\), and the one-dimensional source term reduces to Eq.~\eqref{eq:qheat-static-along-strip-thermal}. The temperature then evolves according to Eq.~\eqref{eq:thermal-balance-1d-conservative} until conduction to the strip ends and radiative losses balance the retained beam-heating source. The steady-state condition is obtained by setting \(\partial T/\partial t=0\),
\begin{equation}
	0
	=
	\frac{\dd}{\dd s}
	\left(
	\kappa A \frac{\dd T}{\dd s}
	\right)
	-
	\varepsilon \sigma_{\text{SB}} p_{\text{rad}}
	\left[
	T^4(s)-T_0^4
	\right]
	+
	A Q_{\text{heat}}(s)\,.
	\label{eq:static-steady-state-heat -balance}
\end{equation}

The static solution develops a symmetric temperature profile with the maximum at the beam position. At early times the temperature rise is localized near the Gaussian heat source, while at later times heat conduction broadens the profile toward the fixed-temperature strip ends. The final profile represents the continuous-exposure equilibrium. For moving targets, the exposure time of any given material element is finite, so the peak temperature is expected to be lower than this static benchmark.

\begin{figure*}[htb]
	\centering
	\begin{subfigure}[t]{0.325\textwidth}
		\centering
		\safeincludegraphics[width=\textwidth]{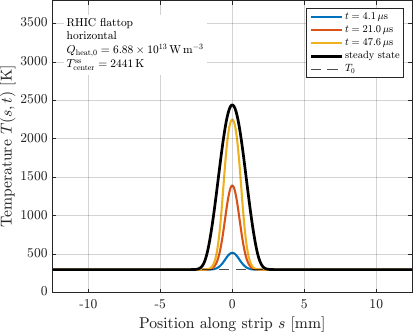}
		\caption{RHIC flattop, horizontal target.}
		\label{fig:static_temperature_profile_rhic}
	\end{subfigure}
	\begin{subfigure}[t]{0.325\textwidth}
		\centering
		\safeincludegraphics[width=\textwidth]{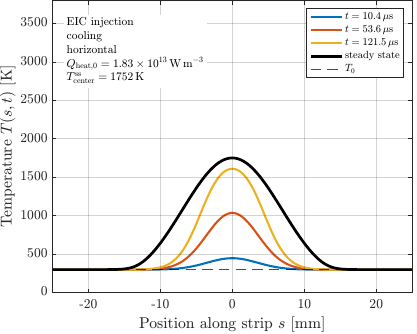}
		\caption{EIC injection, cooling, diagonal target.}
		\label{fig:static_temperature_profile_eic_inj}
	\end{subfigure}
	\begin{subfigure}[t]{0.325\textwidth}
		\centering
		\safeincludegraphics[width=\textwidth]{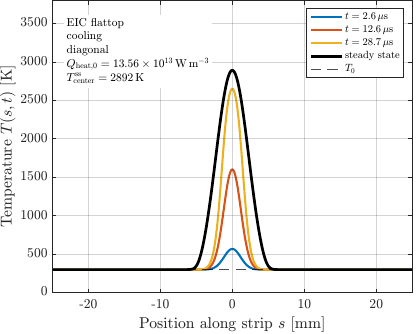}
		\caption{EIC flattop, cooling, diagonal target.}
		\label{fig:static_temperature_profile_eic_flat}
	\end{subfigure}
\caption{Static centered-source temperature profiles for three representative beam conditions. The beam-heating source is held fixed at the strip center, and the strip ends are kept at \(T_0=\SI{300}{K}\). Intermediate times and the final steady-state profile are shown.  Panel (a) shows the RHIC flattop reference case, panel (b) the EIC injection case after cooling, and panel (c) the EIC flattop cooled-emittance case.}
	\label{fig:static_temperature_profiles_selected}
\end{figure*}

Figure~\ref{fig:static_temperature_profiles_selected} compares the static beam-centered benchmark for three representative cases. In this limit the beam-heating source is held fixed at the strip center, so the same material element remains continuously exposed to the maximum source. The result is therefore a conservative continuous-exposure reference, not the normal operating situation of a moving target.

The RHIC flattop case in Fig.~\ref{fig:static_temperature_profiles_selected}(a) serves as the empirical reference and reaches \(T_{\text{center}}^{\text{ss}}\simeq\SI{2441}{K}\). The EIC injection case after cooling, shown in panel (b), remains lower, with \(T_{\text{center}}^{\text{ss}}\simeq\SI{1752}{K}\). The EIC flattop cooled-emittance case in panel (c) gives the largest static temperature rise, \(T_{\text{center}}^{\text{ss}}\simeq\SI{2892}{K}\), and is therefore the relevant static stress case among the examples shown.

For a specific beam condition, the static peak temperature is controlled by the local peak source \(Q_{\text{heat},0}\) and is essentially independent of the strip orientation. Changing the target inclination changes the projected beam width along the strip and therefore the width of the temperature profile, but not the local maximum source.

To complement the static temperature-profile comparison, it is useful to examine the temporal response for the strongest heating case, namely the EIC flattop cooled-emittance case with diagonal target orientation. In the static centered-source benchmark, the same strip element remains continuously exposed to the beam-heating source. This makes it possible to separate the time scale for heating up toward steady state from the subsequent cooling once the source is removed.

\begin{figure*}[htb]
	\centering
	\safeincludegraphics[width=\textwidth]{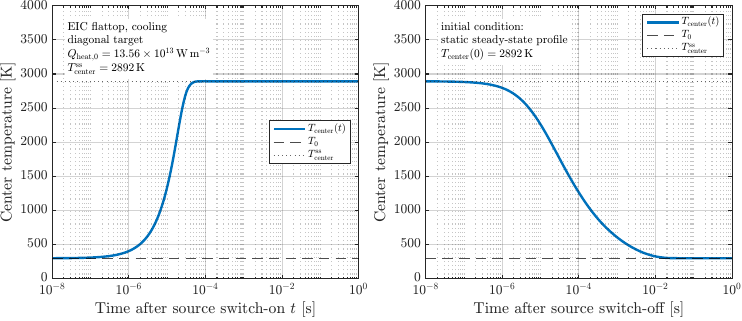}
	\caption{Center-temperature evolution for the EIC flattop cooled-emittance case in the static centered-source calculation shown in Fig.\,\ref{fig:static_temperature_profile_eic_flat}. The left panel shows the temperature rise after the beam-heating source is switched on at $t=0$, starting from $T_0=\SI{300}{K}$. The right panel shows the subsequent cooling after the source is switched off, starting from the static steady-state profile.}
	\label{fig:static_temperature_time_evolution}
\end{figure*}

Figure~\ref{fig:static_temperature_time_evolution} shows the corresponding center-temperature evolution for the static centered-source case. The left panel shows heating after the source is switched on, while the right panel shows cooling after the source is switched off, starting from the steady-state profile. The heating rise is much faster than the cooling tail, by roughly two orders of magnitude. This reflects the nonlinear radiative term: radiative losses grow rapidly at high temperature, but become weak again as the strip cools back toward \(T_0\).

\subsection{Temperature profile with moving target}

The static centered-source calculation does not describe normal target operation. In operation, the carbon strip moves through the fixed beam, so a given strip element is exposed to the beam-heating source only for a finite time. The motion is described by the target-position coordinate $n(t)$, which specifies the displacement of the strip normal to its long axis relative to the beam center. The instantaneous local source strength $Q_{\text{heat},0}$ at beam center is not reduced by the motion. Instead, target motion changes the time interval during which a particular material element overlaps the high-intensity part of the beam and therefore changes the amount of heat retained during one passage through the beam.

The coordinate $s$ remains the coordinate along the carbon strip. The heat-source distribution is given by Eq.~\eqref{eq:qheat-moving-strip-thermal}, with the constant-speed target motion defined in Eq.~\eqref{eq:constant-speed-target-motion}. The widths $\sigma_s$ and $\sigma_n=\sigma_{\perp}(\theta)$ are the projected rms beam widths parallel and normal to the strip, respectively. The maximum local source remains $Q_{\text{heat},0}$ and occurs at $s=0$ and $n(t)=0$. Thus target motion changes the time dependence of the local heating, but not the peak heating rate at beam center.

With this time-dependent source, the one-dimensional heat-balance equation remains
\begin{widetext}
\begin{equation}
	\rho c_p A \frac{\partial T(s,t)}{\partial t}
	=
	\frac{\partial}{\partial s}
	\left(
	\kappa A \frac{\partial T(s,t)}{\partial s}
	\right)
	-
	\varepsilon \sigma_{\text{SB}} p_{\text{rad}}
	\left[
	T^4(s,t) - T_0^4
	\right]
	+
	A Q_{\text{heat}}(s,t).
	\label{eq:thermal-balance-moving-target}
\end{equation}
\end{widetext}
The boundary conditions at the strip ends remain fixed at the holder temperature $T_0$, as in the static calculation.

The relevant exposure-time estimate is given by Eq.~\eqref{eq:thermal-exposure-time-estimate}. The moving-target problem therefore compares this exposure time with the intrinsic thermal response time of the strip. For slow target motion, the local temperature can approach the static centered-source result. For faster motion, the material element leaves the high-intensity part of the beam before reaching the static temperature, and the maximum temperature is reduced. The corresponding narrowing of the heating pulse with increasing $v_{\text{t}}$ is illustrated in Fig.~\ref{fig:heat_source_pulse_eic_flattop_cooled}.

The following calculations use the same material parameters, heat-source normalization, boundary conditions, and radiative cooling model as the static calculation. The new control parameter is the target speed $v_{\text{t}}$. The moving-target results are used to determine how the maximum temperature depends on target motion and to identify the range of speeds for which the EIC flattop cooled-emittance case remains below the chosen thermal limit.

\begin{figure*}[htb]
	\centering
	\safeincludegraphics[width=\textwidth]{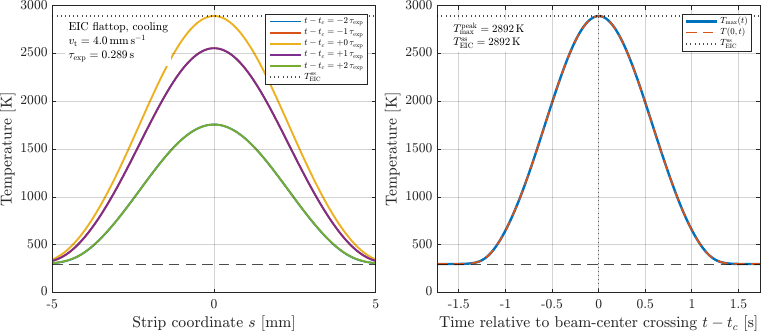}
	\caption{Moving-target temperature response for the EIC flattop cooled-emittance case using the reference target scan speed from Table~\ref{tab:thermal_model_inputs}. The left panel shows instantaneous temperature profiles \(T(s,t)\) along the strip at selected times relative to the beam-center crossing time \(t_c\). The exposure scale $\tau_{\mathrm{exp}}$ [Eq.\,\eqref{eq:thermal-exposure-time-estimate}] denotes the characteristic time over which a strip element traverses the rms beam width normal to the target.  The right panel shows the corresponding evolution of the maximum and strip-center temperatures.}
	\label{fig:moving_target_temperature_profile_eic_flattop}
\end{figure*}

Before discussing the moving-target solution in detail, it is useful to compare the relevant time scales. For the video-derived target speed of order $v_{\text{t}}\simeq\SI{4}{mm.s^{-1}}$, the exposure time from Eq.~\eqref{eq:thermal-exposure-time-estimate} is much longer than the intrinsic thermal rise time of the strip, illustrated in Fig.~\ref{fig:static_temperature_time_evolution}. One therefore expects the moving-source calculation to remain close to the static centered-source limit. Figure~\ref{fig:moving_target_temperature_profile_eic_flattop} confirms this expectation: for such slow motion, the peak temperature is essentially unchanged compared with the static case. This also shows that reducing $T_{\max}$ significantly by target motion alone would require scan speeds far above the video-derived scale.

This result should not be interpreted as a failure of the moving-target model. Rather, it shows that the present slow target motion is thermally quasi-static. In the EIC flattop cooled-emittance case, the static centered-source calculation gives \(T_{\text{center}}^{\text{ss}}\simeq\SI{2892}{K}\), which remains about \(\SI{760}{K}\) below the adopted carbon sublimation reference \(T_{\text{sub}}=\SI{3650}{K}\). The moving-target calculation at the video-estimated speed therefore does not significantly change the peak temperature, but it also is not required merely to keep the temperature below \(T_{\text{sub}}\).

For completeness, the calculation can be inverted to estimate the speed required for target motion to reduce the EIC peak temperature to the RHIC static reference level. For the EIC flattop cooled-emittance case considered here, this requires \(v_{\text{t}}\simeq\SI{8.2e4}{mm.s^{-1}}\), corresponding to \(\tau_{\text{exp}}\simeq\SI{1.4e-5}{s}\). This speed is many orders of magnitude above the video-estimated slow target motion and would require a dedicated high-speed target concept rather than a modest change of the present target drive.

Another possible mitigation would be to increase the transverse beam size at the target, for example by using larger beta functions. Since \(Q_{\text{heat},0}\propto 1/(\sigma_x\sigma_y)\), reducing the EIC cooled flattop peak heat source to the RHIC static reference level requires an increase of the transverse beam area by approximately a factor of \(1.97\). If both beta functions are increased together, this corresponds approximately to \(\beta_x\rightarrow2\beta_x\) and \(\beta_y\rightarrow2\beta_y\).

\section{Wake-field coupling, RF-induced end heating, and resistance changes}
\label{sec:wakefield-coupling-rf-heating-resistance}

The direct beam-heating model in Sec.~\ref{sec:thermal-response-carbon-strip} describes energy deposited near the beam-intercept point of the carbon strip. It does not include RF or wakefield coupling to the target holder and strip-end geometry. This additional mechanism is required by RHIC observations: the carbon-strip ends can glow before direct beam interception, the effect depends on the \SI{200}{\mega\hertz} cavity voltage, rounded metal fins near the target ends improve target lifetime, and the strip resistance changes substantially after beam exposure~\cite{Steski2014CarbonMicroRibbons,Steski2018TargetLifetime}. The present section collects these observations, converts the observed end glow into an approximate heating-power scale, and uses wakefield and impedance calculations to compare the performance of aluminum and alumina EIC target-holder geometries as an example to estimate whether realistic chamber and holder wakefields can provide the RF-drive scale required for the observed end glow.

\subsection{RHIC evidence for RF-induced end heating}
\label{sec:rhic-evidence-rf-end-heating}

RHIC video observations show two related but physically distinct effects. First, the carbon strip can deform strongly during a scan, with loose target length displaced toward the beam center. Second, the strip ends can glow before the target intercepts the beam. Earlier RHIC studies reported both of these effects, attributed to the electric field of the beam~\cite{Steski2014CarbonMicroRibbons,Steski2018TargetLifetime}.

The end-glow observation is especially important because it occurs before direct beam interception of the strip center. It therefore cannot be explained by the direct beam-heating source described in Sec.~\ref{sec:thermal-response-carbon-strip} alone. Steski et al. associated the end glow with beam-induced high-frequency electric fields that move electrons along the target, causing resistive heating~\cite{Steski2018TargetLifetime}. They also reported that reducing the voltage of the \SI{200}{\mega\hertz} cavity suppressed the end glow, which directly links the effect to RF fields rather than to local stopping-power heating.

The color of the end glow provides a useful qualitative constraint. In the RHIC video observations, the carbon-strip ends appear to glow with an approximately yellow color. This cannot be converted into a precise temperature without camera calibration, emissivity, exposure settings, optical transmission, and knowledge of the local emitting area. Nevertheless, it shows that the end regions reach a substantial temperature and that the effect is not a small perturbation.

The design of the local end geometry provides an additional constraint. Rounded metal fins, or ``donuts'', installed near the carbon-strip ends reduced the induced electric field and improved target lifetime~\cite{Steski2018TargetLifetime}. The fact that end heating remained a relevant issue even with such field-shaping elements indicates that the target-holder geometry remains an essential part of the problem. The fins reduce the local electric-field enhancement and provide a more favorable boundary condition, but they do not remove the need to model RF or wakefield coupling to the holder and strip-end region.

These observations motivate treating the strip-end heating as a separate source term from direct beam interception. In the response model, the direct beam source determines the temperature rise near the beam crossing point, while an additional RF or wakefield source is localized near the target attachments and depends on the holder geometry, electrical contact, strip resistance profile, and bunch spectrum.

\subsection{Power scale implied by the observed end glow}
\label{sec:end-glow-power-scale}

The observed end glow can be used to estimate the minimum power that must be deposited near a strip end. The estimate is approximate, because the temperature is inferred only from visible incandescence rather than from calibrated pyrometry. However, it is sufficient to determine whether the required RF-induced power is at the microwatt, milliwatt, or watt level.

Inspection of the RHIC video frames indicates that the glowing region near a target end extends over a length of order
\begin{equation}
	\ell_{\text{end}}
	\simeq
	\SI{8}{mm}.
	\label{eq:end-glow-length-estimate}
\end{equation}
The apparent color of the end glow was compared qualitatively with the visible glow of an oven at known temperature. This comparison suggests an end temperature of order
\begin{equation}
	T_{\text{end}}
	\simeq
	\SI{1200}{\celsius}
	\simeq
	\SI{1470}{K}.
	\label{eq:end-glow-temperature-estimate}
\end{equation}
This is not a calibrated pyrometric measurement, because the camera response, exposure, emissivity, optical transmission, and viewing geometry are not known. Nevertheless, it provides a physically useful order-of-magnitude scale for the RF-induced end-heating power.

Consider an end region of length $\ell_{\text{end}}$ heated to an effective temperature $T_{\text{end}}$. The radiated power from this region is approximately
\begin{equation}
	P_{\text{rad,end}}
	\simeq
	\varepsilon \sigma_{\text{SB}} p_{\text{rad}} \ell_{\text{end}}
	\left(
	T_{\text{end}}^4 - T_0^4
	\right),
	\label{eq:end-glow-radiative-power-estimate}
\end{equation}
where $\varepsilon$ is the effective emissivity, $p_{\text{rad}}$ is the radiating perimeter of the strip, and $T_0$ is the ambient temperature. For a thin strip with width $w$ much larger than its thickness, $p_{\text{rad}}\simeq 2w$ is a useful approximation.

Using $w\simeq\SI{10}{\micro\meter}$, $p_{\text{rad}}\simeq 2w$, $\varepsilon\simeq0.85$, $\ell_{\text{end}}\simeq\SI{8}{mm}$, $T_{\text{end}}\simeq\SI{1470}{K}$, and $T_0=\SI{300}{K}$ gives
\begin{equation}
	P_{\text{rad,end}}
	\simeq
	\SI{36}{mW}.
	\label{eq:end-glow-radiative-power-1200C}
\end{equation}
Equivalently,
\begin{widetext}
\begin{equation}
	P_{\text{rad,end}}
	\simeq
	\SI{36}{mW}
	\left(
	\frac{\varepsilon}{0.85}
	\right)
	\left(
	\frac{w}{\SI{10}{\micro\meter}}
	\right)
	\left(
	\frac{\ell_{\text{end}}}{\SI{8}{mm}}
	\right)
	\left[
	\frac{
		T_{\text{end}}^4 - T_0^4
	}{
		(\SI{1470}{K})^4 - T_0^4
	}
	\right].
	\label{eq:end-glow-radiative-power-scale}
\end{equation}
\end{widetext}
Thus the observed end glow implies an RF-induced end-heating power of order a few \SI{10}{mW} per glowing end region, with the dominant uncertainty coming from the effective emitting length, emissivity, and the uncalibrated optical temperature estimate.

The same estimate can be converted into an RF-current scale. If the local end resistance is $R_{\text{end}}$, then
\begin{equation}
	P_{\text{end}}
	\simeq
	I_{\text{RF,rms}}^2 R_{\text{end}},
	\label{eq:end-heating-current-scale}
\end{equation}
and therefore
\begin{equation}
	I_{\text{RF,rms}}
	\simeq
	\left(
	\frac{P_{\text{end}}}{R_{\text{end}}}
	\right)^{1/2}.
	\label{eq:end-heating-current-required}
\end{equation}
RHIC measurements found resistances of order \SIrange{200}{800}{M\ohm} for unexposed carbon micro-ribbons, while beam-exposed targets had resistances of order \SI{1}{M\ohm}, possibly due to graphitization or other beam-induced modification~\cite{Steski2014CarbonMicroRibbons}. For a strip length of order \SI{25}{mm}, the unexposed resistance range corresponds to
\begin{equation}
	R'
	\simeq
	\SIrange{8}{32}{M\ohm.mm^{-1}}.
	\label{eq:strip-resistance-per-length-scale}
\end{equation}
For a glowing length $\ell_{\text{end}}\simeq\SI{8}{mm}$, this gives
\begin{equation}
	R_{\text{end}}
	\simeq
	R'\ell_{\text{end}}
	\simeq
	\SIrange{64}{256}{M\ohm}.
	\label{eq:end-resistance-scale}
\end{equation}

If a power of order $P_{\text{end}}\simeq\SI{36}{mW}$ is deposited in this high-resistance end region, the required RF current is only
\begin{equation}
	I_{\text{RF,rms}}
	\simeq
	\SIrange{12}{24}{\micro A},
	\label{eq:end-heating-current-numerical}
\end{equation}
where the range reflects the resistance interval in Eq.~\eqref{eq:end-resistance-scale}. The corresponding effective RF voltage scale is
\begin{equation}
	V_{\text{RF,rms}}
	\simeq
	I_{\text{RF,rms}} R_{\text{end}}
	\sim
	\SIrange{1.5}{6}{kV}.
	\label{eq:end-heating-voltage-scale}
\end{equation}

This estimate gives a quantitative target for the wakefield calculation. The question is not whether the wakefield produces large macroscopic currents, but whether the holder and strip-end geometry can drive currents at the level of order \SI{10}{\micro A} through a high-resistance carbon end segment, or equivalently produce effective RF voltages at the kilovolt scale in the local end circuit. This is the scale against which chamber and target-holder wakefield calculations should be compared when assessing whether RF-induced end heating is quantitatively plausible.

\subsection{Example estimate of the RF drive from wakefield calculations}
\label{sec:wakefield-voltage-scale-target-holders}

The purpose of the wakefield calculation in the present model is not to compute the microscopic current distribution inside the carbon micro-ribbon. Instead, it provides the RF or wakefield drive produced by the chamber and target-holder geometry. This distinction is essential because the relevant electromagnetic resonances are determined by the beam chamber, target holder, dielectric pieces, gaps, coatings, ports, and boundary conditions, whereas the carbon strip has dimensions far below the chamber scale. In particular, resolving a strip thickness of order \SI{50}{nm} in a millimeter-to-centimeter-scale time-domain wakefield calculation would require an impractically fine mesh. In addition, the CST wakefield solver used for the simulation does not admit any object that directly intercepts the particle beam. The carbon strip is therefore treated later as an effective electrical load, characterized by a strip-end resistance, contact impedance, and possible surface or coating current paths. The calculation is therefore used here as a scale-setting example, not as a complete electromagnetic-thermal model of the carbon-strip target.

\begin{figure*}[t]
	\centering
	
	\begin{subfigure}[t]{0.323\textwidth}
		\centering
		\includegraphics[width=\textwidth]{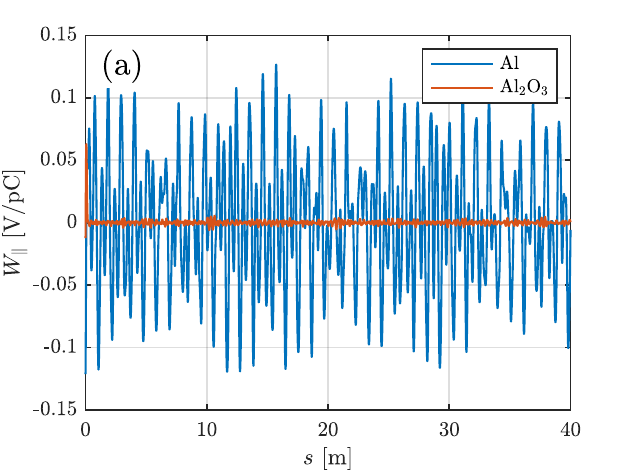}
		\caption{Longitudinal wake potential $W_{\parallel}(s)$.}
		\label{fig:medani_wake_panel_wake}
	\end{subfigure}
	\hspace{0.01cm}
	\begin{subfigure}[t]{0.323\textwidth}
		\centering
		\includegraphics[width=\textwidth]{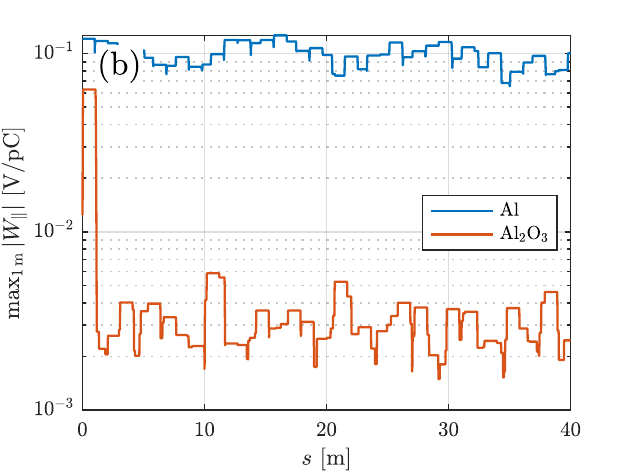}
		\caption{Moving maximum of $|W_{\parallel}(s)|$ over a \SI{1}{m} window.}
		\label{fig:medani_wake_panel_envelope}
	\end{subfigure}
	\hspace{0.01cm}
	\begin{subfigure}[t]{0.323\textwidth}
		\centering
		\includegraphics[width=\textwidth]{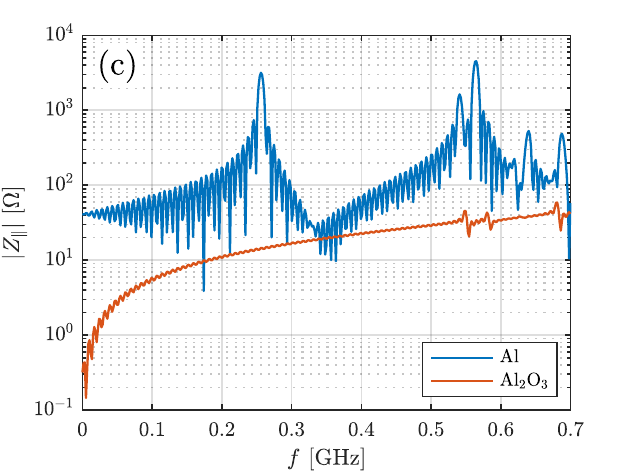}
		\caption{Low-frequency longitudinal impedance magnitude $|Z_{\parallel}(f)|$.}
		\label{fig:medani_wake_panel_impedance}
	\end{subfigure}
	
	\caption{
		CST wakefield comparison for aluminum and Al$_2$O$_3$ target-holder geometries.
		The coordinate $s$ denotes the longitudinal wake coordinate relative to the source bunch, with $s=0$ at the bunch position and $s>0$ corresponding to trailing positions behind the bunch.
		The aluminum holder produces both large local wake-potential excursions and a persistent long-range wake envelope.
		The Al$_2$O$_3$ holder does not eliminate the largest local wake excursion, but strongly suppresses the long-range wake envelope and reduces the low-frequency resonant impedance structure.
	}
	\label{fig:medani_cst_wakefield_comparison}
\end{figure*}

Figure~\ref{fig:medani_cst_wakefield_comparison} shows wakefield calculations for candidate EIC target-holder geometries, comparing an aluminum holder with an Al$_2$O$_3$ holder. The wakefield and impedance comparison uses an rms bunch length of $\sigma_L=\SI{60}{mm}$ and $N_b=290$ bunches, consistent with the EIC entries in Table~\ref{tab:target_parameters}, together with a bunch charge of \SI{30.5}{nC}. The three panels show the longitudinal wake potential $W_{\parallel}(s)$, the running maximum of $|W_{\parallel}(s)|$ taken over a sliding \SI{1}{m} window and shown across the full \SI{40}{m} wake length, and the corresponding low-frequency longitudinal impedance magnitude. The comparison is directly relevant to the end-heating estimate of Sec.~\ref{sec:end-glow-power-scale}, because the wake potential gives the voltage scale available to drive RF current through the strip-end region.

From Fig.~\ref{fig:medani_cst_wakefield_comparison}, the aluminum-holder wake potential reaches local peak values of order
\begin{equation}
	\max_{s>0}|W_{\parallel}^{\text{Al}}|
	\simeq
	\SI{0.13}{V/pC}.
	\label{eq:wake-potential-al-peak-scale}
\end{equation}
For the Al$_2$O$_3$ holder, the largest local wake excursion is smaller, but not eliminated,
\begin{equation}
	\max_{s>0}|W_{\parallel}^{\text{Al}_2\text{O}_3}|
	\simeq
	\SI{0.063}{V/pC}.
	\label{eq:wake-potential-alumina-peak-scale}
\end{equation}
Thus, the peak wake potential is reduced only by about a factor of two when the aluminum holder is replaced by Al$_2$O$_3$. This is important because the largest local wake excursion remains at the level of several $10^{-2}\,\si{V/pC}$, large enough to provide a substantial RF voltage scale for EIC bunch charges.

For an EIC bunch charge of order
\begin{equation}
	q_b
	\simeq
	\SIrange{10}{20}{nC}
	=
	\SIrange{1e4}{2e4}{pC},
	\label{eq:eic-bunch-charge-scale}
\end{equation}
the corresponding single-bunch wake-voltage scale is
\begin{equation}
	|V_{\text{wake}}|
	=
	|W_{\parallel}| q_b .
	\label{eq:wake-voltage-scale}
\end{equation}
Using the peak wake-potential values gives
\begin{equation}
\begin{split}	
	|V_{\text{wake}}^{\text{Al}}|
	\simeq
	\SIrange{1.3}{2.6}{kV}, \\
	|V_{\text{wake}}^{\text{Al}_2\text{O}_3}|
	\simeq
	\SIrange{0.6}{1.3}{kV}.
	\label{eq:wake-voltage-peak-scale}
\end{split}
\end{equation}
These values should not be interpreted as DC voltages across the carbon strip. They define the available RF wake-voltage scale that can couple to the strip ends through the local contact impedance, return-current geometry, metallization, holder structure, and surrounding chamber boundary conditions.

The more significant difference appears in the long-range wake envelope. For $s>0$, the RMS wake potential is approximately
\begin{equation}
	\begin{split}
	W_{\text{rms}}^{\text{Al}}
	& \simeq
	\SI{0.052}{V/pC},
	\\
	W_{\text{rms}}^{\text{Al}_2\text{O}_3}
	& \simeq
	\SI{0.003}{V/pC}.
	\label{eq:wake-rms-scale}
	\end{split}
\end{equation}
The Al$_2$O$_3$ holder therefore suppresses the persistent wake drive by more than an order of magnitude, even though it does not remove the largest local wakefield excursion. The impedance panel in Fig.~\ref{fig:medani_cst_wakefield_comparison} shows the same qualitative behavior: the aluminum holder has pronounced low-frequency resonant structure, while the Al$_2$O$_3$ holder reduces this structure substantially.

The conclusion from this scale comparison is that realistic chamber and target-holder wakefields can plausibly provide an RF drive large enough to contribute to strip-end heating. The aluminum holder gives both large local wake excursions and a persistent long-range wake. The Al$_2$O$_3$ holder is preferable because it suppresses the long-range wake envelope and the low-frequency impedance structure, but residual strip-end glow remains possible if the local strip-end coupling is efficient.

This comparison should not be interpreted as a complete prediction of the carbon-strip end temperature. The wakefield calculation gives the available electromagnetic drive. The local temperature rise depends on how that drive couples to the carbon strip, the strip-holder contact, the thin (\(\sim\SI{10}{\nano\meter}\)) Au coating on the Al$_2$O$_3$ holder surface, and the return-current path. This remaining step is described by the effective RF-current model in Sec.~\ref{sec:effective-rf-end-heating}.

\subsection{Effective RF-current model for strip-end heating}
\label{sec:effective-rf-end-heating}

To include RF-induced end heating in the thermal model, the wakefield drive is represented by an effective local Joule-heating term. For a one-dimensional strip coordinate $s$, the additional end-localized heat source can be written as
\begin{equation}
	Q_{\text{RF}}(s,t)
	=
	\frac{p'_{\text{RF}}(s,t)}{A},
	\label{eq:qrf-effective-source}
\end{equation}
where $A$ is the carbon-strip cross-sectional area and $p'_{\text{RF}}$ is the RF-induced power deposited per unit strip length. A minimal parametrization is
\begin{equation}
	p'_{\text{RF}}(s,t)
	=
	I_{\text{RF}}^2(t)\,R'(s,t),
	\label{eq:rf-heating-current-form}
\end{equation}
where $I_{\text{RF}}(t)$ is an effective RF or wakefield-driven current and $R'(s,t)$ is the local resistance per unit length of the carbon strip.

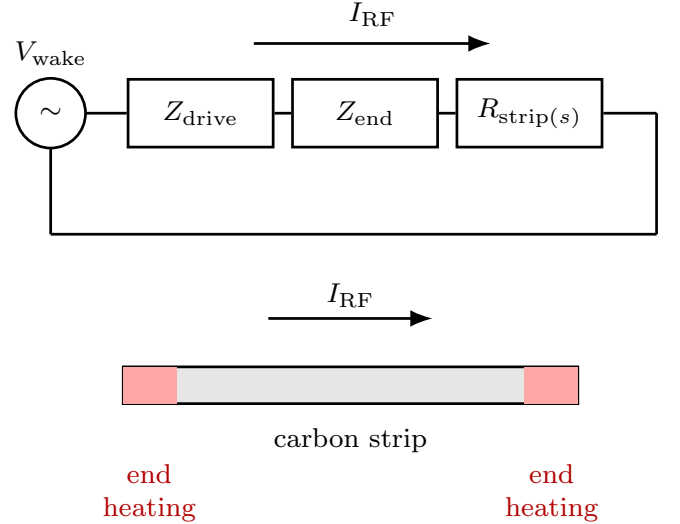
\begin{figure}[t]
	\centering
	\resizebox{\columnwidth}{!}{%
		\begin{tikzpicture}[
			>=Latex,
			font=\footnotesize,
			box/.style={
				draw,
				thick,
				align=center,
				minimum width=15mm,
				minimum height=7mm,
				inner sep=2pt
			},
			arrow/.style={-Latex, thick},
			line/.style={thick}
			]
			
			
			\draw[line] (0,0) circle (0.36);
			\node at (0,0) {$\sim$};
			\node at (0,0.62) {$V_{\mathrm{wake}}$};
			
			\node[box] (zdrive) at (1.55,0) {$Z_{\mathrm{drive}}$};
			\node[box] (zend)   at (3.25,0) {$Z_{\mathrm{end}}$};
			\node[box] (rstrip) at (4.95,0) {$R_{\mathrm{strip}(s)}$};
			
			\draw[line] (0.36,0) -- (zdrive.west);
			\draw[line] (zdrive.east) -- (zend.west);
			\draw[line] (zend.east) -- (rstrip.west);
			
			\coordinate (retA) at ($(rstrip.east)+(0.55,0)$);
			\coordinate (retB) at ($(rstrip.east)+(0.55,-1.25)$);
			\draw[line] (rstrip.east) -- (retA);
			\draw[line] (retA) -- (retB);
			\draw[line] (retB) -- (0,-1.25);
			\draw[line] (0,-1.25) -- (0,-0.36);
			
			\draw[arrow] (2.10,0.72) -- (4.55,0.72);
			\node at (3.33,1.02) {$I_{\mathrm{RF}}$};
			
			\draw[fill=gray!20,thick] (0.75,-3.00) rectangle (5.45,-2.62);
			
			\draw[fill=red!35,draw=none] (0.75,-3.00) rectangle (1.30,-2.62);
			\draw[fill=red!35,draw=none] (4.90,-3.00) rectangle (5.45,-2.62);
			
			\node at (3.10,-3.38) {carbon strip};
			
			\node[red!70!black,align=center] at (1.02,-3.92) {end\\heating};
			\node[red!70!black,align=center] at (5.18,-3.92) {end\\heating};
			
			\draw[arrow] (2.25,-2.12) -- (3.95,-2.12);
			\node at (3.10,-1.88) {$I_{\mathrm{RF}}$};
			
		\end{tikzpicture}%
	}
		\caption{Equivalent-circuit sketch for RF-induced strip-end heating. The wakefield drive is represented by $V_{\mathrm{wake}}$, coupled through $Z_{\mathrm{drive}}$ and an effective strip-end impedance $Z_{\mathrm{end}}$ to the strip resistance. The RF current $I_{\mathrm{RF}}$ can produce localized Joule heating near the strip attachments, with the largest dissipation where the local resistance per unit length, represented by $R_{\mathrm{strip}}(s)$, is largest. }
	\label{fig:rf_end_heating_circuit}
\end{figure}

This form makes explicit why RF-induced end heating depends sensitively on the local electrical state of the strip. The deposited power is largest where the induced RF current passes through a high-resistance section, so the relevant quantity is not only the total strip resistance but the local resistance near the strip attachment. RHIC carbon-target studies reported that the resistance of unexposed carbon micro-ribbons was typically much larger than the resistance measured after beam exposure, with changes attributed to beam-induced structural modification of the carbon, including graphitization, crystallization, or damage~\cite{Steski2014CarbonMicroRibbons,Steski2018TargetLifetime}. The resistance should therefore be treated as spatially nonuniform and history dependent,
\begin{equation}
	R_{\text{strip}}
	\rightarrow
	R(s,t).
	\label{eq:strip-resistance-history-dependent}
\end{equation}
In this picture (Fig.~\ref{fig:rf_end_heating_circuit}), a beam-modified central section and less-modified strip ends can have very different resistance per unit length. RF-current dissipation can therefore be concentrated near the attachments even when the wakefield drive is generated by the surrounding chamber and holder geometry. The observed targets are not virgin strips but have accumulated extensive prior beam exposure, so their central section is already graphitized to a low resistance while the ends remain comparatively high-resistance. RF dissipation is then largest at the ends, consistent with the observation that the strip ends glow before the beam intercepts the lower-resistance center.

Equivalently, if the wakefield calculation is reduced to an effective RF voltage at the strip-end region, the deposited power at frequency $\omega_m$ may be written in circuit form as
\begin{equation}
	P_{\text{end},m}
	=
	\frac{1}{2}
	\frac{
		\left|V_{\text{RF},m}\right|^2
		\operatorname{Re} Z_{\text{end}}(\omega_m)
	}{
		\left|Z_{\text{drive}}(\omega_m)+Z_{\text{end}}(\omega_m)\right|^2
	},
	\label{eq:rf-heating-voltage-form}
\end{equation}
where $V_{\text{RF},m}$ is the effective wake-induced voltage at harmonic $\omega_m$, $Z_{\text{drive}}$ represents the source, chamber, and holder coupling, and $Z_{\text{end}}$ represents the carbon-strip end, contact, coating, and return path. The total end-heating power is then obtained by summing over the bunch-spectrum harmonics,
\begin{equation}
	P_{\text{end}}
	=
	\sum_m P_{\text{end},m}.
	\label{eq:rf-end-power-sum}
\end{equation}

The total heat source in the thermal equation is then generalized from the direct beam-heating term alone to
\begin{equation}
	Q_{\text{tot}}(s,t)
	=
	Q_{\text{heat}}(s,t)
	+
	Q_{\text{RF}}(s,t).
	\label{eq:qtot-direct-plus-rf}
\end{equation}
The first term describes stopping-power energy retained in the carbon strip near the beam-intercept point, while the second term describes RF or wakefield-induced Joule heating near the strip attachments. The two terms are physically distinct and can have different spatial and temporal dependences.

For the present model, $Q_{\text{RF}}$ should not be treated as an arbitrary fit parameter. Its normalization is constrained by the wakefield or impedance calculation, its spatial distribution is constrained by the holder and contact geometry, and its temperature response is constrained by the visible end glow and resistance changes observed in RHIC operation. The comparison of different chamber and holder geometries can then be made at two levels: first by comparing the available wake-voltage scale, long-range wake envelope, and impedance structure, as in Fig.~\ref{fig:medani_cst_wakefield_comparison}, and then by evaluating how much of that RF drive is dissipated near the carbon-strip ends through Eqs.~\eqref{eq:rf-heating-current-form} or \eqref{eq:rf-heating-voltage-form}.

\subsection{Role of holder geometry and surface conductivity}
\label{sec:holder-geometry-surface-conductivity}

The estimates above show how holder geometry and surface conductivity enter the carbon-strip response model. The holder should reduce the RF or wakefield drive available to the strip ends and should avoid creating a persistent resonant wake environment that can maintain end heating over many bunch passages. In current terms, the end-heating risk corresponds to effective RF currents through short high-resistance carbon end regions at the tens-of-microamps scale, unless the local end resistance is reduced by a controlled electrical path.

The aluminum-holder calculations are therefore relevant as a conservative or unfavorable comparison case. A metallic holder can provide strong local coupling and well-defined conducting boundaries, but it can also enhance wakefield amplitudes, long-range wake envelopes, and impedance peaks near the target ends. The Al$_2$O$_3$ holder concept reduces the persistent wake envelope and the low-frequency impedance structure. However, the local wake excursion is not eliminated, so residual strip-end heating remains possible if the strip-end coupling is efficient.

The relevant model conclusion is therefore not simply that the holder should be insulating or conducting. What matters is the electromagnetic boundary condition seen by the strip ends. A weakly wake-coupled dielectric body with a controlled surface conductivity is one possible way to reduce the persistent RF drive. In the example above, the Al$_2$O$_3$ holder is chosen primarily to reduce the wakefield coupling and the low-frequency impedance structure. The thin (\(\sim\SI{10}{\nano\meter}\)) Au coating on the ceramic body is then a construction detail that helps in additional ways. It keeps the ceramic surface conducting enough that the holder does not charge up, yet a thin metal film on a dielectric body presents a much smaller stored-charge reservoir than a solid metallic holder. Because the charge involved in the end-region response is itself small, as estimated in Eq.~\eqref{eq:required-elementary-charges}, the limited charge that this thin surface can supply restricts the transient charge buildup at the strip ends, and hence the associated end heating, more than a solid metal holder would. The ceramic body with a thin conducting coating therefore tends to be more favorable at the strip ends than a solid metal holder. This is consistent with the RHIC experience that local end-field shaping, using rounded metal fins, improved target lifetime but did not eliminate the need to control the electromagnetic environment of the strip ends.

In the complete response model, the EIC holder choice enters through both the thermal and mechanical parts of the calculation. It changes the available RF drive for end heating, and it also changes the local electrostatic boundary conditions that determine charge buildup and beam-induced force. In such a model, the relevant quantities are not only total wake power, but also the effective strip-end voltage, long-range wake envelope, end-current path, charge relaxation time, and force on a weakly tensioned carbon strip.

\section{Beam-induced force and mechanical response}
\label{sec:beam-induced-forces}

The thermal model in Sec.~\ref{sec:thermal-response-carbon-strip} describes the temperature rise caused by direct energy deposition in the carbon strip. RHIC video observations, however, show an additional effect: the carbon strip can be attracted toward the beam center. This section estimates the force scale required for visible strip motion and compares possible beam-induced force mechanisms. The central question is which mechanism can produce a time-averaged transverse force on the movable carbon strip. The target holder and fork are mechanically fixed, but they enter the problem as electrical boundary conditions that can supply, remove, or redistribute charge in the strip.

\subsection{Observed force symmetry and range}
\label{sec:observed-force-symmetry}

As summarized in Fig.~\ref{fig:carbon_strip_scan_schematic} and Table~\ref{tab:video_observation_parameters}, the observed deformation has the symmetry of an attractive force toward the beam center. When the strip is on one side of the beam, it is displaced toward the beam; after the target has crossed to the other side, the sign of the displacement reverses and the strip is again deflected toward the beam center. The force is therefore not a fixed-direction mechanical push in the laboratory frame.

The apparent range is also important. The deformation is visible over many millimeters during a target scan, and the inferred force direction remains toward the beam center. This range provides an important constraint on possible force mechanisms. A mechanism that is strong only very close to the beam may contribute near the beam center, but it is unlikely to explain the full observed deformation pattern by itself.

\subsection{Mechanical force scale for visible strip displacement}
\label{sec:mechanical-force-scale}

The force required for visible strip displacement is set mainly by the effective axial tension of the mounted carbon strip. A simple estimate is obtained by treating the displaced strip as two straight string segments joined at the point of maximum transverse displacement. For a midpoint displacement $\Delta x$, each half of the strip has length $L/2$ and makes a small angle $\theta$ with the original strip direction. For $\Delta x\ll L$,
\begin{equation}
	\tan\theta
	\simeq
	\sin\theta
	\simeq
	\frac{\Delta x}{L/2}
	=
	\frac{2\Delta x}{L}\,.
	\label{eq:slack-string-angle}
\end{equation}
The transverse restoring force is the sum of the transverse components of the tension from the two strip halves,
\begin{equation}
	F_{\text{mech}}
	\simeq
	2T\sin\theta
	\simeq
	\frac{4T}{L}\,\Delta x\,,
	\label{eq:slack-string-force}
\end{equation}
where $T$ is the axial tension force. Writing $T=\sigma_{\text{eff}}A$, with effective mounted-strip stress $\sigma_{\text{eff}}$ and strip cross-sectional area $A$, gives the required force scale. Using the RHIC carbon-strip length $L=\SI{25}{mm}$ and the video-derived projected displacement scale $\Delta x_{\mathrm{proj}}\simeq\SI{3}{mm}$ from Table~\ref{tab:video_observation_parameters},
\begin{equation}
	F_{\text{mech}}
	\simeq
	0.48\,T
	=
	0.48\,\sigma_{\text{eff}}A\,.
	\label{eq:slack-string-force-numeric}
\end{equation}
With the strip width and thickness from Table~\ref{tab:thermal_model_inputs}, the cross-sectional area is $A=wt=\SI{5e-13}{m^2}$, and Eq.~\eqref{eq:slack-string-force-numeric} gives
\begin{equation}
	F_{\text{mech}}
	\simeq
	\SI{2.4e-7}{N}
	\left(
	\frac{\sigma_{\text{eff}}}{\SI{1}{MPa}}
	\right).
	\label{eq:mechanical-force-scale}
\end{equation}

The effective mounted-strip stress $\sigma_{\mathrm{eff}}$ is not the same as the intrinsic residual film stress $\sigma_{\mathrm{res}}$ listed in Table~\ref{tab:carbon_properties}. The latter provides an orientation scale for deposited carbon films, while the former is the actual tension-carrying stress of the mounted, possibly twisted, heated, or relaxed free strip. We therefore treat $\sigma_{\mathrm{eff}}$ as a model parameter and use the representative range $\sigma_{\mathrm{eff}}\sim\SI{0.1}{MPa}$ to $\SI{10}{MPa}$. For this range $\sigma_{\text{eff}}\sim\SI{0.1}{MPa}$ to $\SI{10}{MPa}$, millimeter-scale displacement requires only
\begin{equation}
	F_{\text{mech}}
	\sim
	\SI{1e-8}{N}
	\text{ to }
	\SI{1e-6}{N}.
	\label{eq:mechanical-force-range}
\end{equation}

Although this force range is small on macroscopic scales, it is large for a freely suspended nanometer-scale carbon strip: for the nominal RHIC strip geometry, the strip weight is only \(mg\simeq\SI{2e-10}{N}\), well below the inferred nanonewton-to-micronewton force scale.

The video observations also constrain the strip shape geometrically. The camera records the projection onto the transverse plane, and in the beam-interception frames the strip appears essentially straight across the bright central spot (Fig.~\ref{fig:carbon_strip_scan_schematic}, $t=\SI{10}{s}$ and $\SI{17}{s}$). Since the mounted strip is not under high tension, its slack cannot vanish; if the projection is straight, the slack must lie out of the camera plane, along the beam axis, caused by the magnetic Lorentz force on a radial strip current, which points along $\pm z$ [Eq.~\eqref{eq:radial-current-longitudinal-force}].

\subsection{Magnetic Lorentz-force geometry}
\label{sec:magnetic-lorentz-force-geometry}

A magnetic Lorentz force from current in the free carbon strip is not the most natural explanation for the observed transverse attraction. Let the beam direction be $\vec{e}_z$. The beam magnetic field is azimuthal,
\begin{equation}
	\vec{B}_{\text{beam}}(r,t)
	=
	B_{\phi}(r,t)\,\vec{e}_{\phi}\,.
	\label{eq:beam-azimuthal-magnetic-field}
\end{equation}
The force on a target current element is
\begin{equation}
	d\vec{F}(t)
	=
	I_t(t)\,d\vec{l}\times \vec{B}_{\text{beam}}(r,t)\,.
	\label{eq:lorentz-force-current-element}
\end{equation}
For current mainly along the free strip, $d\vec{l}$ is mainly transverse to the beam. A radial current element gives
\begin{equation}
	d\vec{l}\parallel\vec{e}_r
	\quad\Rightarrow\quad
	d\vec{l}\times\vec{B}_{\phi}
	\parallel
	\vec{e}_z\,,
	\label{eq:radial-current-longitudinal-force}
\end{equation}
and an azimuthal current element gives
\begin{equation}
	d\vec{l}\parallel\vec{e}_{\phi}
	\quad\Rightarrow\quad
	d\vec{l}\times\vec{B}_{\phi}
	\simeq
	0\,.
	\label{eq:azimuthal-current-force}
\end{equation}
A radial force requires a current component parallel to the beam,
\begin{equation}
	d\vec{l}\parallel\vec{e}_z
	\quad\Rightarrow\quad
	d\vec{l}\times\vec{B}_{\phi}
	\parallel
	-\vec{e}_r\,,
	\label{eq:beam-parallel-current-radial-force}
\end{equation}
with the sign depending on the current direction. Thus, unless the movable carbon segment carries a significant beam-parallel current component, the Lorentz force does not explain the observed beam-centered attraction. 

\subsection{Net or local charge force}
\label{sec:net-local-charge-force}

A simple electrostatic force scale is obtained by considering a net or locally redistributed charge on the target in the external electric field of the beam. In the following estimate, the beam is approximated as a line charge and the target response is reduced to the force on that charge~\cite{JacksonClassicalElectrodynamics}. The radial electric field of the beam is
\begin{equation}
	E(r)
	=
	\frac{\lambda}{2\pi\epsilon_0 r},
	\qquad
	\lambda
	=
	\frac{I_{\text{pk}}^{b}}{\beta c}\,,
	\label{eq:beam-electric-field-line-charge}
\end{equation}
where $I_{\text{pk}}^{b}$ is the peak bunch-current scale. For the RHIC flattop reference case, Table~\ref{tab:beam_parameters_rhic_eic} gives $I_{\text{pk}}^{b}=\SI{6.97}{A}$. With $\beta\simeq 1$, Eq.~\eqref{eq:beam-electric-field-line-charge} gives
\begin{equation}
	\begin{split}
	E(\SI{1}{mm})
	\simeq
	4.2\times10^5\,\si{V.m^{-1}}, \\
	E(\SI{5}{mm})
	\simeq
	8.4\times10^4\,\si{V.m^{-1}}\,.
	\label{eq:beam-electric-field-rhic-reference}
	\end{split}
\end{equation}
A net or local charge $q$ on the movable part of the carbon strip experiences\,\cite{JacksonClassicalElectrodynamics,LandauLifshitzPitaevskii1984Electrodynamics}
\begin{equation}
	F_q
	=
	qE\,.
	\label{eq:electrostatic-force-net-charge}
\end{equation}
Using the lower end of the mechanical reference scale, $F_q=\SI{1e-8}{N}$, the required charge at $r=\SI{5}{mm}$ is
\begin{equation}
	q_{\text{req}}
	=
	\frac{F_q}{E}
	\simeq
	1.2\times10^{-13}\,\si{C}\,.
	\label{eq:required-charge-for-visible-force}
\end{equation}
In elementary charges, this corresponds to
\begin{equation}
	N_e
	=
	\frac{|q_{\text{req}}|}{e}
	\simeq
	7.4\times10^5\,.
	\label{eq:required-elementary-charges}
\end{equation}
This is still a very small electron number for a connected strip-holder system. A negative charge on the movable strip is attracted to the positive bunch field, giving the correct force direction from either side of the beam. The estimate is therefore only a charge-scale benchmark, not a mechanism.

\subsection{Polarization force of an isolated carbon strip}
\label{sec:polarization-force-isolated-strip}

A neutral but polarizable strip can also be attracted toward the beam because the beam electric field is spatially nonuniform. For an induced dipole in an electric field, the force scale is
\begin{equation}
	F_{\text{pol}}
	\simeq
	\frac{1}{2}\alpha\nabla E^2\,,
	\label{eq:polarization-force-estimate}
\end{equation}
where $\alpha$ is the effective polarizability of the movable object. For the line-charge field of Eq.~\eqref{eq:beam-electric-field-line-charge}, $E\propto r^{-1}$, so the isolated-object polarization force scales approximately as
\begin{equation}
	F_{\text{pol}}
	\sim
	\alpha\frac{E^2}{r}
	\propto
	\frac{1}{r^3}\,.
	\label{eq:polarization-force-line-charge-scale}
\end{equation}

The estimate depends strongly on the transverse geometry assigned to the strip. A round-filament approximation based on the strip cross-sectional area gives
\begin{equation}
	a_{\text{eff}}
	=
	\sqrt{\frac{A}{\pi}}
	\simeq
	\SI{0.4}{\micro\meter}\,,
	\label{eq:effective-filament-radius}
\end{equation}
using the area $A$ defined in Table~\ref{tab:thermal_model_inputs}. The corresponding transverse polarizability scale of a long conducting filament is
\begin{equation}
	\alpha_{\text{round}}
	\sim
	2\pi\epsilon_0 a_{\text{eff}}^2 L
	\sim
	2\times10^{-25}\,
	\si{C.m^2.V^{-1}}\,.
	\label{eq:round-filament-polarizability-numerical}
\end{equation}
This is a conservative lower-scale estimate because it replaces the flat ribbon by a much narrower effective round filament. A flat-ribbon estimate uses the half-width as the relevant transverse polarization scale,
\begin{equation}
\begin{split}
	a_w & = \frac{w}{2} \simeq \SI{5}{\micro\meter}\,, \\
	\alpha_w & \sim 2\pi\epsilon_0 a_w^2 L \sim 3.5\times10^{-23}\, \si{C.m^2.V^{-1}}\,.
\end{split}
	\label{eq:flat-ribbon-width-polarizability}
\end{equation}
Thus the flat-ribbon scale is about two orders of magnitude larger than the round-filament scale.

Using the RHIC peak-field estimate from Eq.~\eqref{eq:beam-electric-field-rhic-reference}, the broad-side isolated-strip polarization force at $r=\SI{1}{mm}$ is of order
\begin{equation}
	F_{\text{pol}}(\SI{1}{mm})
	\sim
	\alpha_w\frac{E^2}{r}
	\sim
	6\times10^{-9}\,\si{N}\,.
	\label{eq:polarization-force-1mm-flat-ribbon}
\end{equation}
At $r=\SI{5}{mm}$, the same estimate is reduced by about a factor $5^3$,
\begin{equation}
	F_{\text{pol}}(\SI{5}{mm})
	\sim
	5\times10^{-11}\,\si{N}\,.
	\label{eq:polarization-force-5mm-flat-ribbon}
\end{equation}
The isolated-strip polarization force therefore has the correct attractive symmetry and may contribute close to the beam, especially for a flat or twisted ribbon. However, its approximate $1/r^3$ falloff makes it too short-ranged to be the primary explanation for deformation visible over many millimeters.

\subsection{Bunch-driven charge redistribution and RF current}
\label{sec:bunch-driven-charge-redistribution}

A positive bunch passing near the C strip produces a strong time-dependent electric field. This field can displace mobile electrons along the connected C strip, contacts, and target holder. The relevant microscopic process is therefore beam-induced charge redistribution in the C-strip and target holder. The charge motion can include both displacement current and real conduction through the C strip, contacts, and holder.

The charge on the movable C strip can be decomposed schematically as
\begin{equation}
	q(t)
	=
	q_0
	+
	\tilde{q}(t)\,.
	\label{eq:charge-decomposition}
\end{equation}
Here $q_0$ denotes a slowly relaxing or residual charge offset, while $\tilde{q}(t)$ denotes the bunch-synchronous charge displacement. A negative $q_0$ gives an attractive force toward the positive beam. The oscillating part produces an RF current,
\begin{equation}
	I_{\text{RF}}(t)
	=
	\frac{d\tilde{q}}{dt}\,.
	\label{eq:rf-current-charge-motion}
\end{equation}
Thus the same bunch-driven charge motion can contribute to transverse attraction through $q(t)E(t)$ and to strip-end heating through $I_{\text{RF}}^2R$.

The electrical relaxation time of the strip-holder circuit is
\begin{equation}
	\tau_{\text{RC}}
	=
	R_{\text{eff}}C_{\text{eff}}\,.
	\label{eq:strip-holder-rc-time}
\end{equation}

A crude lower-scale estimate for the capacitance of the free C strip can be obtained by approximating it as a long thin conductor. The estimate follows the standard logarithmic line-charge result for cylindrical conductors~\cite{JacksonClassicalElectrodynamics},
\begin{equation}
	C_{\text{strip}}
	\sim
	\frac{2\pi\epsilon_0 L}
	{\ln(2L/a_{\text{eff}})}\,,
	\label{eq:strip-capacitance-estimate}
\end{equation}
where $a_{\text{eff}}=\sqrt{A/\pi}$ is the radius of a round conductor with the same cross-sectional area. Using the RHIC active strip length $L=\SI{25}{mm}$ and the area from Table~\ref{tab:thermal_model_inputs} gives $C_{\text{strip}}\sim\SI{0.1}{pF}$. This should be interpreted only as an order-of-magnitude capacitance scale for the suspended strip; the effective circuit capacitance also includes the strip ends, contacts, holder surfaces, and stray capacitance to nearby conductors.

For $R_{\text{eff}}\sim\SIrange{1}{10}{\mega\ohm}$, one obtains $\tau_{\text{RC}}\sim\SIrange{0.1}{10}{\micro s}$. For high-resistance C-strip end regions, $R_{\text{eff}}\sim\SIrange{100}{300}{\mega\ohm}$ gives $\tau_{\text{RC}}\sim\SIrange{10}{300}{\micro s}$. These values are much longer than the bunch spacing $\tau_b$ listed in Table~\ref{tab:beam_parameters_rhic_eic}, especially for the high-resistance end regions. The charge distribution therefore need not return to equilibrium between bunches.

The RF-current scale required for the observed strip-end glow is modest. From the end-glow estimate in Sec.~\ref{sec:end-glow-power-scale}, $P_{\text{end}}\sim\SI{3e-2}{W}$ per glowing end region. For $R_{\text{end}}\sim\SIrange{60}{300}{\mega\ohm}$,
\begin{equation}
	I_{\text{RF}}^{\text{rms}}
	=
	\sqrt{\frac{P_{\text{end}}}{R_{\text{end}}}}
	\sim
	\SIrange{10}{25}{\micro A}\,.
	\label{eq:rf-current-required-end-glow}
\end{equation}
At the EIC flattop bunch frequency listed in Table~\ref{tab:beam_parameters_rhic_eic}, this current corresponds to a charge amplitude
\begin{equation}
	\begin{split}
	Q_0
	& =
	\frac{\sqrt{2}\,I_{\text{RF}}^{\text{rms}}}{2\pi f_b}
	\sim
	\SIrange{2.5e-14}{6.2e-14}{C},
	\\
	N_0
	& =
	\frac{Q_0}{e}
	\sim
	\numrange{1.5e5}{4e5}\,.
	\label{eq:eic-charge-amplitude-required}
	\end{split}
\end{equation}
Here $N_0$ is the corresponding number of electrons displaced at the RF frequency. The corresponding capacitive voltage scale is
\begin{equation}
	V_{\text{RF}}^{\text{rms}}
	=
	\frac{I_{\text{RF}}^{\text{rms}}}{2\pi f_b C_{\text{eff}}}
	\sim 
	\SIrange{20}{400}{V}\,.
	\label{eq:rf-voltage-required-eic}
\end{equation}
This is a realistic voltage scale for a bunch-driven C-strip and holder circuit. The estimate shows that only a small RF charge displacement is needed to produce the current scale required for strip-end heating.

\subsection{Time-averaged force and mechanical response}
\label{sec:time-averaged-force-mechanical-response}

The C strip cannot respond mechanically to individual bunches. Its transverse displacement is therefore governed by the bunch-averaged force. A minimal slow-response model is
\begin{equation}
	m_{\text{eff}}\ddot{x}
	+
	\Gamma\dot{x}
	+
	kx
	=
	F_{\text{eff}}(x)\,,
	\label{eq:slow-mechanical-response-equation}
\end{equation}
where $m_{\text{eff}}$ is the effective moving mass, $\Gamma$ is a damping coefficient, and $k$ is the effective transverse spring constant. The effective force is the average over one bunch period,
\begin{equation}
	F_{\text{eff}}(x)
	=
	\frac{1}{\tau_b}
	\int_{\text{one bunch period}}
	F(x,t)\,dt\,,
	\label{eq:bunch-averaged-force}
\end{equation}
where $\tau_b$ is the bunch spacing listed in Table~\ref{tab:beam_parameters_rhic_eic}.

For an electrostatic charge force, $F(x,t)=q(t)E(x,t)$, so that
\begin{equation}
	F_{\text{eff}}(x)
	=
	\left\langle q(t)E(x,t)\right\rangle\,.
	\label{eq:averaged-charge-force}
\end{equation}
Using the charge decomposition in Eq.~\eqref{eq:charge-decomposition}, this becomes
\begin{equation}
	F_{\text{eff}}(x)
	=
	q_0\left\langle E(x,t)\right\rangle
	+
	\left\langle \tilde{q}(t)E(x,t)\right\rangle\,.
	\label{eq:averaged-force-offset-and-rf}
\end{equation}
The first term is the force from a slowly relaxing charge offset. The second term is the force from bunch-synchronous charge displacement. It can be nonzero if the charge motion is correlated with the bunch field.

The relevant mechanical response time is much longer than the bunch spacing,
\begin{equation}
	\tau_{\text{mech}}
	\gg
	\tau_b\,.
	\label{eq:mechanical-time-much-longer-than-bunch}
\end{equation}
Thus the C strip responds to the time-averaged force, not to individual bunch passages. A pulsed attractive force can therefore appear as a quasi-static attraction in video observations. A residual charge offset is not strictly required for such an averaged force, although it naturally gives the longer-range $1/r$ behavior discussed above.

\subsection{Resulting force picture}
\label{sec:resulting-force-picture}

The force estimates indicate that the most plausible leading explanation for the observed transverse target deformation is electrostatic charge displacement or charge redistribution on the strip. The required force scale is set by the visible displacement of the mechanically compliant carbon strip, as estimated in Eq.~\eqref{eq:mechanical-force-range}. The electrostatic charge estimate in Eqs.~\eqref{eq:beam-electric-field-rhic-reference}--\eqref{eq:required-elementary-charges} shows that only a small local imbalance, of order $10^5$ to $10^6$ elementary charges, is sufficient to reach the lower part of this required force range. By contrast, the isolated induced-polarization force has the correct attractive symmetry, but its approximate $1/r^3$ falloff makes it too small and too short-ranged to explain the observed deformation by itself.

\begin{figure*}[t]
	\centering
	\includegraphics[width=0.85\textwidth]{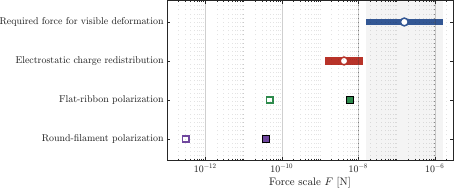}
	\caption{
		Comparison of representative force scales relevant to carbon-strip deformation.
		The upper two rows show horizontal ranges: the force scale required for visible deformation and the force expected from electrostatic charge redistribution corresponding to $10^5$ to $10^6$ elementary charges at $r=\SI{5}{mm}$.
		The open circles mark representative central values for these two ranges.
		The lower two rows show isolated induced-polarization force estimates for flat-ribbon and round-filament approximations.
		Filled squares correspond to $r=\SI{1}{mm}$, and open squares correspond to $r=\SI{5}{mm}$.
	}
	\label{fig:carbon_strip_force_scale_comparison}
\end{figure*}

Figure~\ref{fig:carbon_strip_force_scale_comparison} summarizes this hierarchy. The required force for visible deformation is shown first as the reference scale. The electrostatic charge-redistribution force is shown next because it is the leading candidate mechanism for the beam-directed deformation. The flat-ribbon and round-filament polarization estimates are included only as comparison scales. The comparison supports the interpretation that the observed motion is more naturally associated with net charge, charge redistribution, and RF-driven currents in the strip-holder system than with the static polarization force of an isolated neutral strip alone.

The transverse attraction and the end glow are related, but they are not the same local effect. The transverse attraction reflects a net or time-averaged force on the flexible strip, whereas the end glow reflects localized power dissipation near the strip attachments. Both effects can arise from the same underlying electrical response of the target: beam-induced charging, charge redistribution, RF-current drive, and a resistance profile that changes with beam exposure.

The resulting picture is therefore coupled electrical, mechanical, and thermal. Charge displacement or redistribution can provide the transverse force needed for visible deformation, while RF currents driven through nonuniform resistance can deposit heat preferentially near the strip ends. This combined interpretation is consistent with the observed beam-directed deformation, the strip-end glow before direct beam interception, the sensitivity to RF conditions, and the strong history dependence of the carbon-strip resistance.

\section{Carbon-target temperature estimates for proton and $^{3}\mathrm{He}$ operation in the EIC injector chain and collider}
\label{sec:carbon-estimates-protons_light-ions}

The preceding sections developed the carbon-strip response model for RHIC and EIC benchmark cases. Here the same framework is applied to Booster and AGS CNI polarimeter targets for proton and $^{3}\mathrm{He}^{++}$ operation. The material parameters are kept identical to the graphitic/CVD baseline used in Sec.~\ref{sec:thermal-response-carbon-strip} and Appendix~\ref{sec:carbon-material-properties}. The thermal-model inputs and carbon-strip dimensions are summarized in Table~\ref{tab:booster_ags_thermal_model_inputs}. The beam parameters used to construct the heating source are listed in Table~\ref{tab:booster_ags_beam_conditions}.

\begin{table*}[htb]
	\centering
	\small
	\renewcommand{\arraystretch}{1.15}
	\setlength{\tabcolsep}{6pt}
	\begin{tabular}{llll}
		\hline
		\textbf{Quantity} &
		\textbf{Symbol} &
		\textbf{Value} &
		\textbf{Unit} \\
		\hline
		Carbon scenario &
		-- &
		graphitic/CVD carbon &
		-- \\
		Density &
		$\rho$ &
		\num{1700} &
		\si{kg.m^{-3}} \\
		Specific heat capacity &
		$c_p$ &
		\num{750} &
		\si{J.kg^{-1}.K^{-1}} \\
		Thermal conductivity &
		$\kappa$ &
		\num{5} &
		\si{W.m^{-1}.K^{-1}} \\
		Emissivity &
		$\varepsilon$ &
		\num{0.85} &
		-- \\
		Booster strip widths &
		$w_{\text{Booster}}$ &
		\num{50}, \num{75}, \num{125} &
		\si{\micro\meter} \\
		Booster strip thickness &
		$t_{\text{Booster}}$ &
		\num{30} &
		\si{\nano\meter} \\
		Booster active length &
		$\ell_{\text{Booster}}$ &
		\num{50} &
		\si{\milli\meter} \\
		AGS strip widths &
		$w_{\text{AGS}}$ &
		\num{50}, \num{75}, \num{125} &
		\si{\micro\meter} \\
		AGS strip thickness &
		$t_{\text{AGS}}$ &
		\num{30} &
		\si{\nano\meter} \\
		AGS active length &
		$\ell_{\text{AGS}}$ &
		\num{50} &
		\si{\milli\meter} \\
		Holder temperature &
		$T_0$ &
		\num{300} &
		\si{K} \\
		Effective heat-source scale &
		$\eta_{\text{heat}}$ &
		from Sec.~\ref{sec:heat-retained-target} &
		-- \\
		\hline
	\end{tabular}
	\caption{Thermal-model input parameters for the Booster and AGS carbon-strip temperature estimates. The material values are kept identical to the graphitic/CVD carbon baseline used for the RHIC and EIC calculations in Table~\ref{tab:thermal_model_inputs}. The AGS target dimensions follow target-production information from D.~Steski~\cite{SteskiPrivateCommunication}.}
	\label{tab:booster_ags_thermal_model_inputs}
\end{table*}

\subsection{Carbon sublimation and vacuum mass loss}
\label{sec:carbon-sublimation-vacuum-mass-loss}

The temperature at which carbon loss becomes relevant for the present target problem is not a single sharp material constant. Near one atmosphere, graphite sublimation is commonly quoted near $T\simeq\SI{3915}{K}$~\cite{Stewart2015CermetReview}. Abrahamson found that the one-atmosphere graphite sublimation temperature inferred from graphite sublimation, carbon-arc, and crystallite-erosion data lies in the range $\SI{3895}{K}$--$\SI{4020}{K}$~\cite{Abrahamson1974GraphiteSublimation}. These values provide useful thermodynamic reference points, but they are not by themselves the relevant operational criterion for a nanometer-scale carbon strip in high or ultra-high vacuum.

In vacuum, the carbon mass-loss rate is controlled by the vapor pressure $P_{\mathrm{vap}}(T)$ at the hot target surface and by the resulting molecular flux away from the surface. The vapor pressure and heat of sublimation of graphite were measured and analyzed by Brewer, Gilles, and Jenkins~\cite{Brewer1948GraphiteVaporPressure}. Graphite-target erosion estimates for accelerator applications have also used vapor-pressure parameterizations and high-vacuum mass-loss measurements, for example in the ORNL graphite sublimation study by Haines and Tsai~\cite{HainesTsai2002GraphiteSublimation}. Related accelerator-target studies of thin carbon stripper foils have shown that sublimation in vacuum, thermal stresses, and fatigue can all contribute to beam-induced foil damage~\cite{Tahir2014CarbonFoilStripper}.

The corresponding kinetic-theory description is the Hertz-Knudsen evaporation flux~\cite{OHanlon2003VacuumTechnology},
\begin{equation}
	J
	=
	\alpha
	\frac{P_{\mathrm{vap}}(T)-P_{\mathrm{amb}}}
	{\sqrt{2\pi m k_{\mathrm{B}}T}},
	\label{eq:hertz-knudsen-flux}
\end{equation}
where $J$ is the particle flux away from the surface, $\alpha$ is the evaporation or accommodation coefficient, $m$ is the mass of the evaporating species, $k_{\mathrm{B}}$ is the Boltzmann constant, and $P_{\mathrm{amb}}$ is the ambient partial pressure of the same vapor species. In high or ultra-high vacuum the ambient partial pressure of carbon vapor is negligible compared with the vapor pressure at the hot surface, so that $P_{\mathrm{amb}}\simeq 0$.

It is useful to rewrite Eq.~\eqref{eq:hertz-knudsen-flux} in terms of the positive mass-loss rate per unit surface area,
\begin{equation}
	\Gamma_m(T)
	=
	\frac{1}{A}\frac{\dd m_{\mathrm{loss}}}{\dd t}
	=
	Jm ,
	\label{eq:mass-loss-rate-area}
\end{equation}
so that
\begin{equation}
	\Gamma_m(T)
	\simeq
	\alpha
	P_{\mathrm{vap}}(T)
	\sqrt{\frac{m}{2\pi k_{\mathrm{B}}T}} .
	\label{eq:hertz-knudsen-mass-loss-uhv}
\end{equation}
Equivalently, using the molar mass $M$ and the gas constant $R$, Eq.~\eqref{eq:hertz-knudsen-mass-loss-uhv} may be written as
\begin{equation}
	\Gamma_m(T)
	\simeq
	\alpha
	P_{\mathrm{vap}}(T)
	\sqrt{\frac{M}{2\pi R T}} .
	\label{eq:hertz-knudsen-molar-form}
\end{equation}

The corresponding surface recession speed is
\begin{equation}
	v_{\mathrm{sub}}(T)
	=
	\frac{\Gamma_m(T)}{\rho},
	\label{eq:sublimation-recession-speed}
\end{equation}
where $\rho$ is the carbon density. For a temperature history $T(t)$, the removed thickness is then
\begin{equation}
	\Delta h_{\mathrm{sub}}
	=
	\int v_{\mathrm{sub}}(T(t))\,dt ,
	\label{eq:sublimation-removed-thickness}
\end{equation}
and the relevant dimensionless damage measure for a strip of thickness $t$ is
\begin{equation}
	f_{\mathrm{sub}}
	=
	\frac{\Delta h_{\mathrm{sub}}}{t}.
	\label{eq:sublimation-fractional-loss}
\end{equation}
Thus, in the present application the relevant limit is not a thermodynamic sublimation temperature alone, but an allowed fractional thickness loss over the target exposure time.

\begin{figure*}[t]
	\centering
	\begin{subfigure}[t]{0.483\textwidth}
		\centering
		\includegraphics[width=\linewidth]{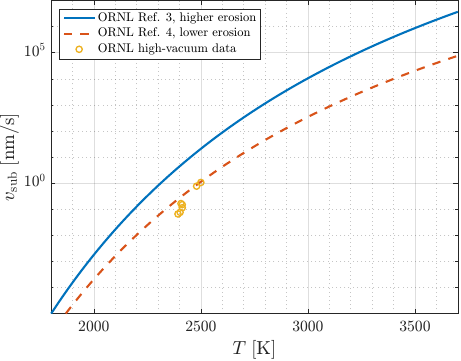}
		\caption{Graphite recession speed in vacuum.}
		\label{fig:graphite_sublimation_recession_speed}
	\end{subfigure}\hfill
	\begin{subfigure}[t]{0.49\textwidth}
		\centering
		\includegraphics[width=\linewidth]{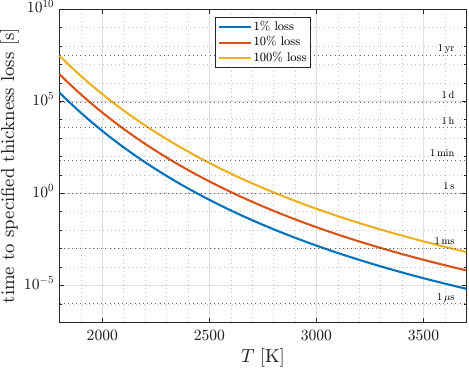}
		\caption{Time scale for specified fractional thickness loss.}
		\label{fig:graphite_sublimation_time_to_loss}
	\end{subfigure}
	\caption{Graphite sublimation estimates in vacuum. Panel (a) shows the recession speed obtained from two ORNL graphite vapor-pressure parameterizations and from ORNL high-vacuum mass-loss data converted to an effective recession speed~\cite{HainesTsai2002GraphiteSublimation}. Panel (b) shows the time required to remove specified fractions of a \SI{50}{\nano\meter} strip.}
	\label{fig:graphite_sublimation_margin}
\end{figure*}

Figure~\ref{fig:graphite_sublimation_margin} summarizes the sublimation estimate used below. The vapor-pressure input is taken from the two graphite vapor-pressure parameterizations compared in the ORNL graphite sublimation study~\cite{HainesTsai2002GraphiteSublimation}. The ORNL report measured high-vacuum graphite mass loss over the range $\SI{2393}{K}$--$\SI{2500}{K}$ and found that the measured erosion rates were closer to the lower-erosion vapor-pressure curve. All quantitative estimates below therefore use the lower-erosion curve as the data-driven reference; the higher-erosion curve is shown only for context, since it overpredicts the measured rates and is not used to set the loss estimates.

The time-to-loss curves in Fig.~\ref{fig:graphite_sublimation_margin}(b) express the same recession-speed scale as an exposure-time limit for a \SI{50}{\nano\meter} strip. They show how long the strip can remain at a fixed temperature before sublimation removes 1\%, 10\%, or 100\% of its thickness, using the lower-erosion curve from Fig.~\ref{fig:graphite_sublimation_margin}(a). This panel is therefore the link between the temperature estimates and the fractional-loss criterion $\Delta h_{\mathrm{sub}}/t$.

\begin{figure}[t]
	\centering
	\includegraphics[width=\linewidth]{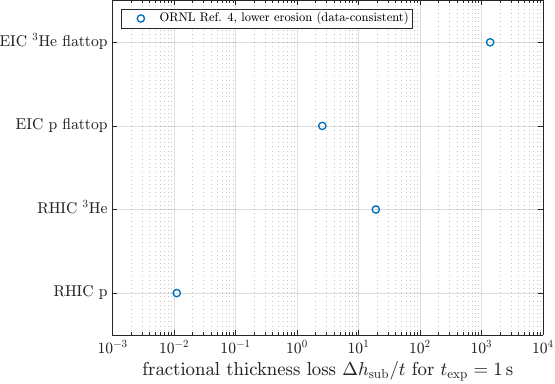}
	\caption{Fractional carbon-strip loss after a \SI{1}{\second} exposure at the calculated peak temperatures.}
	\label{fig:graphite_sublimation_case_loss_1s}
\end{figure}

Figure~\ref{fig:graphite_sublimation_case_loss_1s} compares the calculated temperature cases on a common \SI{1}{\second} exposure scale. The horizontal axis directly gives the fractional thickness loss $\Delta h_{\mathrm{sub}}/t$ for a \SI{50}{\nano\meter} strip, evaluated at the peak temperature of each case. The lower and upper estimates use the two ORNL vapor-pressure parameterizations shown in Fig.~\ref{fig:graphite_sublimation_margin}(a). This one-second normalization is a severity scale; it does not imply that a moving target remains at $T_{\max}$ for \SI{1}{\second}.

\subsection{Stopping-power treatment}

The direct heating term is based on the stopping power of the beam particles in carbon. The general Bethe-Bloch formalism is standard and is not repeated here; see, for example, Leo~\cite{Leo1994}. In the present implementation the stopping power is evaluated from the projectile charge state, mass, and velocity listed in Table~\ref{tab:booster_ags_beam_conditions}, using the same carbon material assumptions as in the rest of this paper.

This is a useful cross-check of the heating normalization used earlier. In Sec.~\ref{sec:energy-loss-carbon-strip}, the energy deposited in the carbon strip was introduced as the input to the retained-heat calculation. The retained fraction was then defined in Sec.~\ref{sec:heat-retained-target}, and the resulting heat source was propagated through the thermal-response model in Sec.~\ref{sec:thermal-response-carbon-strip}. For the proton cases considered here, the Bethe-Bloch calculation reproduces the earlier proton stopping-power input to good accuracy. The Booster and AGS proton estimates therefore use the same direct-heating scale as the benchmark calculations, but now obtained self-consistently within the calculation.

The same stopping-power implementation is then applied to $^{3}\mathrm{He}^{++}$ by changing the projectile charge state, mass, and velocity. This avoids introducing a separate normalization for the light-ion cases and keeps the comparison between proton and $^{3}\mathrm{He}^{++}$ operation internally consistent. Tabulated $^{3}\mathrm{He}$ stopping powers are not available over the required energy range (the ASTAR database covers only $^{4}\mathrm{He}$), so the Bethe-Bloch stopping power is evaluated directly for the light-ion cases; for protons it reproduces the tabulated normalization behind the retained-heat scale of Table~\ref{tab:retained-heat-scale} to within one to two percent, which accounts for the small offset between the $\eta_{\text{heat}}$ values quoted there and the per-case heating used in the temperature estimates.

\begin{table*}[htb]
	\centering
	\renewcommand{\arraystretch}{1.15}
	\setlength{\tabcolsep}{4pt}
	\begin{tabular}{lllcccc}
		\hline
		\textbf{Quantity} &
		\textbf{Symbol} &
		\textbf{Unit} &
		\multicolumn{2}{c}{\textbf{Booster}} &
		\multicolumn{2}{c}{\textbf{AGS}} \\
		\cline{4-7}
		&
		&
		&
		\textbf{p} &
		\textbf{$^{3}$He$^{++}$} &
		\textbf{p} &
		\textbf{$^{3}$He$^{++}$} \\
		\hline
		
		Rest mass &
		$m$ &
		\si{MeV}/$c^2$ &
		\num{938.272} &
		\num{2808.391} &
		\num{938.272} &
		\num{2808.391} \\
		
		Charge state &
		$z$ &
		$e$ &
		\num{1} &
		\num{2} &
		\num{1} &
		\num{2} \\
		
		Ring circumference &
		$C$ &
		\si{m} &
		\num{201.78} &
		\num{201.78} &
		\num{807.12} &
		\num{807.12} \\
		
		Injection kinetic energy &
		$T_{\text{inj}}$ &
		\si{MeV} (\si{MeV/u}) &
		\num{200} &
		\num{6} (\num{2}) &
		\num{1500} &
		\num{2496} (\num{832}) \\
		
		Extraction kinetic energy &
		$T_{\text{extr}}$ &
		\si{MeV} (\si{MeV/u}) &
		\num{1500} &
		\num{2496} (\num{832}) &
		\num{23500} &
		\num{46113} (\num{15371}) \\
		
		Injection magnetic rigidity &
		$B\rho_{\text{inj}}$ &
		\si{T.m} &
		\num{2.150} &
		\num{0.306} &
		\num{7.506} &
		\num{7.506} \\
		
		Extraction magnetic rigidity &
		$B\rho_{\text{extr}}$ &
		\si{T.m} &
		\num{7.506} &
		\num{7.506} &
		\num{81.46} &
		\num{81.46} \\
		
		Injection velocity factor &
		$\beta_{\text{inj}}$ &
		1 &
		\num{0.5662} &
		\num{0.0653} &
		\num{0.9230} &
		\num{0.8484} \\
		
		Extraction velocity factor &
		$\beta_{\text{extr}}$ &
		1 &
		\num{0.9230} &
		\num{0.8484} &
		\num{0.99926} &
		\num{0.99835} \\
		
		Injection revolution frequency &
		$f_{\text{rev,inj}}$ &
		\si{MHz} &
		\num{0.841} &
		\num{0.097} &
		\num{0.343} &
		\num{0.315} \\
		
		Extraction revolution frequency &
		$f_{\text{rev,extr}}$ &
		\si{MHz} &
		\num{1.371} &
		\num{1.260} &
		\num{0.371} &
		\num{0.371} \\
		
		Bunches per cycle &
		$n_b$ &
		1 &
		\num{1} &
		\num{1}\textsuperscript{a} &
		\num{1} &
		\num{1}\textsuperscript{a} \\
		
		Particles per bunch &
		$N_b$ &
		$10^{11}$ &
		\num{1.5} &
		\num{1.0} (\num{2.5} design) &
		\num{1.5} &
		\num{1.0} (\num{2.5} design) \\
		
		Particles per cycle &
		$N_{\text{cyc}}=n_bN_b$ &
		$10^{11}$ &
		\num{1.5} &
		\num{1.0} (\num{2.5} design) &
		\num{1.5} &
		\num{1.0} (\num{2.5} design) \\
		
		Normalized rms emittance &
		$\epsilon_N$ &
		\si{\micro m} &
		\num{1.0} &
		\num{1.0} &
		\num{1.0} &
		\num{1.0} \\
		
		Horizontal beta function at target &
		$\beta_x$ &
		\si{m} &
		\num{7.75} &
		\num{7.75} &
		\num{20} &
		\num{20} \\
		
		Vertical beta function at target &
		$\beta_y$ &
		\si{m} &
		\num{7.75} &
		\num{7.75} &
		\num{20} &
		\num{20} \\
		
		Injection horizontal rms beam size &
		$\displaystyle \sigma_{x,\text{inj}}$ &
		\si{mm} &
		\num{3.36} &
		\num{10.89} &
		\num{2.89} &
		\num{3.53} \\
		
		Injection vertical rms beam size &
		$\displaystyle \sigma_{y,\text{inj}}$ &
		\si{mm} &
		\num{3.36} &
		\num{10.89} &
		\num{2.89} &
		\num{3.53} \\
		
		Extraction horizontal rms beam size &
		$\displaystyle \sigma_{x,\text{extr}}$ &
		\si{mm} &
		\num{1.80} &
		\num{2.20} &
		\num{0.88} &
		\num{1.07} \\
		
		Extraction vertical rms beam size &
		$\displaystyle \sigma_{y,\text{extr}} $ &
		\si{mm} &
		\num{1.80} &
		\num{2.20} &
		\num{0.88} &
		\num{1.07} \\
		
		Target scan speed &
		$v_t$ &
		\si{mm.s^{-1}} &
		\num{4} &
		\num{4} &
		\num{4} &
		\num{4} \\
		\hline
	\end{tabular}
	\caption{Beam-condition inputs for the Booster and AGS carbon-strip temperature estimates. The listed quantities determine the beam-heating source and the static or moving-target temperature response. The Booster values are taken from Ref.~\cite{BoosterPolarimeterTechNote}. The beam sizes are computed from Eq.\,\eqref{eq:beam-sigmas} using the normalized rms emittance and target-location beta functions listed in the table. The target scan speed $v_t$ is based on the video-derived near-beam scale summarized in Table~\ref{tab:video_observation_parameters}.}
	\label{tab:booster_ags_beam_conditions}
	\begin{minipage}{0.95\linewidth}
		\footnotesize
		\textsuperscript{a}The $^{3}$He beam is injected into the Booster as four bunches and merged into a single bunch during a momentum-hold porch before the one-bunch estimate is applied. The same one-bunch beam is then transported from Booster to AGS.
	\end{minipage}
\end{table*}

\subsection{Booster and AGS estimates}

\begin{figure*}[htbp]
	\centering
	\includegraphics[width=\textwidth]{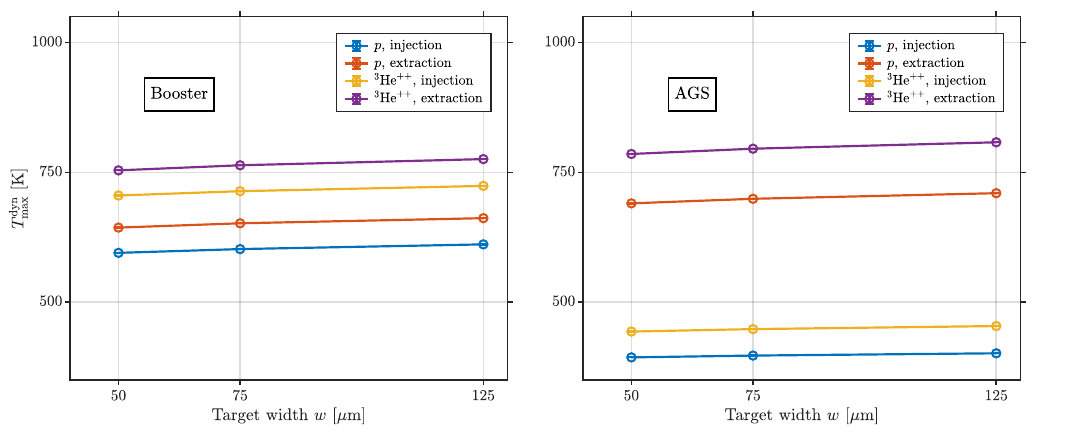}
	\caption{Calculated moving-transient peak temperature as a function of carbon-strip width for Booster and AGS CNI polarimeter targets, shown for proton and $^{3}\mathrm{He}^{++}$ beams at injection and extraction. }
	\label{fig:booster_ags_target_temperature_vs_width}
\end{figure*}

Figure~\ref{fig:booster_ags_target_temperature_vs_width} shows the calculated moving-transient peak temperature as a function of carbon-strip width for Booster and AGS proton and $^{3}\mathrm{He}^{++}$ beams at injection and extraction.

All Booster and AGS cases shown in Fig.~\ref{fig:booster_ags_target_temperature_vs_width} remain below about \SI{2000}{K} for the target-width range considered here. The largest temperatures occur for the $^{3}\mathrm{He}^{++}$ extraction cases. From the one-second normalization in Fig.~\ref{fig:graphite_sublimation_case_loss_1s}, even the worst AGS extraction case gives $\Delta h_{\mathrm{sub}}/t < 2\times 10^{-5}$ for a \SI{1}{s} exposure at the calculated peak temperature. Vacuum sublimation is therefore negligible for the Booster/AGS cases considered here.

All Booster and AGS cases shown in Fig.~\ref{fig:booster_ags_target_temperature_vs_width} remain below about \SI{2000}{K} for the target-width range considered here. The largest temperatures occur for the $^{3}\mathrm{He}^{++}$ extraction cases. In the sublimation-rate interpretation of Sec.~\ref{sec:carbon-sublimation-vacuum-mass-loss}, these injector-chain cases are well separated from the temperature range where vacuum recession becomes a dominant limitation on sub-second time scales.

The width dependence is noticeable but moderate. Over the range $w=50$ to $125\,\mu\mathrm{m}$, the calculated peak temperature changes by only a few hundred kelvin. Thus, within the present moving-target model, changing the carbon-strip width affects the thermal margin, but it is not the dominant parameter controlling the peak temperature.

\subsection{Temperature comparison and sublimation-loss scale}

The stationary centered-source estimate is denoted by $T_{\max}^{\mathrm{stat}}$, while the moving-target transient estimate is denoted by $T_{\max}^{\mathrm{dyn}}$. Table~\ref{tab:carbon_target_temperature_comparison} summarizes representative values of $T_{\max}^{\mathrm{dyn}}$ from the proton and $^{3}\mathrm{He}^{++}$ calculations. The Booster and AGS entries use the largest target width included in the scan, $w=\SI{125}{\micro\meter}$, because this gives the largest temperature for each injection or extraction case. The RHIC and EIC entries use the hottest geometry for each beam condition.

The difference between $T_{\max}^{\mathrm{stat}}$ and $T_{\max}^{\mathrm{dyn}}$ is small for the cases considered here, and is usually comparable to or smaller than the numerical convergence estimate. The dynamic value is therefore used as the representative temperature in Table~\ref{tab:carbon_target_temperature_comparison}. The numerical convergence estimate is usually below \SI{1}{K}, increasing to about \SI{6}{K} only for the narrowest EIC cooled-emittance flattop case.

\begin{table*}[htb]
	\centering
	\small
	\renewcommand{\arraystretch}{1.15}
	\setlength{\tabcolsep}{4.5pt}
	\begin{tabular}{llllc}
		\hline
		\textbf{Machine} &
		\textbf{Species} &
		\textbf{Beam condition} &
		\textbf{Target condition} &
		\textbf{$T_{\max}^{\mathrm{dyn}}$ [\si{K}]} \\
		\hline
		
		Booster &
		p &
		injection &
		$w=\SI{125}{\micro\meter}$ &
		\num{611} \\
		Booster &
		p &
		extraction &
		$w=\SI{125}{\micro\meter}$ &
		\num{662} \\
		Booster &
		$^{3}\mathrm{He}$ &
		injection &
		$w=\SI{125}{\micro\meter}$ &
		\num{724} \\
		Booster &
		$^{3}\mathrm{He}$ &
		extraction &
		$w=\SI{125}{\micro\meter}$ &
		\num{775} \\
		\hline
		AGS &
		p &
		injection &
		$w=\SI{125}{\micro\meter}$ &
		\num{401} \\
		AGS &
		p &
		extraction &
		$w=\SI{125}{\micro\meter}$ &
		\num{710} \\
		AGS &
		$^{3}\mathrm{He}$ &
		injection &
		$w=\SI{125}{\micro\meter}$ &
		\num{454} \\
		AGS &
		$^{3}\mathrm{He}$ &
		extraction &
		$w=\SI{125}{\micro\meter}$ &
		\num{808} \\
		\hline
		RHIC &
		p &
		flattop &
		vertical &
		\num{2449} \\
		RHIC &
		$^{3}\mathrm{He}$ &
		flattop reference &
		horizontal &
		\num{3111} \\
		\hline
		EIC &
		p &
		injection before cooling &
		diagonal &
		\num{1321} \\
		EIC &
		p &
		injection after cooling &
		horizontal &
		\num{1782} \\
		EIC &
		p &
		flattop large emittance &
		diagonal &
		\num{2149} \\
		EIC &
		p &
		flattop cooled emittance &
		diagonal &
		\num{2900} \\
		EIC &
		$^{3}\mathrm{He}$ &
		injection before cooling &
		diagonal &
		\num{1669} \\
		EIC &
		$^{3}\mathrm{He}$ &
		injection after cooling &
		horizontal &
		\num{2252} \\
		EIC &
		$^{3}\mathrm{He}$ &
		flattop large emittance &
		vertical &
		\num{2731} \\
		EIC &
		$^{3}\mathrm{He}$ &
		flattop cooled emittance &
		vertical &
		\num{3685} \\
		\hline
	\end{tabular}
	\caption{Representative calculated dynamic peak temperatures $T_{\max}^{\mathrm{dyn}}$ for proton and $^{3}\mathrm{He}$ carbon-target cases in the EIC injector chain and comparison collider conditions. The dynamic calculation assumes a target scan speed $v_t=\SI{4}{\milli\meter\per\second}$. For Booster and AGS, the listed rows use the largest target width in the scan, $w=\SI{125}{\micro\meter}$, which gives the largest temperature in Fig.~\ref{fig:booster_ags_target_temperature_vs_width}. For RHIC and EIC, the listed row is the hottest geometry for each beam condition.}
	\label{tab:carbon_target_temperature_comparison}
\end{table*}

\begin{figure}[htb]
	\centering
	\includegraphics[width=\columnwidth]{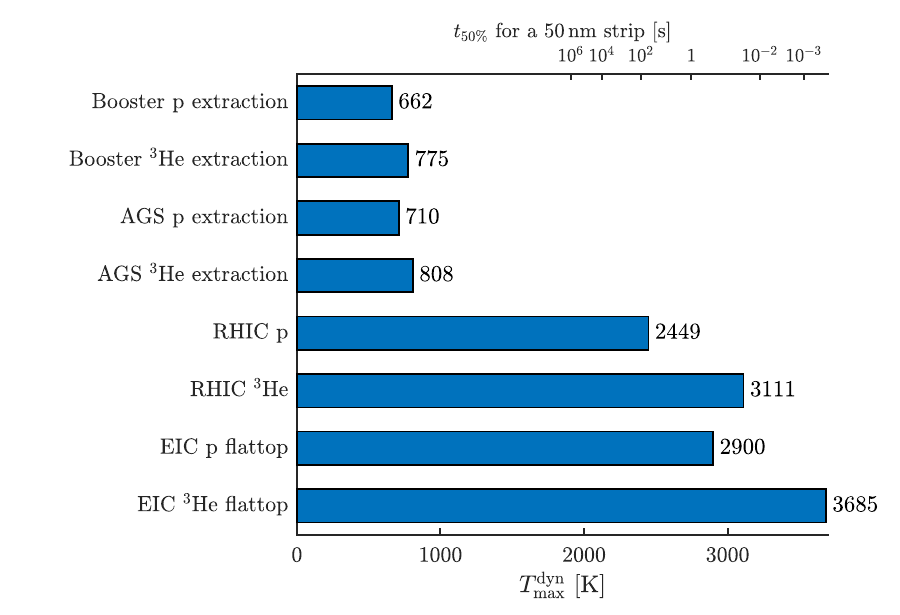}
	\caption{Representative dynamic peak temperatures for selected Booster, AGS, RHIC, and EIC carbon-target cases. The lower axis gives $T_{\max}^{\mathrm{dyn}}$, and the upper axis gives the estimated time to remove 50\% of a \SI{50}{\nano\meter} strip at the corresponding temperature.}
	\label{fig:carbon_target_thermal_risk_summary}
\end{figure}

The representative cases from Table~\ref{tab:carbon_target_temperature_comparison} are summarized in Fig.~\ref{fig:carbon_target_thermal_risk_summary}. The lower axis shows the calculated dynamic peak temperature, while the upper axis converts the same temperature into the estimated time $t_{50\%}$ required to remove 50\% of a \SI{50}{\nano\meter} strip by vacuum sublimation. This conversion uses the lower-erosion graphite recession-speed curve from Fig.~\ref{fig:graphite_sublimation_recession_speed} and should be interpreted as a damage-time scale at fixed temperature, not as a separate thermal calculation.

The Booster and AGS cases remain below about \SI{810}{K} for the assumptions used here. The largest injector-chain value is obtained for AGS $^{3}\mathrm{He}$ extraction, with $T_{\max}^{\mathrm{dyn}}\simeq\SI{808}{K}$. As discussed above, Fig.~\ref{fig:graphite_sublimation_case_loss_1s} gives $\Delta h_{\mathrm{sub}}/t<2\times10^{-5}$ for this worst Booster/AGS case on a one-second exposure scale, so vacuum sublimation is negligible for the injector-chain cases considered here.

The RHIC proton entry provides a calculated comparison to the established RHIC pC operating regime, not a temperature benchmark, since the target temperature was not measured directly. The RHIC $^{3}\mathrm{He}$ entry is shown only as a flattop reference calculation. The EIC proton flattop cooled-emittance case, with $T_{\max}^{\mathrm{dyn}}\simeq\SI{2900}{K}$, is already close to a destructive sublimation regime for a \SI{50}{\nano\meter} strip: on the one-second scale of Fig.~\ref{fig:graphite_sublimation_case_loss_1s}, the estimated fractional thickness loss is of order unity. The EIC $^{3}\mathrm{He}$ cooled-emittance flattop case remains the most severe case shown, with $T_{\max}^{\mathrm{dyn}}\simeq\SI{3685}{K}$ and an estimated one-second loss far above the full strip thickness.

These results indicate that the relevant design criterion is the integrated sublimation loss over the actual target temperature history, not a single fixed sublimation temperature.

Target thinning alone is not an efficient mitigation for the EIC $^{3}\mathrm{He}$ stress case. In the thin-target limit, the deposited energy, thermal mass, and conductive cross section all scale approximately with the target thickness, so reducing the thickness does not lower the peak temperature in proportion to the removed material. At the same time, a fixed sublimation recession depth corresponds to a larger fractional thickness loss for a thinner strip. Width, scan speed, and reduced dwell time are therefore more relevant operational control parameters than thickness reduction alone.

\subsection{Target-lifetime scale and extrapolation to EIC operation}
\label{sec:target_lifetime_scale_eic_extrapolation}

\subsubsection{RHIC lifetime scale}

The RHIC p calculation provides an empirical scale check for the present model. For the RHIC p reference case, Fig.~\ref{fig:graphite_sublimation_case_loss_1s} gives the one-second fractional sublimation loss
\begin{equation}
	\left(f_{\mathrm{sub}}\right)_{\SI{1}{s}}
	\simeq 0.011 ,
\end{equation}
where \(f_{\mathrm{sub}}\) is defined in Eq.~\eqref{eq:sublimation-fractional-loss}. This value uses the lower-erosion graphite recession curve of Haines and Tsai~\cite{HainesTsai2002GraphiteSublimation}, taken from the saturation-pressure data of Darken and Gurry~\cite{Darken1953PhysicalChemistry}. Their high-vacuum weight-loss measurements fall at or slightly below this curve, while the higher-erosion curve from the Union Carbide handbook~\cite{UnionCarbide1969Graphite} overpredicts the measured rates by more than an order of magnitude and is therefore not used here. The spread between the two curves reflects the mix of monatomic and polyatomic carbon vapor species, which the measurements resolve toward the lower curve.

Using the video-derived beam-visible interval from Table~\ref{tab:video_observation_parameters},
\begin{equation}
	\Delta t_{\mathrm{beam}}\simeq \SI{1.5}{s},
\end{equation}
a full measurement cycle, consisting of one pass into the beam and one pass out of the beam, corresponds to
\begin{equation}
	\Delta t_{\mathrm{cycle}}
	\simeq
	2\,\Delta t_{\mathrm{beam}}
	\simeq
	\SI{3}{s}.
\end{equation}
A conservative estimate evaluates this loss at the peak-intensity temperature \(T_{\max}^{\mathrm{dyn}}\simeq\SI{2449}{K}\), which gives the fractional sublimation loss per measurement cycle
\begin{equation}
	\left(
	f_\mathrm{sub}
	\right)_\mathrm{cycle}
	\simeq
	\Delta t_\mathrm{cycle}
	\left(
	f_\mathrm{sub}
	\right)_{\SI{1}{s}}
	\simeq
	0.033 ,
\end{equation}
so that beam-center sublimation alone would remove about \(3\%\) of a \SI{50}{\nano\meter} strip per full in-and-out cycle. The corresponding number of cycles for \(50\%\) and complete thickness loss is
\begin{equation}
	\begin{split}
	N_{\mathrm{cycle}}(50\%)
	& \simeq
	\frac{0.5}{(f_\mathrm{sub})_{\mathrm{cycle}}}
	\simeq
	15 ,
	\\
	N_{\mathrm{cycle}}(100\%)
	& \simeq
	\frac{1}{(f_\mathrm{sub})_{\mathrm{cycle}}}
	\simeq
	30 .
	\end{split}
\end{equation}
Because \(f_{\mathrm{sub}}\) depends exponentially on temperature, this peak-intensity estimate is a lower bound on target lifetime: the beam intensity is not constant during a store but decays over the fill, so a polarization measurement performed at injection, mid-store, and end-of-store samples a cycle-averaged intensity below the start-of-store peak. For a beam lifetime of order \SI{16}{h} and a typical \SI{8}{h} store, the cycle-averaged intensity is about \(80\%\) of the peak. Since the peak heat source scales with beam intensity and the radiatively limited peak temperature scales as \(T\propto I^{1/4}\), this reduces the effective temperature to about \SI{2316}{K}, lowering \(f_{\mathrm{sub}}\) by roughly a factor of eight and giving an upper bound
\begin{equation}
	\begin{split}
	N_{\mathrm{cycle}}(50\%)
	& \simeq
	120 ,
	\\
	N_{\mathrm{cycle}}(100\%)
	& \simeq
	240 .
	\end{split}
\end{equation}
The estimated target lifetime from beam-center sublimation therefore spans roughly \(15\)--\(120\) cycles for \(50\%\) thickness loss, or \(30\)--\(240\) cycles for complete removal, with the lower bound corresponding to continuous peak-intensity operation and the upper bound to the realistic cycle-averaged intensity.

This range brackets the RHIC target-lifetime statistics reported by Steski et al.~\cite{Steski2018TargetLifetime}. In that study, the average target lifetime was about 27 measurements in 2013, while targets equipped with rounded metal fins in 2015 reached an average lifetime of about 142 measurements, with one fin-equipped target surviving 396 measurements. The un-finned 2013 lifetime sits near the lower, peak-intensity bound, while the finned 2015 lifetimes approach and somewhat exceed the upper, cycle-averaged bound. The comparison should not be interpreted as a sublimation-only failure model. Instead, it shows that the present beam-center thermal calculation reproduces the RHIC p target-lifetime scale at the order-of-magnitude level. The improvement from metal fins, which lifts the realized lifetime from the lower toward and beyond the upper bound, then validates the broader coupled picture developed here: beam-center heating sets a destructive thermal scale, while RF-induced end heating, electrostatic deformation, and electromagnetic boundary conditions near the target ends strongly affect the realized target lifetime.

\subsubsection{EIC extrapolation}

Applying the same conservative cycle estimate to the EIC p flattop cooled-emittance case gives a much more restrictive result. For
\begin{equation}
	T_{\max}^{\mathrm{dyn}}\simeq\SI{2900}{K},
\end{equation}
Fig.~\ref{fig:graphite_sublimation_case_loss_1s} gives, on the lower-erosion curve,
\begin{equation}
	\left(f_\mathrm{sub}\right)_{\SI{1}{s}}
	\simeq
	2.6 .
\end{equation}
For a RHIC-like full cycle with
\begin{equation}
	\Delta t_{\mathrm{cycle}}\simeq\SI{3}{s},
\end{equation}
this corresponds to
\begin{equation}
	\left(f_\mathrm{sub}\right)_{\mathrm{cycle}}
	\simeq
	7.7 .
\end{equation}
Thus, already on the data-driven lower-erosion curve, the loss corresponds to more than one full strip thickness per RHIC-like measurement cycle. The EIC p flattop cooled-emittance case therefore cannot be treated as a straightforward extrapolation of RHIC p operation at the same dwell time.

The EIC \(^{3}\mathrm{He}\) cooled-emittance flattop case is more severe. With
\begin{equation}
	T_{\max}^{\mathrm{dyn}}\simeq\SI{3685}{K},
\end{equation}
Fig.~\ref{fig:graphite_sublimation_case_loss_1s} indicates a one-second fractional loss well above \(10^3\). The corresponding peak temperature, \SI{3685}{K}, now lies about \SI{35}{K} above the adopted graphite sublimation reference \(T_{\mathrm{sub}}=\SI{3650}{K}\). Under the present assumptions, this carbon-strip operating point is likely not viable with RHIC-like scan times. It would require a major reduction of dwell time near the beam core, a reduction of the peak heat source, a different target geometry or operating concept, or a new target technology.

\subsubsection{Beam-loss penalty of increasing the carbon cross section}
\label{sec:beam-loss-penalty}

The preceding comparison shows that the EIC proton and especially the EIC $^{3}\mathrm{He}$ cooled-emittance flattop cases would require much longer sublimation-loss times than obtained with the nominal carbon strip. One possible approach would be to increase the amount of carbon in the target. This option, however, must be evaluated against the stored-beam loss caused by the target itself. The relevant machine quantity is not the raw number of target passages, but the luminosity-weighted beam loss caused by the polarimeter measurements during a fill.

For this estimate we use the same roll-averaged rectangular-strip geometry as in Sec.~IV~E. The roll-averaged chord length is given in Eq.~\eqref{eq:roll-averaged-chord-nominal}. For the nominal RHIC/EIC carbon strip,
\begin{equation}
	A_{\mathrm{C},0}
	=
	w_0 t_0
	=
	(10\,\mu\mathrm{m})(50\,\mathrm{nm})
	=
	0.50\,\mu\mathrm{m}^2
	\, .
	\label{eq:nominal-RHIC-cross-sectional-area}
\end{equation}
The corresponding effective projected scan width in the roll-averaged
description is
\begin{equation}
	w_{\mathrm{eff}}
	=
	\frac{A_{\mathrm{C},0}}
	{\langle \ell_{\mathrm{chord}}\rangle_{\psi}}
	\simeq
	2.62\,\mu\mathrm{m}
	\, .
\end{equation}
Thus, a stored beam particle sees fewer target encounters than in a flat
$50\,\mathrm{nm}$ bookkeeping picture, but each encounter has a larger
roll-averaged carbon chord. For a target speed $v_t$ and revolution frequency
$f_{\mathrm{rev}}$, the number of roll-averaged target encounters during one
complete passage is
\begin{equation}
	N_{\mathrm{hit,roll}}
	=
	f_{\mathrm{rev}}
	\frac{w_{\mathrm{eff}}}{v_t}
	\, .
\end{equation}
Using $f_{\mathrm{rev}}$ (Table~\ref{tab:beam_parameters_rhic_eic}) and
$v_t$ (Table~\ref{tab:video_observation_parameters}) gives
\begin{equation}
	N_{\mathrm{hit,roll}}
	\simeq
	51.2
	\, .
\end{equation}
The carbon areal density for one roll-averaged encounter is
\begin{equation}
	n_{\mathrm{C}}^{\mathrm{roll}}
	=
	\frac{
		\rho
		\langle \ell_{\mathrm{chord}}\rangle_{\psi}
		N_{\mathrm{A}}
	}
	{A_{\mathrm{mol}}}
	\, ,
\end{equation}
where $A_{\mathrm{mol}}=12.01\,\mathrm{g\,mol}^{-1}$ is the molar mass of carbon. With $\rho=1.7\,\mathrm{g\,cm}^{-3}$ (Table~\ref{tab:carbon_properties}) this gives
\begin{equation}
	n_{\mathrm{C}}^{\mathrm{roll}}
	\simeq
	1.63\times 10^{18}\,\mathrm{cm}^{-2}
	\, .
\end{equation}
The single-passage removal exponent is
\begin{equation}
	\mu_{\mathrm{pass}}
	=
	N_{\mathrm{hit,roll}}\,
	n_{\mathrm{C}}^{\mathrm{roll}}\,
	\sigma_{\mathrm{loss}}=
	f_{\mathrm{rev}}
	\frac{\rho A_{\mathrm{C}}}{v_t}
	\frac{N_{\mathrm{A}}}{A_{\mathrm{mol}}}
	\sigma_{\mathrm{loss}}
	\, .
	\label{eq:single-passage-removal-exponent}
\end{equation}
The second form shows explicitly that the integrated beam-removal probability for a full passage scales with the carbon cross section $A_{\mathrm{C}}$. We define the carbon cross-section scale factor
\begin{equation}
	F_A
	=
	\frac{A_{\mathrm{C}}}{A_{\mathrm{C},0}}
	\, ,
	\label{eq:carbon-cross-section-scale-factor}
\end{equation}
where $A_{\mathrm{C}}$ is the carbon cross-sectional area transported through the beam and $A_{\mathrm{C},0}$ is the nominal RHIC reference value from Eq.~\eqref{eq:nominal-RHIC-cross-sectional-area}.
One obtains, for the nominal RHIC/EIC scan speed,
\begin{equation}
	\mu_{\mathrm{pass}}
	=
	8.33\times 10^{-8}\,
	\sigma_{\mathrm{loss}}[\mathrm{mb}]\,
	F_A
	\, .
	\label{eq:single-passage-removal-exponent-numeric}
\end{equation}
The actual fractional beam loss in one passage is
\begin{equation}
	\left(\frac{\Delta N}{N}\right)_{\mathrm{pass}}
	=
	1-\exp\left(-\mu_{\mathrm{pass}}\right)
	\, .
	\label{eq:single-passage-removal-probability}
\end{equation}

Recent RHIC operation used, to good approximation, three in-out carbon-target measurements during a fill: near the beginning, near the middle, and near the end. The end-of-fill scan is useful for the polarization and polarization lifetime determination, but it has essentially no remaining luminosity penalty. For a constant unperturbed luminosity during the fill, the luminosity-weighted penalty from the beginning and middle measurements is therefore
\begin{equation}
	\Delta_{\mathcal{L}}
	=
	1
	-
	\frac{1}{2}
	\left[
	\exp\left(-2\mu_{\mathrm{pass}}\right)
	+
	\exp\left(-4\mu_{\mathrm{pass}}\right)
	\right]
	\, .
	\label{eq:luminosity-weighted-target-loss}
\end{equation}
For small target-induced losses this reduces to
\begin{equation}
	\Delta_{\mathcal{L}}
	\simeq
	3\mu_{\mathrm{pass}}
	\, .
\end{equation}
If the luminosity decreases during the fill, the middle-scan weight is smaller than one half, so this expression is a conservative estimate of the integrated luminosity penalty.

Table~\ref{tab:beam_loss_thick_targets} gives representative values. The carbon-area scale factors $F_A$ are order-of-magnitude factors obtained by asking how much carbon would be required to make the sublimation $t_{50\%}$ of the hotter cases comparable to the RHIC proton reference in Fig.~\ref{fig:carbon_target_thermal_risk_summary}. They are not proposed design values.

\begin{table*}[htb]
	\centering
	\caption{Beam-removal estimate for nominal and survival-equivalent carbon targets. The survival-equivalent EIC entries have their carbon cross section scaled up by the factor $F_A$ so that the target provides a comparable number of measurements to the RHIC p nominal strip before being consumed, offsetting the much faster EIC erosion. The table shows that the luminosity penalty of this approach is prohibitive. The carbon cross-section scale factor $F_A$ is defined in Eq.~\eqref{eq:carbon-cross-section-scale-factor}. The single-passage removal exponent $\mu_{\mathrm{pass}}$ is defined in Eq.~\eqref{eq:single-passage-removal-exponent}, with the numerical form used here given in Eq.~\eqref{eq:single-passage-removal-exponent-numeric}. The removal per pass is defined in Eq.~\eqref{eq:single-passage-removal-probability}. The luminosity-weighted loss per fill, $\Delta_{\mathcal{L}}$, is defined in Eq.~\eqref{eq:luminosity-weighted-target-loss}, assuming three in-out measurements per fill with scans near the beginning, middle, and end. The end scan is assigned no future-luminosity penalty.}
	\label{tab:beam_loss_thick_targets}
	\begin{ruledtabular}
		\footnotesize
		\setlength{\tabcolsep}{4pt}
		\begin{tabular}{p{4.1cm}cccc}
			Case
			& $F_A$
			& $\sigma_{\mathrm{loss}}$
			& Removal/pass
			& $\Delta_{\mathcal{L}}$/fill \\
			& [1]
			& [mb]
			& [1]
			& [1] \\
			\hline
			RHIC $p$, nominal strip
			& 1
			& $250$--$300$
			& $(2.1$--$2.5)\times 10^{-5}$
			& $(6.3$--$7.5)\times 10^{-5}$ \\
			EIC $p$ flattop, survival-equiv.
			& $2.1\times 10^{2}$
			& $250$--$300$
			& $0.44$--$0.52\%$
			& $1.3$--$1.6\%$ \\
			RHIC $^{3}\mathrm{He}$, survival-equiv.
			& $2.0\times 10^{3}$
			& $1000$
			& $15.3\%$
			& $38.5\%$ \\
			EIC $^{3}\mathrm{He}$ flattop, survival-equiv.
			& $1.7\times 10^{5}$
			& $1000$
			& $\simeq 100\%$
			& $\simeq 100\%$ \\
		\end{tabular}
	\end{ruledtabular}
\end{table*}

The nominal RHIC proton case gives a luminosity-weighted loss of only about $6\times 10^{-5}$ to $8\times 10^{-5}$ per fill for the historical scan pattern, consistent with RHIC operation. By contrast, increasing the carbon cross section enough to recover a RHIC-like sublimation $t_{50\%}$ for the EIC proton cooled-emittance flattop case would already cost of order one percent of the integrated luminosity per fill. For the corresponding $^{3}\mathrm{He}$ cases, the beam removal becomes very large or essentially complete. Increasing the target cross section is therefore not a viable mitigation path for the severe EIC light-ion cases.

\subsection{Implications for target-survival mitigation}
\label{sec:target_survival_mitigation}

The preceding comparison shows that EIC carbon-target survival cannot be judged by a single temperature marker. The relevant quantity is the fractional sublimation loss \(f_{\mathrm{sub}}\), defined in Eq.~\eqref{eq:sublimation-fractional-loss} from the removed thickness \(\Delta h_{\mathrm{sub}}\) in Eq.~\eqref{eq:sublimation-removed-thickness}. Mitigation must therefore either reduce the temperature history \(T(t)\), reduce the dwell time at high temperature, suppress additional nonlocal failure mechanisms, or replace the target technology. The main options are as follows.

\begin{enumerate}\item \textbf{Reduce target dwell time.}
	
	Faster motion through the beam core does not reduce the instantaneous peak heat source, but it reduces the exposure time and therefore the fractional sublimation loss \(f_{\mathrm{sub}}\). To first approximation,
	\begin{equation}
		f_\mathrm{sub}
		\propto
		\frac{1}{v_t}.
	\end{equation}
	The video analysis in Table~\ref{tab:video_observation_parameters} gives a near-beam speed scale of $v_t\simeq\SI{4}{mm.s^{-1}}$. The EIC p flattop cooled-emittance estimate in Sec.~\ref{sec:target_lifetime_scale_eic_extrapolation} shows that an order-of-magnitude extrapolation to $v_t\simeq\SI{4}{m.s^{-1}}$ would reduce the dwell-time contribution to the sublimation loss by about a factor $10^3$, bringing the EIC p case much closer to the RHIC p loss scale.
	
	The mechanical implication of such a speed can be checked with a simple symmetric acceleration-deceleration estimate. If the target reaches an average insertion speed $v_{\mathrm{avg}}$ over a representative travel distance $s$, then
	\begin{equation}
		a \simeq \frac{4v_{\mathrm{avg}}^2}{s}.
	\end{equation}
	Using $s\simeq\SI{30}{mm}$ and $v_{\mathrm{avg}}\simeq\SI{4}{m.s^{-1}}$ gives
	\begin{equation}
		a \simeq \SI{2.1e3}{m.s^{-2}}.
	\end{equation}
	This is still far below the simple tensile-inertia breaking scale
	\begin{equation}
		a_{\mathrm{break}}
		\sim
		\frac{\sigma_{\mathrm{break}}}{\rho \ell}
		\sim
		\SI{1e6}{m.s^{-2}},
	\end{equation}
	obtained for the order-of-magnitude choices  $\sigma_{\mathrm{break}}=\SI{100}{MPa}$, $\rho=\SI{1700}{kg.m^{-3}}$, and $\ell=\SI{50}{mm}$. The practical limits are therefore more likely actuator dynamics, jerk, vibration, slack-strip modes, holder/contact stresses, and beam-induced forces than the self-inertia of the strip itself.	
	
	\item \textbf{Compensate reduced dwell time with detector acceptance.} Shorter dwell time reduces the number of detected events per pass. Since fast beam-polarization and beam-profile feedback remain required for machine operation, this loss must be compensated by larger detector area, larger angular acceptance, improved readout efficiency, or a modified scan strategy.
	
	\item \textbf{Change beam  optics.}
	The peak particle flux and projected beam size at the target enter directly into the thermal source. Larger beam size, lower peak flux, reduced bunch intensity, or different optics at the polarimeter location would reduce the target temperature. These parameters are constrained by  accelerator operation  and cannot be assumed as polarimeter-only mitigation.
	
	\item \textbf{Suppress RF-induced end heating.}
	The RHIC target-survival study~\cite{Steski2018TargetLifetime} shows that modifying the electromagnetic environment near the target ends can strongly improve lifetime. For EIC operation, alumina holders or other low-conductivity holder concepts will suppress RF-induced end heating and target-end failures, as discussed in Sec.~\ref{sec:effective-rf-end-heating}. This addresses end losses and holder-connection failures, but not the beam-center sublimation constraint quantified above.
	
	\item \textbf{Develop alternative target or diagnostic technologies.}
	For the most severe EIC cases, improved carbon-strip operation may no longer be sufficient. The EIC p flattop cooled-emittance case may still be extended through faster scans, larger detector acceptance, and improved electromagnetic boundary conditions, but only as an interim measure rather than a scalable solution. By contrast, the EIC \(^{3}\mathrm{He}\) cooled-emittance flattop case appears beyond practical carbon-strip operation under the present assumptions: on the data-driven lower-erosion curve, Fig.~\ref{fig:graphite_sublimation_case_loss_1s} gives a one-second fractional loss of about \(1400\), i.e., of order \(10^3\) strip thicknesses. For still heavier polarized species, such as \(^{6,7}\mathrm{Li}\), conventional carbon-strip targets are therefore very unlikely to remain viable. New target or diagnostic technologies will be needed, such as renewable or continuously replenished targets, pellet-style targets, moving tape or ribbon targets, alternative low-\(Z\) materials, or non-invasive spin-diagnostic approaches.
	
	Among these, pellet-based concepts developed in other contexts are directly relevant. In the charged-particle EDM storage-ring program, pellet beam-sample extraction has been proposed as a polarimetry route that does not rely on a scanned, consumable strip~\cite[Appendix\,K]{CPEDM:2019nwp}. Adapted to proton-carbon polarimetry, the corresponding target would be a continuously replenished stream of carbon pellets, i.e., diamond micropellets of order \(10\,\mu\mathrm{m}\), so that erosion or sublimation of any individual pellet no longer sets the target lifetime. Optical pellet-tracking systems based on lasers and fast line-scan cameras can in principle recover the transverse position of individual pellets at the time of interaction~\cite{Khoukaz2011PANDAInternalTargets,Pyszniak2014PelletTracking}, preserving the beam-profile and polarization-profile information  presently obtained by scanning a carbon strip, although tracking at this pellet size has not yet been demonstrated.
\end{enumerate}

These options should not be viewed as independent fixes. For the EIC p flattop cooled-emittance case, a viable carbon-strip upgrade would likely require a combined approach: shorter dwell time, detector acceptance matched to faster scans, and suppression of RF-induced end heating at the target ends. For the most severe light-ion cases, especially cooled-emittance \(^{3}\mathrm{He}\), the present estimates indicate that conventional carbon-strip operation is unlikely to remain viable under RHIC-like scan conditions. These cases should therefore motivate parallel development of new target or spin-diagnostic technologies rather than relying on incremental improvements of the existing carbon-strip concept alone.

\section{Conclusions}
\label{sec:conclusion}

This paper developed a physical response model for ultra-thin carbon-strip targets in relativistic bunched beams, motivated by RHIC observations and by the requirements of future EIC hadron polarimetry. The model separates the beam-target interaction geometry, particle flux, stopping-power energy loss, secondary-electron escape, retained heat, target motion, thermal response, wakefield coupling, RF-induced strip-end heating, resistance evolution, electrostatic forces, and slack-strip mechanical response. This organization provides a common framework for comparing RHIC operation with EIC proton and light-ion benchmark cases.

The RHIC observations show that direct beam heating at the strip center is not sufficient to describe the full target response. The observed strip-end glow before direct beam interception requires an additional RF or electrical heating channel, while the beam-directed deformation requires a transverse force acting on the flexible strip. The force estimates indicate that electrostatic charge displacement or charge redistribution on the strip-holder system is the most plausible leading explanation for the observed target attraction. The isolated induced-polarization force has the correct attractive symmetry, but its short range and smaller magnitude make it unlikely to explain the full deformation pattern by itself.

The thermal and sublimation calculations show that the Booster and AGS cases considered here are well separated from vacuum-sublimation loss on the time scales relevant for target scans. For the worst injector-chain case, AGS \(^{3}\mathrm{He}\) extraction, the one-second fractional-loss estimate remains below \(2\times10^{-5}\). The RHIC p calculation provides an important empirical scale check: the estimated beam-center sublimation loss corresponds to roughly \(15\)--\(120\) full measurement cycles for 50\% thickness loss, or \(30\)--\(240\) cycles for complete removal by sublimation alone, where the lower bound assumes continuous peak-intensity operation and the upper bound the realistic cycle-averaged intensity. This range brackets the RHIC target-lifetime statistics. This agreement should not be interpreted as a sublimation-only failure model. Rather, it shows that the beam-center thermal calculation gives the correct order-of-magnitude damage scale, while the observed improvement from metal fins demonstrates that RF-induced end heating and electromagnetic boundary conditions near the target ends strongly affect the realized lifetime.

The EIC extrapolation is substantially more restrictive. For the EIC p flattop cooled-emittance case, the calculated peak temperature gives a one-second fractional-loss range of \(2.6\)--\(71\). A RHIC-like in-and-out measurement cycle would therefore remove more than one strip thickness even on the lower-erosion estimate. This case cannot be treated as a straightforward extrapolation of RHIC p operation at the same dwell time. The EIC \(^{3}\mathrm{He}\) cooled-emittance flattop case is more severe: on the data-driven lower-erosion curve, the one-second fractional-loss estimate is of order \(10^3\), placing it beyond practical conventional carbon-strip operation under RHIC-like scan conditions.

Several mitigation paths follow directly from the model. Reducing the dwell time near the beam core is the most direct way to reduce the integrated fractional loss \(f_{\mathrm{sub}}\), but the shorter dwell time must be compensated by larger detector acceptance, improved readout, or modified scan strategy to preserve fast machine feedback. A simple acceleration estimate indicates that much faster target motion is not excluded by carbon-strip self-inertia alone; practical limits are more likely to come from the actuator, holder, contact region, jerk, vibration, slack-strip modes, and beam-induced forces. RF-induced end heating must also be suppressed, for example through improved holder geometry and low-conductivity holder concepts, but this addresses a different failure channel from beam-center sublimation. EIC proton operation may remain feasible with aggressive optimization of dwell time, detector acceptance, and electromagnetic boundary conditions. However, these measures appear to be mitigation strategies rather than a scalable long-term solution. For cooled-emittance \(^{3}\mathrm{He}\), incremental improvements of the existing carbon-strip concept are unlikely to be sufficient, and simply increasing the carbon cross section is closed off by the resulting luminosity penalty (Sec.~\ref{sec:beam-loss-penalty}).

The result also has implications beyond protons and \(^{3}\mathrm{He}\). The broader polarized-ion science case for the EIC anticipates future operation with polarized deuteron, \(^{3}\mathrm{He}\), and eventually \(^{6,7}\mathrm{Li}\) beams. At fixed velocity, the dominant stopping-power scaling for a fully stripped ion is approximately
\begin{equation}
	\frac{\dd E}{\dd x}
	\propto
	z_{\mathrm{ion}}^2 ,
\end{equation}
apart from the weaker logarithmic dependence in the Bethe-Bloch expression. Thus, relative to protons, fully stripped \(^{3}\mathrm{He}^{++}\) gives an energy-loss scale larger by about a factor of four per intercepted ion, while fully stripped \(^{6,7}\mathrm{Li}^{3+}\) gives a scale larger by about a factor of nine. Relative to \(^{3}\mathrm{He}^{++}\), the lithium case is therefore larger by approximately
\begin{equation}
	\frac{
		(\dd E/\dd x)_{{}^{6,7}\mathrm{Li}^{3+}}
	}{
		(\dd E/\dd x)_{{}^{3}\mathrm{He}^{++}}
	}
	\simeq
	\frac{9}{4}
	\simeq
	2.25 .
\end{equation}
Since the cooled-emittance \(^{3}\mathrm{He}\) benchmark already appears beyond RHIC-like carbon-strip operation, lithium beams with comparable intercepted ion flux and beam size would be still more demanding.

The main conclusion is therefore twofold. First, the RHIC experience can be understood only as a coupled thermal, RF, electrical, and mechanical target-response problem, not as local stopping-power heating alone. Second, and most important for the EIC, conventional scanning carbon-strip polarimetry does not extrapolate to the most demanding light-ion cases: the cooled-emittance \(^{3}\mathrm{He}\) benchmark already lies beyond practical operation, and prospective lithium beams are more demanding still. Even for EIC proton flattop operation, viability requires an aggressively optimized system with reduced dwell time, sufficient detector acceptance, and suppressed RF-induced end heating, and such measures are at best a short-term expedient, not a scalable solution. 

The clear implication is that the fast-scan carbon-strip measurements of the transverse polarization profile and the polarization lifetime must be provided by a different technique for the EIC light-ion program. Possible directions, examined further in Sec.~\ref{sec:target_survival_mitigation}, include pellet-style or otherwise replenished targets, moving tape or ribbon targets, alternative low-\(Z\) materials, and non-invasive spin-diagnostic approaches.

\bibliographystyle{apsrev4-2}
\bibliography{carbon_strip_references_v2}

\appendix

\section{Carbon material properties}
\label{sec:carbon-material-properties}

	\begin{table*}[tp]
	\centering
	\scriptsize
	\renewcommand{\arraystretch}{1.12}
	\setlength{\tabcolsep}{3.5pt}
	\begin{tabular}{p{2.8cm} c c p{3.6cm} p{3.6cm} p{3.7cm}}
		\toprule
		\textbf{Property} &
		\textbf{Symbol} &
		\textbf{Unit} &
		\makecell{\textbf{Amorphous}\\\textbf{carbon (a-C)}} &
		\makecell{\textbf{Sputtered or}\\\textbf{evaporated carbon}} &
		\makecell{\textbf{Graphitic or}\\\textbf{CVD carbon}} \\
		\midrule
		Thermal conductivity &
		$\kappa$ &
		$\mathrm{W\,m^{-1}\,K^{-1}}$ &
		0.1--1.0 &
		0.3--1.5 &
		2--10 \\
		Density &
		$\rho$ &
		$\mathrm{kg\,m^{-3}}$ &
		2000\,\cite{NIST_PSTAR} &
		2200\,\cite{robertson2002diamond} / 1900\,\cite{robertson2002diamond} &
		1700\,\cite{NIST_PSTAR} \\
		Specific heat &
		$c_p$ &
		$\mathrm{J\,kg^{-1}\,K^{-1}}$ &
		700--900 &
		700--900 &
		700--800 \\
		Emissivity &
		$\varepsilon$ &
		-- &
		0.75--0.95 &
		0.8--0.95 &
		0.8--0.95 \\
		Young's modulus &
		$E$ &
		$\mathrm{GPa}$ &
		100--250\,\cite{robertson2002diamond,Taylor2003ThinCarbonStress} &
		100--250\,\cite{robertson2002diamond,Taylor2003ThinCarbonStress} &
		strongly anisotropic; effective value model dependent\,\cite{Blakslee1970GraphiteElasticConstants,Taylor2003ThinCarbonStress} \\
		Poisson ratio &
		$\nu$ &
		-- &
		0.15--0.35\,\cite{Cho2005TetrahedralCarbonMEMS,robertson2002diamond} &
		0.15--0.35\,\cite{Cho2005TetrahedralCarbonMEMS,robertson2002diamond} &
		0.1--0.3; anisotropic for crystalline graphite\,\cite{Blakslee1970GraphiteElasticConstants} \\
		Coefficient of thermal expansion &
		$\alpha_T$ &
		$10^{-6}\,\mathrm{K^{-1}}$ &
		1--10\,\cite{Taylor2003ThinCarbonStress} &
		1--10\,\cite{Taylor2003ThinCarbonStress} &
		anisotropic; few to several tens\,\cite{Taylor2003ThinCarbonStress,Blakslee1970GraphiteElasticConstants} \\
		Intrinsic residual film stress &
		$\sigma_{\text{res}}$ &
		$\mathrm{MPa}$ &
		process dependent; $10^2$--$10^3$ typical for dense films\,\cite{robertson2002diamond,Taylor2003ThinCarbonStress} &
		process dependent; $10^2$--$10^3$ typical for dense films\,\cite{Taylor2003ThinCarbonStress,Huff2022ResidualStressReview} &
		process and substrate dependent\,\cite{Taylor2003ThinCarbonStress} \\
		Effective axial stress in mounted strip &
		$\sigma_{\text{eff}}$ &
		$\mathrm{MPa}$ &
		model parameter &
		model parameter &
		model parameter \\
		Carbon-strip electrical resistance &
		$R_{\text{strip}}$ &
		$\mathrm{M\Omega}$ &
		-- &
		200--800 before beam exposure; $\sim$1 after beam exposure\,\cite{Steski2014CarbonMicroRibbons} &
		lower after graphitization; model dependent\,\cite{Steski2014CarbonMicroRibbons} \\
		LET, protons at 250 GeV &
		LET &
		$\mathrm{MeV\,cm^2\,g^{-1}}$ &
		1.871\,\cite{NIST_PSTAR} &
		1.871\,\cite{NIST_PSTAR} &
		1.881\,\cite{NIST_PSTAR} \\
		$\text{LET}_{\text{SI}}$ &
		$\text{LET}_{\text{SI}}$ &
		$\si{J.m^{-1}}$ &
		\num[round-mode=places, round-precision=2]{5.9947e-11} &
		\num[round-mode=places, round-precision=2]{6.5942e-11}/\num[round-mode=places, round-precision=2]{5.6949e-11} &
		\num[round-mode=places, round-precision=3]{5.1227e-11} \\
		\bottomrule
	\end{tabular}
	\caption{Material parameters used for the thermal, electrical, stopping-power, and mechanical modeling of carbon-strip polarimeter targets. The entries are representative ranges for thin carbon films and depend strongly on deposition method, heat treatment, microstructure, hydrogen content, and beam exposure history. The graphitic/CVD carbon column is used as the internally consistent baseline material set for the RHIC and EIC calculations in the present work. For comparison with operational CNI target estimates, D.~Steski quotes a bulk density of approximately $\rho_C=\SI{1.81}{g.cm^{-3}}$ for the deposited amorphous-carbon target material~\cite{SteskiPrivateCommunication}. The Poisson ratio $\nu=-\epsilon_{\perp}/\epsilon_{\parallel}$ describes the transverse contraction under axial strain and is relevant for plate or shell stress estimates, while the present slack-string estimate mainly depends on the effective tension. The intrinsic residual film stress $\sigma_{\text{res}}$ is a material and fabrication property, whereas the effective axial stress $\sigma_{\text{eff}}$ of the mounted suspended strip is treated as a separate model parameter because cutting, transfer, twisting, mounting, heating, graphitization, and relaxation can strongly reduce the tension actually carried by the free strip. The quantity $\text{LET}_{\text{SI}}$ is the line energy loss per incident proton after multiplying the tabulated mass stopping power by the density of the corresponding carbon material; see Sec.~\ref{sec:energy-loss-carbon-strip}.}
	\label{tab:carbon_properties}
\end{table*}

The carbon-strip response depends on thermal, electrical, and mechanical properties. These properties are strongly process dependent. Amorphous carbon lacks long-range crystalline order, while sputtered and evaporated carbon films are produced by physical vapor deposition (PVD). In PVD, material is removed from a solid carbon source and deposited on the substrate, either by energetic-ion sputtering or by thermal evaporation. Chemical vapor deposition (CVD), in contrast, uses gaseous precursors that react or decompose at the substrate surface.

Only carbon film types relevant for thin, solid, high-temperature carbon microstructures are included in Table~\ref{tab:carbon_properties}. Bulk graphite is not used as a default material because its thermal and elastic properties are strongly anisotropic and not directly transferable to thin, disordered, twisted, or beam-modified carbon strips. Diamond-like carbon (DLC), carbon black, and hydrogenated amorphous carbon (a-C:H) are also not used as default materials. DLC properties depend strongly on the sp$^3$/sp$^2$ bonding fraction and hydrogen content, carbon black is a porous nanoparticle aggregate, and a-C:H is generally less suitable for high-temperature operation because hydrogen modifies both thermal stability and transport properties.

The mechanical values in Table~\ref{tab:carbon_properties} should be interpreted as material-property ranges. They are not identical to the effective axial tension of a mounted carbon-strip target. For the mechanical model developed below, the relevant parameter is the effective strip tension,
\begin{equation}
	T_{\text{eff}}
	=
	\sigma_{\text{eff}} A\,,
	\qquad
	A=wt\,,
\end{equation}
where $\sigma_{\text{eff}}$ is the effective axial stress in the suspended strip. This effective stress can be much smaller than the intrinsic residual stress of the deposited film because cutting, transfer, twisting, mounting, heating, graphitization, and partial relaxation can all reduce the tension actually carried by the free strip.

A fixed sublimation temperature is not included in Table~\ref{tab:carbon_properties}. The values often quoted for graphite sublimation depend on pressure, surface state, and the operational criterion used to define material loss. In vacuum, the relevant quantity for the present carbon-strip problem is the recession speed \(v_{\mathrm{sub}}(T)\) and the resulting fractional thickness loss \(\Delta h_{\mathrm{sub}}/t\), discussed in Sec.~\ref{sec:carbon-sublimation-vacuum-mass-loss}.

\end{document}